================ **TRANSLATIONS OF PUBLISHED ARTICLES** ================

# Theoretical and Experimental Investigations of DNA Open States

**Shigaev A.S.*, Ponomarev O.A.†, Lakhno V.D.‡**

*Institute of Mathematical Problems of Biology RAS – the Branch of Keldysh Institute of Applied Mathematics, Russian Academy of Sciences, Pushchino, Russia*

***Abstract.*** Literature data on the properties of DNA open states are reviewed and analyzed. These states are formed as a result of strong DNA fluctuations and have a great impact on a number of biochemical processes; among them is charge transfer in DNA, for example. A comparative analysis of experimental data on the kinetics and thermodynamics of DNA open states for a wide temperature range was carried out. Discrepancies between the results of various experiments have been explained. Three types of DNA open states are recognized based on their differences in thermodynamic properties and other characteristics. Besides, an up-to-date definition of the term "open state" is given. A review is carried out for simple mathematical models of DNA in most of which the state of one pair is described by one or two variables. The main problems arising in theoretical investigations of heterogeneous DNA in the framework of models of this level are considered. The role of each group of models in interpretation of experimental data is discussed. Special consideration is given to the studies of the transfer and localization of the nucleotide pairs oscillations' energy by mechanical models. These processes are shown to play a key role in the dynamics of a heterogeneous duplex. Their theoretical interpretation is proven to be very important for the development of modern molecular biology and biophysics. The main features of the theoretical approaches are considered which enabled describing various experimental data. Prospects of the models' development are described, particular details of their optimization are suggested, and possible ways of modernization of some experimental techniques are discussed.

*shials@yandex.ru
†olegpon36@mail.ru
‡lak@impb.ru

## INTRODUCTION

The main function of DNA is the storage of genetic information of living things at molecular level. DNA molecule has a complex structure. It is composed of two polymer chains that hold together by hydrogen bonds, or H-bonds. This structure is usually called a duplex. Each chain consists of nucleotides bound by phosphodiester bridges. This type of bond is the linkage between the 3' carbon atom of one nucleotide's sugar molecule and the 5' carbon atom of another deoxyribose. The structure of a single chain is shown in Scheme 1.

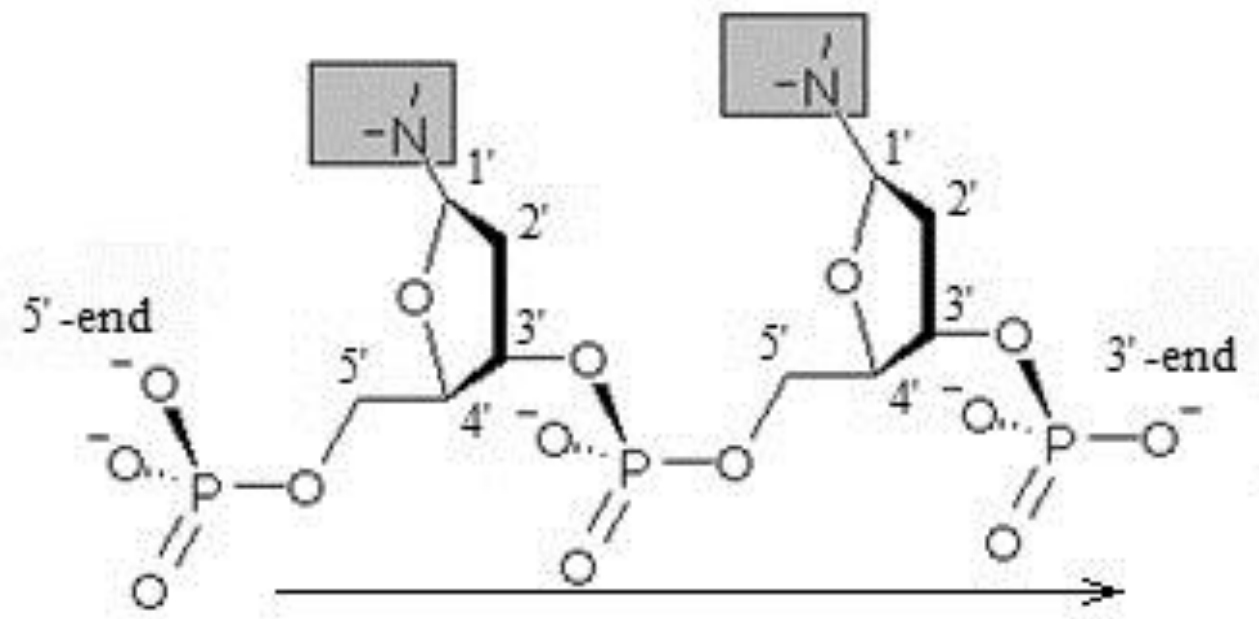

**Scheme 1**. Structure of DNA sugar-phosphate backbone. Digits indicate numbers of carbon atoms of desoxyribose. The arrow indicates the direction of the DNA chain – from 5'-end to 3'-end. Gray rectangles indicate DNA nitrogenous bases: only nitrogen atoms (letters N) attached to atom C-1' of the furanose ring via a glycoside bond are shown.

The part of a polymer molecule shown in Scheme 1 is called a sugar-phosphate backbone. A single DNA chain is widely accepted to be directed from 5'-end to 3'-end. A nucleotide sequence in this direction is called a DNA primary structure. Typical DNA contains four nucleobase types – adenine, thymine, guanine and cytosine. In our paper they will be denoted as A, T, G and C correspondingly.

**Scheme 2**. Interaction of DNA bases according to the complementarity principle. Complementary H-bonds are indicated by dashed lines. As for the sugar-phosphate backbone, only atoms C-1' and C-2' of the furanose rings of desoxyribose and oxygen atoms are shown (in gray rectangles).

Binding of single chains into a duplex occurs via a specific hydrogen binding of adenine with thymine and guanine with cytosine. The principle of this binding is called a complementarity. The corresponding H-bonds are called complementary bonds. A complementary interaction of DNA bases is presented in Scheme 2.

DNA chains are bound in an antiparallel arrangement: a base at 5'-end of one chain is complementarily bound to a base at 3'-end of another chain. A DNA duplex is called a secondary structure. Complementarily bound nucleotide pairs shown in the scheme are called AT- and GC-pairs, respectively. In what follows we will refer to complementary H-bonds simply as to H-bonds.

Besides H-bonds, stability of the secondary structure of a duplex is provided by one more type of a non-covalent binding. These are the so-called stacking interactions, or stacking. This weak interaction arises between neighboring bases of one chain. Arrangement of H-bonds and stacking interactions is shown in Scheme 3.

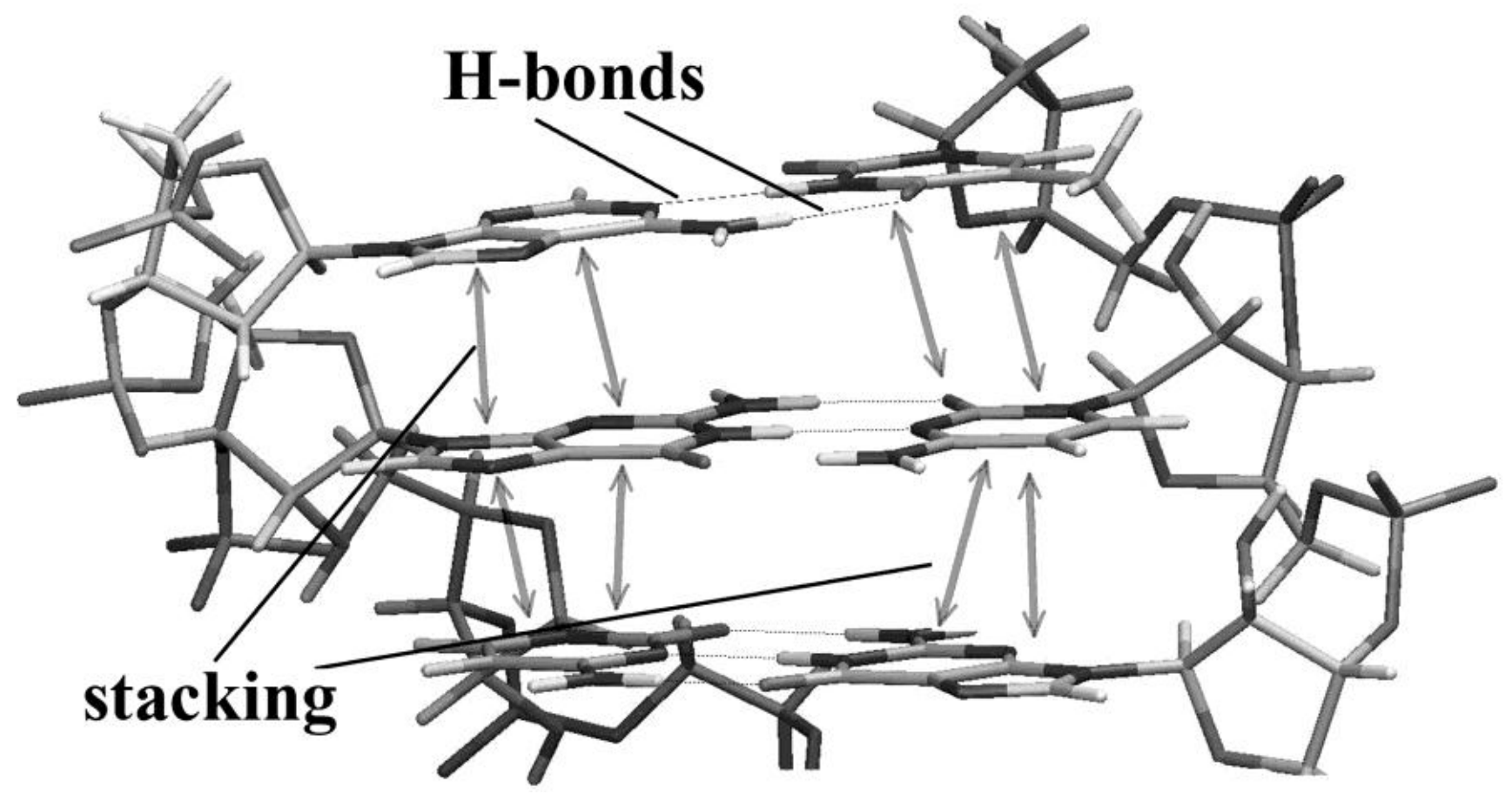

**Scheme 3**. Spatial arrangement of H-bonds and stacking interactions in DNA duplex.

As is seen from Scheme 3, the planes of neighboring bases are arranged nearly in parallel and remind a stack of coins. Therefore the structure of a single chain of a duplex is usually called a “stack”. The name of “stacking interactions” arises from this fact too.

According to quantum-mechanical calculations, the main contribution into the free energy of stacking is made by the so-called aromatic interactions [1–3]. These bonds arise due to partial delocalization of π-electrons, which takes place as a result of overlapping of p-orbitals when nitrogenous bases come in contact [4]. The density of this contact depends on the electrostatic attraction, dipole-quadrupole forces and hydrophobic effects. Nevertheless, all these forces play a supporting role and mainly determine the bases orientation. As a rule, for each pair of heterocycles, there is an optimal orientation (see [5, 6]). The role of stacking interactions in the formation and sustaining of the duplex structure is considered in detail in reviews [7–9].

The chains form a right-handed helix whose structure is shown in diagram 4. The letters S and P denote sugar residues and phosphodiester bridges. A complete circle of a helix includes 10 basepairs in a crystal and 10.5 basepairs in a water solution [10]. However, in a living cell a duplex is often under the influence of an external stress which can change the length of its coils [11]. In this case DNA is said to be supercoiled. If the direction of the external force coincides with that of the helix the supercoiling is said to be positive. In this case the length of a new coil is less than 10.5 basepairs. In the opposite case the supercoiling is said to be negative.

The entire structure which maintains all the bonds is called a native DNA. Covalent bonds in the sugar-phosphate backbone are rather strong – their energy averages out 250 kJ/mol [12]. These bonds can be broken only by exposure to ionizing radiation, action of free radicals or some modifying agents. In the absence of reactive oxygen species a one-chain molecule can survive many hours of boiling up to 200°C without breaking off covalent bonds.

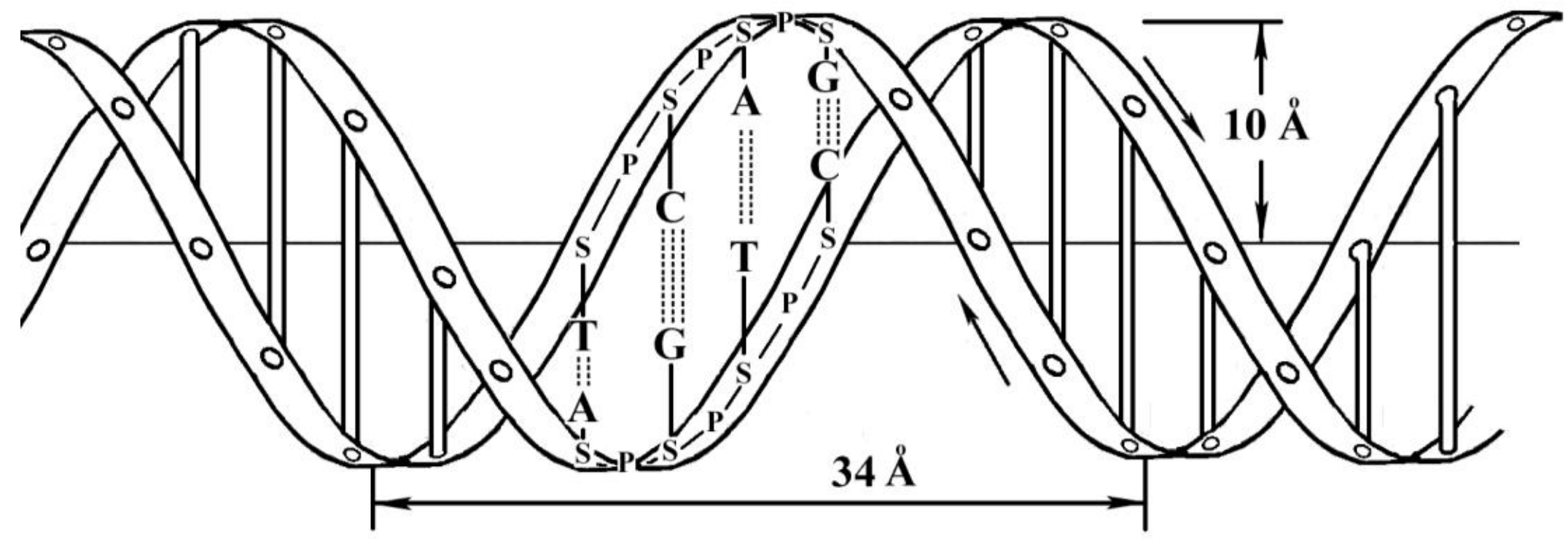

**Scheme 4**. Helical structure of a typical right-handed DNA.

Non-covalent interactions sustaining the secondary structure of the duplex are stable only at moderate temperatures. DNA over 100 basepairs in length dissociates into individual chains in the temperatures range of 70–112 °C. The dissociation temperature depends on the duplex's length and nucleotide composition. For short molecules, the dissociation temperatures are still lower. Moreover, H-bonds and stacking interactions are sensitive to a decrease in the ionic strength of a solution and extreme pH values. Besides, the spectrum of substances capable of breaking the secondary structure of duplex is rather wide.

The term "DNA denaturation" denotes a decay of all complementary H-bonds and the main part of stacking interactions; herewith all the covalent interaction are preserved. The simplest way to break the DNA secondary structure is to heat the DNA solution. For this reason the denaturation is also referred to as melting. In the course of melting DNA dissociates into two individual chains having disordered structures. Therefore the term "denaturation" has one more synonym – "helix-coil transition".

During melting DNA passes through an intermediate state when it consists of denaturated regions neighboring with native ones. The process of melting is shown in Scheme 5 [13].

DNA melting takes place within some temperature range whose boundaries depend on a lot of factors. In this temperature range breaking of H-bonds results in the local unwinding of the chains. The unwinding is usually accompanied by decay of stacking. Therefore for DNA states formed in the course of a helix-coil transition, the terms "local denaturation regions" or "denaturated regions" are appropriate. In literature these areas are often referred to as local DNA unwinding, loops, "denaturation bubbles" or simply "bubbles".

Nevertheless, H-bonds can open even under conditions when most of noncovalent interactions in the DNA molecule remain unbroken. Transient breakings of H-bonds can take place both in an individual basepair and in several neighboring basepairs. In the latter case the DNA duplex usually (locally) separate. Strictly speaking, a fragment where the chains are separated is by no means always a denaturated fragment. In most one-chain helices at low temperatures and normal ionic strength of the DNA solution the bases are bound by stacking interactions. Therefore in the case of a short-lived divergence of the chains stacking can partially persist or reassociate. This problem is discussed in detail in Chapter 5.

It follows from the definition of denaturation that the name "denaturated fragments" is not exactly suitable for DNA fragments with broken H-bonds which are formed at moderate temperatures. For such fragments, a more appropriate term is "DNA open states" – it is a more general term. In what follows, *by the "DNA open state" we will mean any transformation of the duplex's fragment arising as a result of complete or partial breaking of complementary H-bonds in one or several neighboring nucleotide pairs which makes imino protons participating in the formation of these bonds accessible for the molecules of the solution*. As it will be shown in Chapter 5, just this definition is in agreement with most experimental data on low-temperature DNA dynamics.

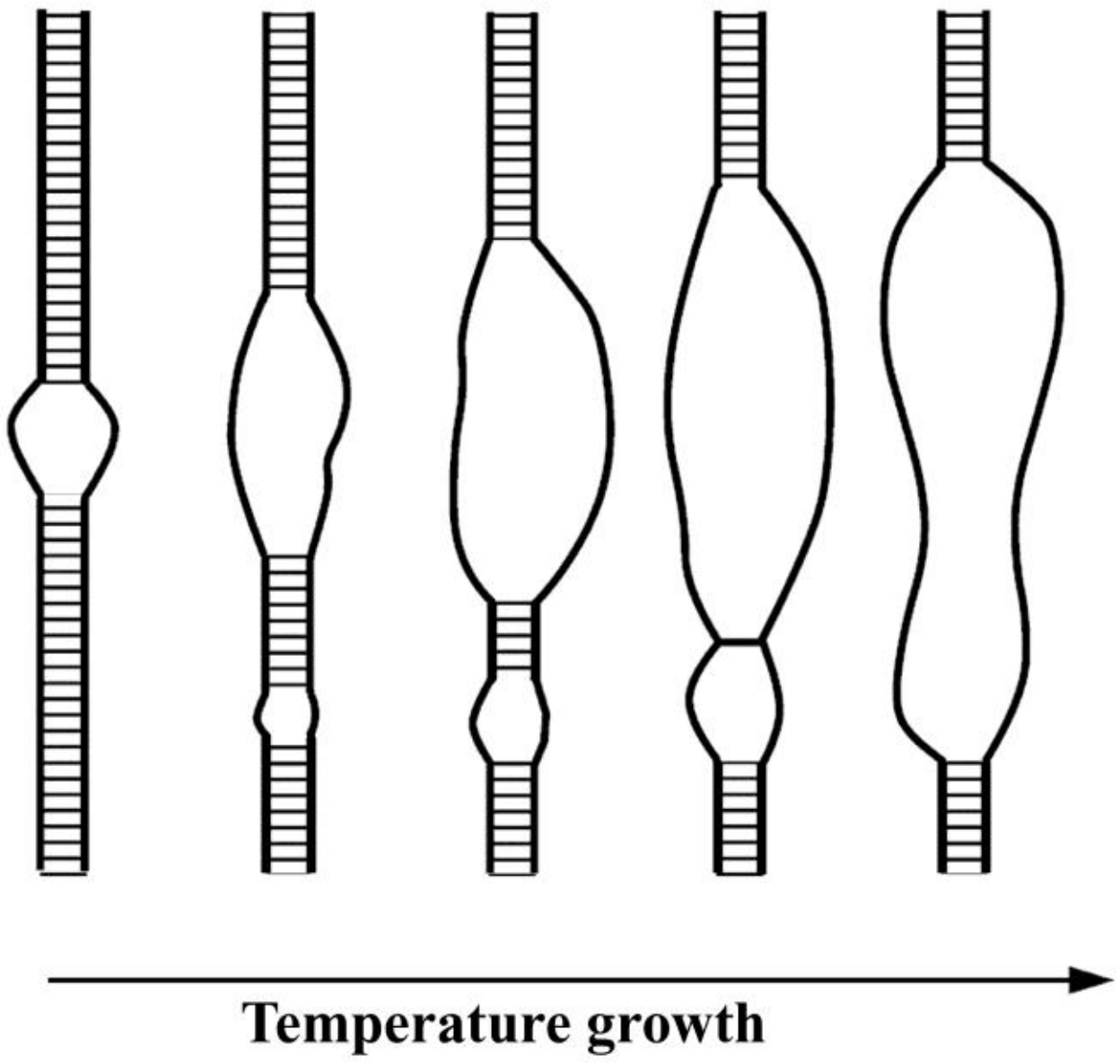

**Scheme 5**. DNA denaturation caused by temperature growth: merging of unwound fragments (denaturation bubbles) leads finally to complete divergence of the duplex chains.

The length of an open state will be held to be the number of neighboring basepairs with open H-bonds. By analogy with the generally accepted term we will call an open state of several basepairs in length a "bubble" taking no care of the degree of deterioration of stacking interactions. However, when it will be necessary to stress that we mean just a denaturated fragment we will use the term "denaturation bubble". The process of formation of an open state will be called "opening". In English-language papers this term is also often used as a nown – a synonym of an open state. This synonym will be used in our work too.

The importance of investigation of DNA open states is caused by a number of factors. The physical laws of the helix-coil transition observed in in vitro experiments turned out to be useful for theoretical investigation of the phase transition problem in quasi-one-dimensional systems. Open states play a great role *in vivo* – even notwithstanding the fact that at moderate temperatures their concentration is very low. For example, opening of non-covalent bonds in a bubble leads to a decrease of the DNA mechanical rigidity. This promotes its flexure, "folding". DNA packing into nucleosomes is shown to occure via its transient opening [14, 15]. The results of experiments on the formation of a circular DNA from oligomers consisting of 100 basepairs suggest the possibility of spontaneous formation of "configurable" fragments there [14, 16, 17]. These fragments are thought to be bubbles arising due to thermal fluctuations [14, 16].

Opening of DNA enables enzymes to get to its bases. Therefore open states take an active part in specific DNA-protein interactions. Recently this viewpoint has been confirmed experimentally [18, 19]. Besides, partial or complete decay of stacking in the course of duplex opening can considerably retard or completely block the process of transfer of cation-radicals (holes) in DNA. A hole can migrate along a stack of basepairs due to the fact that it is a π-conjugated system [20], see above. Charge transfer plays a great role not only in the processes of mutagenesis [21–23] and carcinogenesis [21, 24], but also in the restoration of DNA damages [25–27].

Thermodynamic properties of DNA open states are essential for understanding the underlying physics of DNA behavior. As discussed below, there are several types of openings. Their thermodynamic properties depend greatly on the type of an open state and the primary structure of the fragment where it arises – for details see Chapter 5. There are even openings with a negative activation enthalpy [28]. Moreover, since any DNA opening starts with breaking the complementary H-bonds, open states of different types can probably interact with one another. Therefore calculations of kinetic characteristics of openings as a function of temperature can be rather complicated. Thermodynamic characteristics of the open states of each type should be analyzed so that to estimate:

1) probability of the formation and the averaged lifetime that is characteristic for the open states of each type;

2) degree of their interaction with one another and its influence on the kinetics of each opening in any fragment of a heterogeneous DNA for a wide range of temperatures.

Besides stereochemical factors and the energy of noncovalent interactions, a considerable contribution into the thermodynamics of openings is made by nonequilibrium processes. The most thoroughly studied nonequilibrium processes are transfer and nonlinear localization of the energy of nucleotide pairs' stretched oscillations. A great role in the understanding of these processes has been played by mathematical models where the state of each basepair is described by a few variables – from one to four. Here we mainly concentrate on the models of this level which we figuratively name "simple approaches" or "simple models". Notwithstanding their simplicity these approaches allow studying the physicochemical nature of the peculiarities of DNA dynamics. First, many of these models are analytically tractable. Second, even relevant numerical calculations require relatively few computer resources. This enables one to study the behavior of simple models on a long-term scale, as distinct from molecular dynamics.

Development of DNA models is necessary for solution of a number of problems. One of them is to develop methods for a search for promoter regions in native DNA of a known nucleotide sequence. It seems likely that any DNA has a temperature range in which the probability of its opening is maximal just in biologically-active fragments. Therefore modeling of nonequilibrium effects with regard for thermodynamic peculiarities of the open states can give rather accurate results.

Not less important is the problem of DNA fragments having a special primary structure which can easily become open. As we will show in Chapter 4, the most unstable fragments of a duplex are by no means always characterized by the lowest total energy of H-bonds and stacking interactions [29, 30]. It seems likely that there exists a certain "bubble code" – a specific formula of a nucleotide sequence which promotes opening. At least the dependence of the thermodynamic properties on the sequence context has been revealed for opening of some individual bases, see Chapter 5.

The problem of an interrelation between the thermodynamic properties of the open states and the DNA primary structure has a great applicational significance. Thus, in recent years interest in the development of DNA-based nanoelectronic devices has grown [31, 32]. In modeling DNA sequences possessing maximal electric conductivity, an important criterion is the stability of their secondary structure. Since opening is associated with decay of stacking, high concentration of open states can considerably reduce the conductivity of a duplex and enhances the risk of the bases oxydation. Therefore in the search for stable sequences one should solve an "opposite" problem. It can be called a problem of a search for a "bubble anticode" – a primary structure possessing such properties as:

1) minimum relative probability of the formation of any open state,

2) uniform opening probability along the dulpex,

3) maintenance of the above-indicated properties in a wide temperature range.

The assumption that the DNA openings dynamics influences the charge transfer has determined the specificity of the theoretical investigations considered in this review. First, in the majority of the models described no changes in the duplex's supercoiling can be taken into account. It is evident, that creation of nanobioelectronic devices on the basis of a relaxed DNA is much easier. Besides bubbles opening in itself can lead to a partial release of supercoiling tension [33–35]. This makes comparison of the calculated data with the experimental ones still more difficult. Second, we do not consider investigations of the open states at the DNA ends at moderate temperatures. The length of duplexes *in vivo* is so large that the end effects are negligible. *In vitro* the end basepairs can be easily bound by covalent bonds.

The aims of our work are:

1) to make a brief review of simple models and experimental techniques used for the study of the DNA open states;

2) to review and analyze literature data on the thermodynamical properties of the open states;

3) to recognize the types of openings on the basis of their thermodynamic quantities and other criteria;

4) to explain the apparent contradictions between the results of different experiments;

5) to suggest possible ways of the development of theoretical and experimental methods for the study of the duplex dynamics based on the comparison of currently available data.

In Chapter 1 we describe simple methods for investigation of DNA denaturation. A concept of melting profiles and their dependence on the duplex properties is presented. Consideration is given to Ising-like models, their parameters and the use of these approaches for investigation of helix-coil transitions. In Chapter 2 we make a brief review of some mechanical models where the state of one pair is described by one or a few variables. Using Peyrard-Bishop-Dauxois model as an example we show that localization of the energy of nucleotide pairs' oscillations is mainly responsible for the dynamics of the open states. This is confirmed by the material of Chapter 3 where we review investigations of the denaturation of single duplexes under the influence of an external force. There we demonstrate the contribution of Ising-like models and similar mechanical models into the study of the experimentally observed peculiarities of micromechanical DNA unzipping.

In Chapter 4 we concentrate on the properties of the bubbles in the case when the main part of the duplex remains native. Some experimental data are discussed. Investigations of the Peyrard-Bishop-Dauxois model for the case of moderate temperatures (28 °C) are considered. We describe the molecular beacon spectroscopy method which enables one to investigate the kinetics of bubbles closing. In Chapter 5 we review experimental data on the kinetics of openings at temperatures below 35 °C. We suggest a mechanism of the influence of the DNA primary structure on the activation thermodynamical parameters of single basepairs openings. Discrepancies between the results of molecular beacon spectroscopy and the data of NMR and other methods are explained. The properties of the open states which form in the case of small angular displacements of nucleotide pairs are analyzed.

Chapter 6 summarizes the data on low-temperature DNA dynamics. It is shown that different degrees of freedom of the bases take a major part in the formation of the open states of various types. A necessity is justified to modify mechanical models so that to improve the agreement between the calculated data and the experimental ones. Consideration is given to how the localization of the energy of small angular displacements of the bases affects the process of opening. A hypothesis of an interaction between the open states and various fluctuations of a duplex is considered.

In the conclusions we make the pooled analysis of the material presented in the review. As a result we:

1) suggest the main criteria for classification of the open states;

2) reveal characteristic features of DNA models which are crucial for interpretation of various experimental data;

3) suggest some improvements to simple theoretical approaches and assess the prospects of their development;

4) describe the general scheme of experiments required for further development of one group of simple mechanical models.

# 1. EQUILIBRIUM THERMODYNAMICS OF DNA THERMAL DENATURATION

## 1.1. Early experimental and theoretical investigations of denaturation

Identification of DNA structure in 1953 [36] gave rise to a great many investigations of the molecule's properties. Most of early experiments were devoted to DNA denaturation under the influence of extreme pH values, low ionic strength of the solution, high temperature and different denaturating agents. Early physicochemical investigations in this field are described in detail in reviews [37–39].

Early experiments revealed considerable changes in the physical properties of a DNA solution under the influence of denaturating factors. For example, a relative absorption in the ultraviolet spectrum increased by one third [40], while viscosity of the solution decreased 12 times [41, 42]. According to light scattering data, the molecular mass of the polymer does not change in this case [41]. A conclusion was made that changes in the viscosity and light absorption are caused by a helix-coil transition.

The most popular method for investigating this transition was to measure the absorption of light with the wavelength of 260–268 nm in the course of slow heating of a DNA solution. Unwinding of the duplex chains leads to a considerable increase in the absorption intensity. The effect is based on the loss of contact between neighboring bases in each chain. Dense packing of the bases in a duplex determines its hypochromism – a decrease in the molar coefficient of extinction. As the packing density decreases, the hypochromic effect decreases too.

The hypochromism phenomenon is explained by vector composition of light-induced dipole momenta of a quantum transition of neighboring bases from the ground state to the excited one. In a native double helix, each induced dipole interacts with a dipole of the complementary base occuring at the other chain: dipole moments are opposite. Generally speaking, at low temperatures a one-chain DNA is also twisted into a helix, but its hypochromism is far less pronounced. The reason is that the antiparallel vectors of induced dipoles are separated from one another by a longer distance. For a two-chain DNA, hypochromism is usually 30–40 %, for a one-chain helix it is typically 15–20 %, and for dinucleoside phosphates it does not exceed 10 % [43].

A mathematical model of hypochromism was first suggested by I. Tinoco in 1960 [44]. The author described a decrease in the absorption of antiparallel induced dipole momenta as compared to a set of random orientations arranged side-by-side. In the case of parallel arrangement the absorption, on the contrary, should increase as compared to random arrangement. A similar model was independently introduced by W. Rhodes [45]. Subsequently the theory of hypochromism was further developed and different models were suggested [46–51].

The curve for the temperature dependence of the absorption is called a melting profile of DNA. A typical melting profile of a heterogeneous DNA of several thousand basepairs length is shown in fig. 1.1 [52]. It reproduces cooperative DNA denaturation in the course of which H-bonds and stacking break. The fraction of the open bases $\theta_B$ is calculated from the formula:

$$\theta_B = \frac{A(T) - A(0)}{A(100) - A(0)}, \quad (1.1)$$

where $A(T)$ is the value of the signal at temperature $T$, while $A(0)$ and $A(100)$ are the values at 0 and 100 °C.

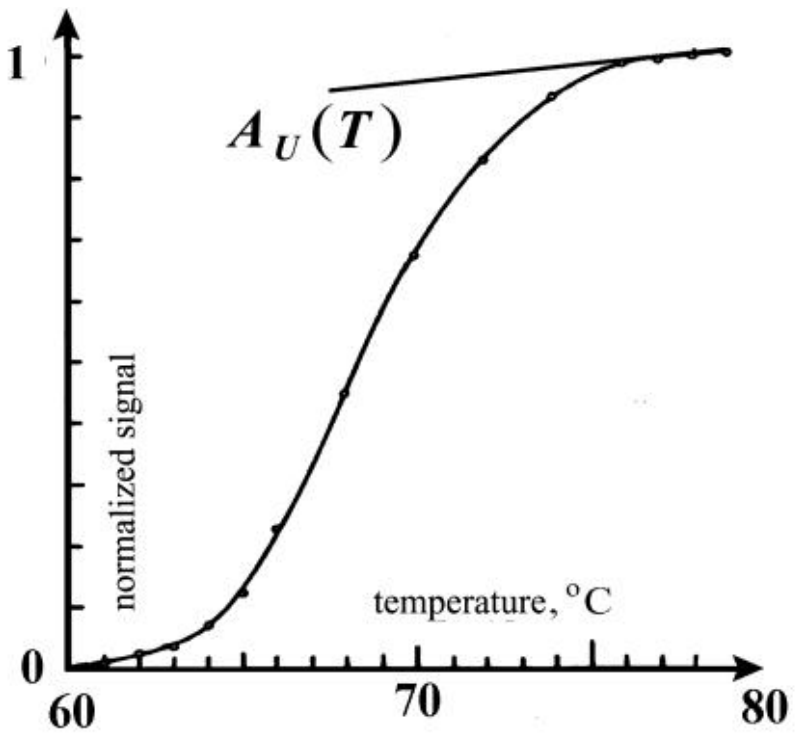

**Fig. 1.1.** Melting profile of DNA consisting of 16 000–32 000 basepairs [52]. The wavelength of light is 260 nm. The function $A_U(T)$ describes a quasilinear increase in the signal after complete cleavage of the chains.

It is easy to see from Fig. 1.1 that the photometric signal grows quasilinearly on the interval 76–80 °C when the chains have already dissociated. This effect is caused by partial retaining of stacking interactions in an unwinded DNA. Upon further heating stacking decays beyond repair. Increase in the absorption after dissociation of the chains is described by the empirical function $A_U(T)$.

According to Wartell and Benight, at low temperatures the photometer signal also grows quasilinearly [53]. The reason is that the contact area of the bases in a stack progressively decreases, though H-bonds do not break in this case. Signal enhancement at low temperatures is approximated by the function $A_L(T)$, and $\theta_B$, according to Wartell and Benight, is described by the expression [53]:

$$\theta_B = \frac{A(T) - A_L(T)}{A_U(T) - A_L(T)}. \qquad (1.2)$$

However, the slope angle of $A_L(T)$ is so small that this function can adequately be taken to be constant and equal to $A(0)$. Moreover, instead of using the function $A_U(T)$ or $A(100)$, use is often made of the signal value at temperature over which $A(T)$ corresponds to $A_U(T)$ (see, e.g., [54, 55]).

The temperature range between the beginning of the strands separation and its ending is called a melting interval. The temperature at which the value of the normalized signal is equal to 0.5 is called a critical temperature or a melting point. It is denoted as $T_m$. This quantity depends mainly on the DNA mass and the ratio between AT- and GC-basepairs in it. Having analyzed melting profiles of several dozens of basepairs Marmur and Doty [56] derived a simple linear relation between the number of GC-basepairs [GC] and $T_m$ in the form: $T_m = 69.3 + 41[GC]$.

The dependence of $T_m$ and melting interval on the pH of a solution, its ionic strength, the DNA molecular mass, and the presence of various stabilizing and destabilizing agents is well undestood. Apart from photometry in the ulrtaviolet spectrum, denaturation was studied by circular dichroism (see [43]), viscosimetry, denaturating gel electrophoresis [57, 58], formaldehyde treatment, microcalorimetry, etc. The results of early investigations of the helix-coil transition in DNA are systematized in the papers by Vedenov et al. [12], Wada et al. [59], Lazurkin and Frank-Kamenetsky [60], and in the book by Bloomfield et al. [43]. More recent material is described in the paper by Wartell and Benight [53].

A good review of theoretical investigations of DNA melting is presented in the paper by O. Gotoh [61]. We will consider in some detail only the early mathematical models of denaturation since their simple analogs have been used so far [62].

Theoretical investigations of DNA denaturation started substantially immediately after obtaining the first experimental data. Rice and Wada [63] studied a homogeneous duplex model in which H-bonds broke independently of one another. They used conventional methods of statistical physics.

It was shown that an equilibrium concentration of open H-bonds corresponding to a minimum of free energy exists for each temperature. In the paper by T. Hill [64] account was taken for the first time of the interactions between neighboring bases and of enhanced enthropy of large denaturated fragments. T. Hill calculated statistical sums for various states of DNA. He showed that denaturation proceeds cooperatively however it has nothing to do with a phase transition, at least, with the phase transition of the first order. B. Zimm, using more rigorous calculations confirmed cooperativity of DNA unwinding and the absence of a phase transition in this model [65].

The common feature of early theoretical investigations was that they all described a basepair with the use of a two-state variable. In other words, for a basepair, only two states were admitted – a closed state and an open one. It should be noted that the definition of an open state in the early models differs from that given in the Introduction. In those models an open state of a pair completely rules out both H-bonds with a complementary partner, and stacking. The energy required for a basepair to open is determined only by the state of the neigboring pairs. Such phenomenological models are called the nearest-neighbor models, their idea is shown in figure 1.2.

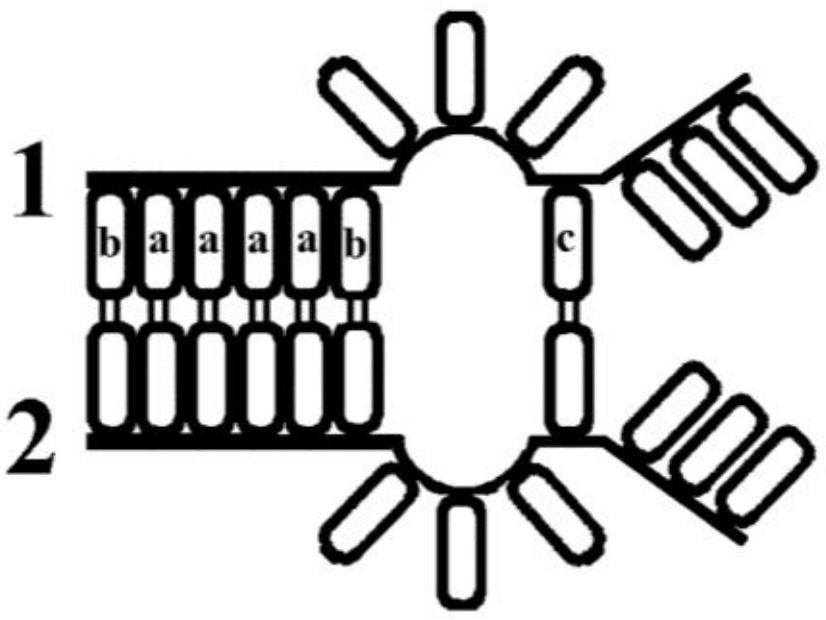

**Fig. 1.2.** Scheme of a nearest-neighbor model. Pairs **a** are the most stable. Pair **c** is not bound by stacking interactions with any of the neighbors. Hence, it is the least stable.

Since this description is similar to Ising model for ferromagnetism [66] this group of models is called Ising-like models.

Ising-like models turned out to be suitable for the description of some principles of denaturation. In particular they reproduce widening of the melting interval and a decrease in $T_m$ when the duplex length decreases [67, 68]. The influence of the ionic strength of the solution on $T_m$ was studied in [69]. The dependence of the number of closed basepairs on the concentration of DNA, its length and averaged energy of H-bonds was investigated in [70]. Due to their simplicity these models made a considerable contribution into the development of a basic theory of a helix-coil transition in DNA [71, 12].

At the same time it was clear that the predictive power of these models in the absence of the exact values of their parameters is insignificant, especially when we deal with a heterogeneous DNA. First of all, the energies of stacking were not known for different pairs of neighboring bases. In the models they were assumed to be similar [72, 73]. Nevertheless, experimental data obtained for RNA dimers demonstrated rather a wide range of the values of these parameters [74, 75].

Moreover, even for the model of DNA with a regular structure, it was shown that the shape of its melting profile greatly depends on the length of alternating fragments of AT- and GC-pairs [72]. For a native DNA, the dependence of the denaturation process on the DNA primary structure seemed to be still more complicated. Subsequently it was shown that their melting profiles may have a fine structure, see Section 1.3.

Hence, to optimize and improve the nearest-neighbor models, determine the exact values of their parameters one needed melting profiles obtained for the duplexes with known primary structures. Therefore, the next stage of the development of these models took place only in the late 1970's with the advent of nucleic acid sequencing methods. The nearest-neighbor models and the influence of the DNA primary structure on the shape of its melting profile will be described in Section 1.3.

Another important field where Ising-like models were used was investigation of the physical nature of denaturation, its mechanism. In early experiments on a homogeneous DNA it was shown that entire dissociation of the chains can take place in a very narror temperature range – not more than 1–2 K [76, 77]. By way of example a melting profile of poly(G):poly(C) several kilobases (thousands of basepairs) in length is demonstrated in figure 1.3.

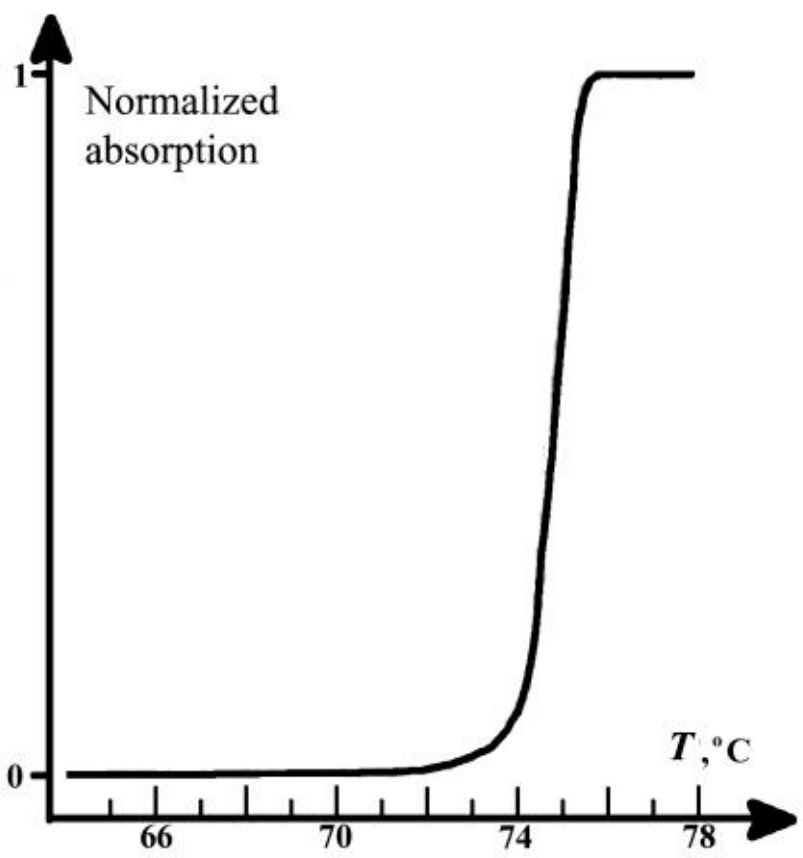

**Fig. 1.3.** Typical example of the melting profile of a homogeneous DNA (adapted from paper [76]). Along y-axis is absorption at 260 nm normalized to unity (cf. fig.1.1).

The sharp cooperative helix-coil transition typical for homogeneous DNA was interpreted by theoreticians differently. Some of them believed that the duplex denaturation corresponded to a phase transition [78, 79], while others had opposite opinion [12, 65, 72].

One of the main problems in the theoretical investigation of the helix-coil transition was calculation of the specific configurational enthropy of large denaturated fragments – loops. The character of changes in this quantity which take place as the size of the loop grows is an important characteristic which indicates the type of a phase transition in a quasi-one-dimensional lattice, such as DNA. In what follows we will consider this problem in greater detail.

## 1.2. Enthropy changes in the helix-coil transition. Poland-Scheraga model

A DNA molecule may be considered to be a quasi-one-dimensional lattice occuring in a three-dimensional space. Investigation of a phase transition is this type of a system is very important for development of theoretical physics. Impossibility of such a transition is a generally known fact [80–82]. However if such systems have infinite length, in the case of infinitely large interaction radius, a phase transition can take place at strictly determined temperature [82]. It is also observed in the case when the interaction potential is multiparticle, i.e. depends on more than one coordinate difference. For the case of DNA in a three-dimensional solution this approximation is quite feasible.

Changes in the duplex enthropy in the course of chain separation were first studied in detail in the Poland-Scheraga model [83, 84]. In this Ising-like model DNA is presented as a native duplex where large loops can be formed. The authors revealed that for infinitely long DNA chains, in the course of DNA denaturation an actual phase transition (second-order) can take place, unlike in the case of infinitely long polypeptide alpha-helices [83]. In paper [84] Poland and Scheraga used Flory formula [85] for the enthropy of a loop consisting of $M$ statistical segments:

$$S(M) = R\left\{ M\ln\Omega - \left[ A_0 + \frac{3}{2}\ln M \right] \right\}. \tag{1.3}$$

Here $A_0$ is a constant depending on the criteria determining the dynamics of denaturation bubble closing (its exact value is unknown), $R$ is a universal gas constant; $M \ln \Omega$ corresponds to the specific conformational entropy of a one-chain DNA fragment in a bubble.

Poland and Scheraga used combinatory analysis to analyze loop conformations in two- and three-dimensional space. In the course of denaruration a fragment consisting of $M$ statistical segments forms a loop of $2M$ statistical segments in length. The length of a statistical segment of a one-chain DNA depends on the properties of the solution and is normally equal to approximately seven nucleotides [12]. Poland and Scheraga modelled a loop as a set of ideal random walks at right angles; coincidence in the coordinates was neglected. The algebraic sum of motions parallel to each coordinate axis was equal to zero so that the loop be closed. For two- and three-dimensional space they got respectively:

$$\ln(No.\,of\ \ confugurations) = M\ln 2 - \left\{ \ln\left(\frac{\pi}{4}\right) + \ln M \right\} \text{ and}$$
$$\ln(No.\,of\ \ confugurations) = M\ln 2 - \left\{ \ln\left(\frac{\pi}{6}\right)^{\frac{3}{2}} + \frac{3}{2}\ln M \right\} \tag{1.4}$$

for loops of long length. By a long length is meant the number of statistical segments which enables one to neglect boundary effects. In other words it was believed that the whole of the loop behaves like a free polymer chain.

Bearing in mind similarity of these formulae to expression (1.3) and introducing a quantity $Z$ proportional to $M$, the authors presented the expression for the statistical weight $\omega_Z$ of a loop composed of $Z$ units ($2\cdot Z$ segments) in the form:

$$\ln\omega_Z = a'Z - b' - c\ln Z, \tag{1.5}$$

where $a'$ and $b'$ are proportionality coefficients depending on the polymer properties. Poland and Scheraga showed that the order of the phase transition (and its existence *per se*) is determined by the quantity $c$ [84]:

for $c \leq 1$ a phase transition does not take place;

for $1 < c \leq 2$ a phase transition of the second order takes place;

for $c > 2$ a phase transition of the first order takes place.

When modeling a loop by ideal random walks, $c$ is equal to $c = d/2$, where $d$ is the number of space dimensions. Thus, for a three-dimensional space $c$ is equal to 3/2 which yields the phase transition of the second order.

Later on M. Fisher improved the calculation results obtained by Poland and Scheraga, having replaced ideal random walks by the so-called self-avoiding walks. This term implies random motions without a chance to get for a second time to a coordinate cell which has already been visited [86].

Walks of this type, imitating the behavior of an elastic polymer chain are used in physics to take into account excluded volume interactions. These interactions reduce the number of possible configurations of the chain.

The values of $c$ obtained by M. Fisher for two- and three-dimensional cases were 1.46 and 1.75, respectively. Hence he was the first to show the possibility of a phase transition in the course of DNA melting in two dimensions. Excluded volume interactions decrease the specific configurational enthropy of a loop when its size grows. Since the length of a loop in the vicinity of $T_m$ sharply increases, this effect can lead to a jump in the heat capacity, i.e. to a phase transition. Effects caused by attraction of the chains were not studied by M. Fisher.

DNA models where DNA chains are considerd as random walks received the name Poland-Scheraga type models or PS-type models. From 1967 till 2000 these models were not developed. Nevertheless the physics of polymers intensively progressed during that time period. Apart from a helix-coil transition, theoreticians investigated one more type of a phase transition – a coil-globule transition [87, 88]. The state of a globule is characterized by certain “structuredness” – it has a denser core while the chains on the periphery are arranged more incoherently.

The theory of a coil-globule transition was developed by I.M. Lifshitz, A.R. Khokhlov et al. for flexible polymers [87–91]. It was shown that a globule is formed in the case of rather a strong attraction between the fragments of a polymer chain in a coil [87, 88]. Investigations of the physical nature of these transitions are reviewed in [90–92]. In heteropolymers the transition has anomalous properties [89]. In the case of DNA, a globule can be formed only in the presence of polyvalent cations [93].

Another important trend in the physics of polymers was investigation of interactions in a pair of directed polymer chains. Using the perturbation theory, cumulant distribution functions taking account of multiparticle interactions were calculated [94, 95] which made possible exact description of random interactions between the polymer chains. Due to success in the physics of polymers and perfection of calculation techniques PS-type models entered a new stage of development.

In 2000 Causo et al. developed a new model for denaturation of a finite-length DNA [96] where two chains of the polymer are described by a sum of $M$ segment-vectors $\boldsymbol{\omega}^1 = \{\boldsymbol{\omega}_0^1, \ldots, \boldsymbol{\omega}_M^1\}$ and $\boldsymbol{\omega}^2 = \{\boldsymbol{\omega}_0^2, \ldots, \boldsymbol{\omega}_M^2\}$. They occur in a 3D coordinate lattice and have a same origin $\boldsymbol{\omega}_0^1 = \boldsymbol{\omega}_0^2 = (0, 0, 0)$.

DNA chains were presented as self-avoiding walks, coincidence of their ends was not considered to be necessary. Overlapping (occurrence in the same cell) was forbidden for both the segments of one chain and the segments of different chains [96]. Only the segments having the same linear coordinate along the chain ($\omega_i^1 = \omega_i^2$) and corresponding to complementary bases could occur in the same cell. These overlappings were energetically advantageous. The binding energies of all the complementary segment pairs were similar, i.e. DNA heterogeneity was not taken into account.

Generally speaking, in the case of rather long DNA chains this approach can lead to appearance of several alternating segments of native and denaturated structures. For a long duplex, this model suits much better than the original Poland-Scheraga model where the enthropy of a single closed polymer chain is calculated. In the paper by Causo et al. it is shown that in the course of thermal denaturation a phase transition of the first order takes place [96]. The result obtained agrees with the experimental data better than the results obtained by M. Fisher which is due to consideration of an interaction between the complementary sites of the chain.

Using analytical methods Kafri et al. confirmed the DNA phase transition to be first order [97]. The authors relied on the results obtained by B. Duplantier who investigated the number of possible configurations for the general case of a polymer chain with different numbers of branches and vertices [98–100]. In the model by Kafri et al. account was taken of the excluded volume interactions not only in the closed loop but also between the loop and the rest of the chain. Having assumed for simplicity that the rest of the chain is in the native form, Kafri et al. modeled DNA with a denaturation bubble as a polymer chain including 4 vertices and 4 legs. It is shown schematically in figure 1.4. Each of two vertices V3 (“vertex 3”) has 3 outgoing legs; each vertex V1 has one outgoing leg.

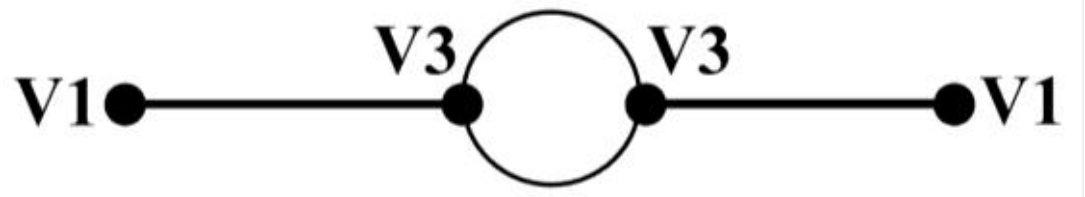

**Fig. 1.4.** Schematic representation of DNA with a denaturation bubble via a polymer network by Kafri et al. [97].

Formula (1.5) was used in the form

$$\omega_k = V\frac{s^k}{Z^c},$$

where $s$ is a non-universal (depending on the polymer properties) constant equal to exp($a'$), and V = exp($-b'$) which is for simplicity taken to be equal to 1.

Kafri et al. obtained a simple expression for $c$ in their work:

$$c = d\nu - 2\sigma_3, \tag{1.6}$$

where $d$ is the number of space dimensions, ν is the correlation length exponent for a self-avoiding walk, and $\sigma_3$ is an index related to the vertix of V3 type. Having taken from the paper by B. Duplantier [98] $\nu = 3/4$ and

$$\sigma_N = \frac{(2-N)(9N+2)}{64}$$

(i. e. $\sigma_3 = -29/64$) for two dimensions, the authors obtained $c \approx 2.4$ for $d = 2$. The expressions

$$\sigma_3 = -\frac{3\varepsilon}{16} + \frac{9\varepsilon}{512} \text{ and } \nu = \frac{1}{2}\left[1 + \frac{\varepsilon}{8} + \frac{15}{4}\left(\frac{\varepsilon}{8}\right)^2\right]$$

in $d = 4 - \varepsilon$ dimensions dimensions yield the value $c \approx 2.115$ for the three-dimensional space. The formula derived by Kafri et al.

$$c = 2 + \frac{\varepsilon}{8} + \frac{5\varepsilon^2}{256} \tag{1.7}$$

in $d = 4 - \varepsilon$ dimensions also yields $c > 2$ both for two- and for three-dimensional space [97].

An analogous result was ontained in the model where no account was taken of excluded volume interactions in one and the same chain [101]. The authors considered only excluded volume interactions between two chains. In their opinion, a ratio between a statistical segment length and a typical bubble length is rather high. Therefore excluded volume interactions between the segments of the same chain should be negligible as compared to excluded volume interactions between the chains.

An important role of excluded volume interactions between the chains was also demonstrated in the paper by Carlon et al. [102]. In this model, as in the paper by Causo et al. [96], the chains were presented as a pair of self-avoiding walks with the same origin point

$$\mathbf{r}_1(0) = \mathbf{r}_2(0) = (0,0,0)$$

and free end points. Overlapping of “complementary” segments in the same cell

$$\mathbf{r}_1(i) = \mathbf{r}_2(i)$$

was associated with an energy gain due to complementary bond. The authors used Monte-Carlo method to study the probability of a bubble formation at a critical temperature depending on the bubble length [102]. To demonstrate the role of excluded volume interactions between the chains two variants of the model were considered.

In the first model the two strands forming a denaturated loop are self-, but not mutually avoiding. In the other words, "non-complementary" segments of different chains can occur in the same cell. Such overlaps do not contribute to the energy. In the second variant account is taken of excluded volume interactions both for one chains and between them and the complementary bonding gives and energy gain. In both the variants the relation

$$Q(M) \sim M^{-c}, \tag{1.8}$$

holds, where $Q(M)$ is the probability of a bubble formation, $M$ is the bubble length measured in segments. Even for DNA of 50 segments length or shorter, expression (1.8) holds valid, at least on the interval $2 \le M \le 10$. Nevertheless the condition for the first-order phase transition is fulfilled only for the second variant where $c = 2.1$. For the first variant, $c$ was equal to 1.73.

The value of $c > 2$ in three dimensions was confirmed in the subsequent works [103–106]. The case of a heterogeneous DNA for which, according to some estimates, $c$ was equal to 2.15 [106], and, by other estimates, $c$ did not exceed 1.91 [107] was also investigated. Decreased $c$ of a heterogeneous DNA suggests a higher enthropy of large loops arising in it, as compared to the case of a homopolymer. This calculation result is in good agreement with a lot of experimental data on denaturation of native DNA, see the following section.

PS-type models were also used to study the finite-size effects [108] and the equilibrium properties of the large loops dynamics [109–111]. However, despite powerful capabilities of these models, the large size of the studied denaturation bubbles gravely limits their application area. Even if the DNA length is tens of kilobases, the temperature interval on which its behavior can be investigated by this approach is rather narrow. The study of a phase transition in short duplexes becomes still more complicated in view of the possibility of an entire separation and reassociation of complementary chains. Therefore to describe melting of a short DNA use was made of the nearest-neighbor phenomenological models.

### 1.3. Nearest-neighbor models and their role in investigations of heterogeneous DNA melting

The main problem solved by PS-type models was to investigate the type of a phase transition taking place in the course of the duplex denaturation. At the same time with the advent of effective methods for nucleic acid sequencing a new problem arose. A necessity emerged to study a relation between the fine structure of the melting profiles of heterogeneous DNA and the nucleotide sequences of these duplexes. The nearest-neighbor models which were described in brief in 1.1 were the most suitable for solving of this problem.

Investigations of the fine structure of melting profiles have a straightforward applicational significance. It is generally known that melting of a heterogeneous DNA starts with the sites that are rich in AT-pairs. The object of early investigations in this field was λ bacteriophage [112, 113]. At that time its 48-kilobase length genome was shown to consist of three large regions [114] which differed considerably in the concentration of AT-pairs [113]. While studying the denaturation behavior of DNA researchers preferred to analyze not the photometry signal *per se*, but its derivative with respect to temperature. When differentiating extinction at 260 nm, Falkow and Cowie revealed several peaks [115]. Each peak corresponded to melting of a large DNA fragment of λ bacteriophage or melting of several lesser fragments with similar melting temperatures. The authors revealed analogous DNA properties for some other bacteriophages [115].

Analysis of differential melting profiles became a simplest and most efficient way for investigating the properties of a genome. For example, in combination with some other methods it was used to study interactions between chromatin DNA and polypeptides: histones [116–118], non-histone proteins [117, 119] and polylysine [120]. Besides, this analysis has long been used in comparative genomics. As early as in the late 1970's it enabled one to find a great similarity between the genomes of mitochondria [121] and chloroplasts [122] in various eukaryotes.

At present, high resolution melting analysis (HRMA) is a powerful tool for comparing short (100–600 basepairs) DNA fragments. A combination of this method with PCR amplification and the use of fluorescent dyes specific to a two-chain form enable one to work with nanogram amounts of DNA. The profiles obtained by HRMA are sensitive even to very few replacements of basepairs. This analysis is used for revealing mutations, genotyping infectious agents and in other fields of medicine [123–126]. An example of differential melting profiles obtained by HRMA is shown in fig. 1.5.

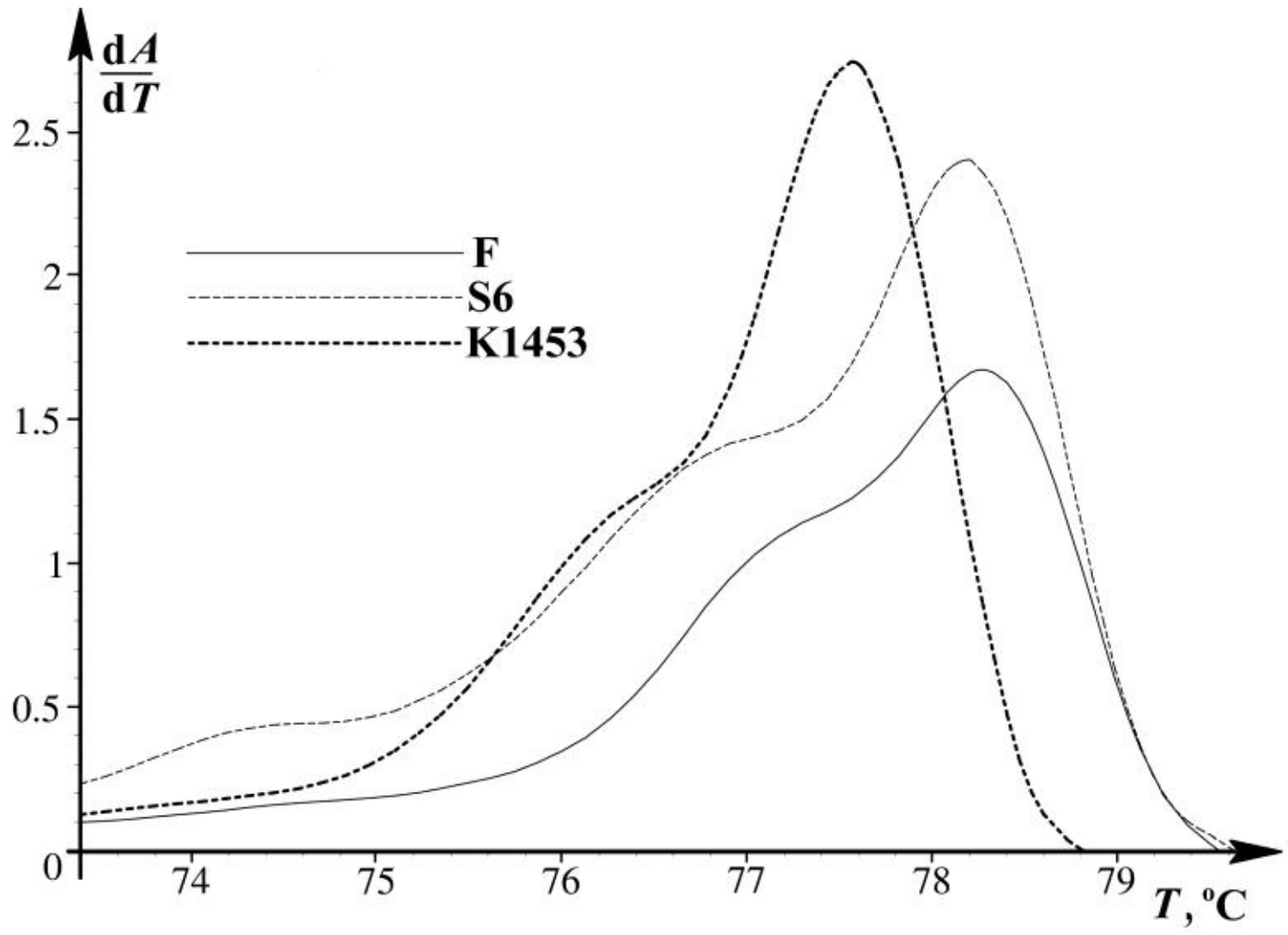

**Fig. 1.5.** Standard differential melting profiles of promotor regions of *vlhA* gene from different strains of *Mycoplasma gallisepticum* [127]. $A(T)$ is a photometer signal. Resolution approximates 0.1 °C.

An important component of HRMA is calculation of melting profiles from a nucleotide sequence. Investigation of a quantitative relation between the denaturation behavior of DNA and its primary structure became possible due to development of molecular biology in the late 1970's. Decoding of a nucleotide sequence promoted determination of the exact values for the nearest-neighbor models. Theoretical description of the melting profiles played an important role in searching for a relation between the thermostability of DNA fragments and their biological functions, see for example [128]. This problem is urgent since it deals with revealing the role of the local denaturation of DNA in its interactions with enzymes.

A detailed description of the nearest-neighbor models and comparison of the calculated data with the experimental ones is given in review by Wartell and Benight [53]. DNA is considered as a quasi-one-dimensional lattice consisting of some basepairs. Dissociation of the chains is described by a simple chemical equation

$$C_2 \leftrightarrow 2C_1,$$

where $C_1$ are single strands, and $C_2$ – is a duplex where at least one nucleotide pair remains bound by H-bonds. Hence, $\theta_{ext}$ – the concentration of DNA duplexes– is determined as

$$\theta_{ext} = \frac{[C_2]}{[C_2] + 0.5[C_1]}. \tag{1.9}$$

In turn, the total fraction of closed basepairs $\theta(T)$ is described by the expression

$$\theta(T) = \theta_{ext}\theta_{int}, \tag{1.10}$$

where $\theta_{int}$ is the fraction of closed basepairs in $C_2$ molecules. In DNA of over 200 basepairs in length, the value of $\theta_{ext}$ is close to 1 even in the vicinity of $T_m$.

In the nearest-neighbor models three parametrs are important: $s_i$ – stability of the $i$-th basepair, $\sigma'$ – parameter of cooperativity and $f_S(\mu)$ – the enthropy contribution of a loop formed of μ basepairs [53]. The first parameter is equilibrium constant for the "closing" reaction of the $i$-th basepair located at the end of a duplex. Since in this case not only H-bonds but also stacking interactions are formed, the expression for $s_i$ is written as [53]

$$s_i = \exp\left(-\left(G_i + \bar{G}^s\right)\left(RT\right)^{-1}\right), \tag{1.11}$$

where $G_i$ is is a difference between the free energies of H-bonds of an open state of the pair and of a closed one, $\bar{G}^s$ is an analogous difference between the stacking energies averaged over all the combinations of the adjacent basepairs, $R$ is a universal gas constant, $T$ is an absolute temperature.

The values of $G_i$ and $\bar{G}^s$ are less than zero, since for $T < T_m$ an open state is energenically less advantageous than a closed one. For the sake of brevity we take the energy of stacking interactions to be the same for all the neighboring basepairs. A more complicated case when $\bar{G}^s$ is geterogeneous is considered in detail in review by Wartell and Benight [53].

Parameter $\sigma'$ in the approximation of homogeneous $\bar{G}^s$ has the form

$$\bar{\sigma}' = \exp\left(\bar{G}^s\left(RT\right)^{-1}\right). \tag{1.12}$$

It is a characteristic of a sequence of open basepairs rather than an individual pair. This quantity takes account of the fact that for a basepair in the middle of a duplex to be open, decay of two stacking interactions is required unlike in the case of a basepair at the end of a duplex. A good illustration of the role of $\sigma'$ can be the expression for the equilibrium constant $K_{d,2}$ in the reaction of simultaneous opening of two neighboring basepairs [129]:

$$K_{d,2} = \frac{\sqrt{\sigma'_{i-1,i}}\, f_S(2)\sqrt{\sigma'_{i+1,i+2}}}{s_i s_{i+1}} = \frac{\bar{\sigma}' f_S(2)}{s_i s_{i+1}}. \tag{1.13}$$

If these pairs reside at an end, i.e. $i$ = 1, then $\sqrt{\sigma'_{i-1,i}} = 1$. Even for $\bar{G}^s = 7.5$ kJ/mol, replacement of $\bar{\sigma}'$ by $\sqrt{\sigma'}$ in the numerator of formula (1.13) increases fivefold the value of $K_{d,2}$ for $T = 300$ K.

Parameter $f_S(\mu)$ takes account of the enthropy difference between an open and a closed fragment of a DNA μ basepairs in length. It reflects the probability of the situation that the spatial arrangement of the complementary chains is advantageous for reduction of the size of an open fragment by one basepair. In the simplest case this parameter is determined by the equation

$$f_S\left(\mu\right) = \left(\mu + \gamma\right)^{-z}$$

where $1 \le z \le 2$, and γ is often taken to be equal to 1 [53]. In general, $f_S(\mu)$ is rather a complicated function and in different works it is determined in different ways [ibid.].

The nearest-neighbor models were mainly used to describe melting of a heterogeneous DNA of different lengthes and to determine the energies of stacking interactions. Later on these models provided the basis for melting profile calculation software: POLAND [130], MELTSIM [131, 132], GeneFizz [133] and others [134]. Having applicational significance (as a component of HRMA, etc.) calculations of melting profiles are also important for the development of evolution genomics [135, 136].

Alhough the nearest-neighbor approaches can describe melting profiles rather exactly they do not suit to describe the physical nature of the helix-coil transition. This drawback is not a common feature of all Ising-like models. For example, PS-approaches enabled one to understand the dependence of melting cooperativity on the changes in the specific configurational enthropy of bubbles which accompany the bubbles' growth. Nevertheless in all the Ising-like models a basepair is described by a binary variable. This does not make possible investigations of the transfer and localization of the nonlinear excitations energy – the key factor of bubbles' behavior.

The processes of energy transfer play a great role in the dynamics of DNA openings at the tempareture of about 300 K. In view of inhomogeneous localization of the energy the probability for a bubble opening in different fragments of a heterogeneous DNA can differ by several orders of magnitude [29]. In the investigations of the duplex dynamics at low temperatures the critical role was played by the so-called mechanical models. In these models, relative lengthening of an H-bond, which takes place when a nucleotide pair opens, is described by a real variable which can be either radial or angular. A detailed description of the mechanical models is given in the next Chapter.

## 2. MECHANICAL MODELS OF DNA. PEYRARD-BISHOP-DAUXOIS MODEL

### 2.1. Early mechanical models of DNA

The history of DNA mechanical models is very rich, but their detailed description is outside the scope of this paper. This theme is clarified in a number of reviews (see, for example [137]), to take an example, this is a monograph by L.V. Yakushevich "Nonlinear physics of DNA" [138]. A common feature of the mechanical models is that they describe the behavior of basepairs by real variables.

Investigations of nucleic acids by mechanical models started in the 1980's when researchers considered a possibility of the energy transfer in a duplex for the first time [139]. S. Englander assumed that the energy of oscillations of the DNA molecular lattice can be concentrated in solitary-wave excitations. His model consists of two parallel axes with torsional pendulums attached to them – mass points on the sticks of a certain length [140]. Hydrogen bonds are described in terms of mutial attraction of mass point (nucleobases). For each point there exists a unique degree of freedom – torsional angle $\varphi$. For this reason a complete denaturation of DNA is impossible in this approach. The solutions of the motion equations are well known, however this model is too rough to describe the dynamics of DNA open states.

A more realistic model which was subsequently called the dynamic plane-base rotator, or DPBR model was suggested by S. Yomosa [141, 142]. In this approach the points $P_n$ and $P_n'$ where the $n$-th complementary bases are attached to the sugar-phosphate backbone reside, so as the bases *per se*, on the $n$-th plane of $OXY$. The direction vectors $B_n$ and $B_n'$ and the segment $P_nP_n'$ make angles $\chi_n = \angle P_n' P_n B_n$ and $\chi_n' = \angle P_n P_n' B_n'$ which are the degrees of freedom. The scheme of this model is presented in figure 2.1.

S. Yomosa suggested the Hamiltonian:

$$\begin{aligned} H = \Big\{ & \frac{1}{2} J\left(\dot{\chi}_n^2 + \dot{\chi}_n'^2\right) + \tilde{X}\left(1-\cos\chi_n\right) + \\ & \tilde{X}\left(1-\cos\chi_n'\right) + \tilde{Y}\left(1-\cos\chi_n \cos\chi_n'\right) \\ & + S_0\left[1-\cos\left(\chi_n - \chi_{n-1}\right)\right] + S_0\left[1-\cos\left(\chi_n' - \chi_{n-1}'\right)\right]\Big\}, \end{aligned} \tag{2.1}$$

where $J$ is the average moment of nucleotides' inertion on the plane around the $P_n$ and $P_n'$ axes, parameters $\tilde{X}$ and $\tilde{Y}$ take account of the influence of a solvent and the interaction of the bases with one atother, $S_0$ is account for the summarized energy of stacking interactions and the torsional stress.

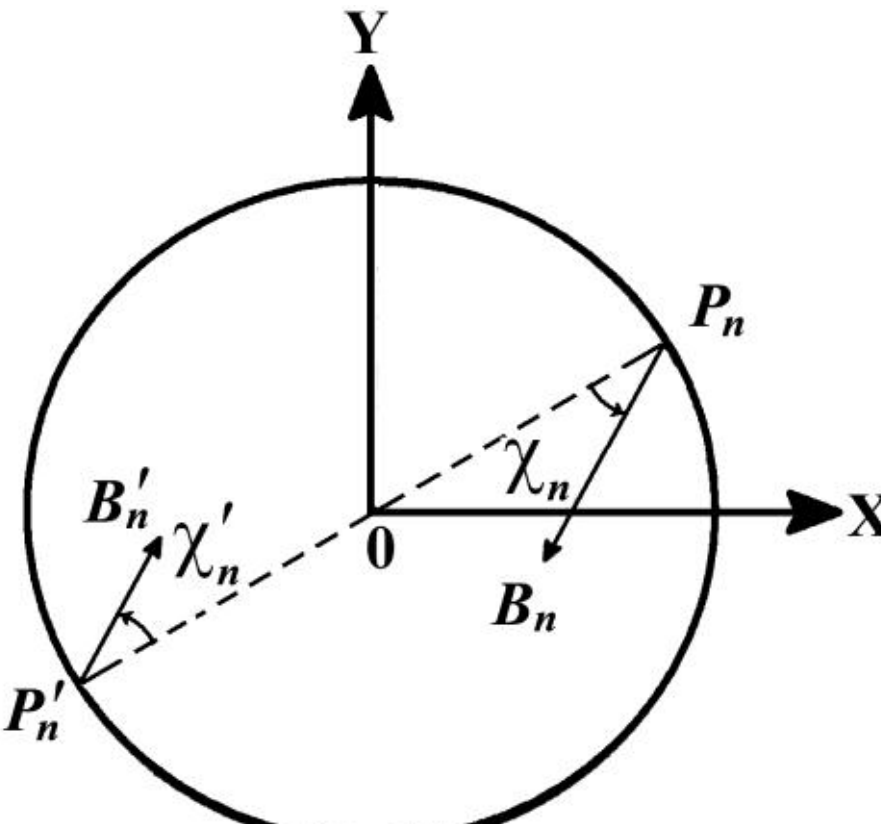

**Fig. 2.1.** Illustration of S. Yomosa model [142]. See text for details.

The model was investigated analytically in the continuum approximation. This approach allows studying the transfer of the energy, but not its localization. As it follows from the solutions obtained by S. Yomosa, “continuum” kinks and antikinks pass easily through one another [142]. Nevertheless for the case of $X \neq 0$, the states when the bases locally come out from the helix were reproduced which made possible comparison of the calculation data with experiments [142].

S. Yomosa investigated statistical mechanics of the model having calculated a specific concentration of solitons in DNA at different temperatures. Considering a basepair in the center of a soliton to be open one can assume the number of open pairs to be equal to the number of solitons.

In this way the thermodynamic parameters of the opening of a single basepair were estimated [142]. The calculation results were in agreement with the data obtained by a proton exchange method in a homopolymer DNA [143–145]. However, it should be noted that back at the time (up to 1989) the experimentalists estimated the lifetime of the DNA open states erroneously. This problem will be considered in 5.2. The modern methods for determining the lifetime of the open states of individual basepairs will be discussed in Section 5.1.

The dynamic plane-base rotator model was further developed in the works by Takeno and Homma [146, 147]. Their Hamiltonian was similar to that by S. Yomosa but in their approach dipole-dipole interactions of the bases were additionally taken into account. The authors studied a particular case of $X = 0$. Similar investigations were also carried out by C. Zhang [148] who assigned the main part to a combination of H-bonds with dipole-dipole interactions.

Thus in the early mechanical models the state of each base was described in terms of a torsional degree of freedom. It should be noted that in the case of DNA (in chemistry) the word *torsional* is typically used as a characteristic of stresses of the sugar-phosphate backbone which are associated with changes in helix pitch distance. Therefore to avoid confusion we will also use this term as applied to the degrees of freedom associated with supercoiling. Displacements of individual bases about the axis of the sugar-phosphare backbone, in turn, will be called *angular* displacements. In our opinion this term is also suitable for the models where these degrees of freedom are used.

Another peculiarity of the approaches of this group was the constant distance between the points of attachment of the bases to the sugar-phosphate backbone. Therefore a complete divergence of the chains is impossible in these models. Hence, angular models can be used only to study the low-temperature dynamics of local excitations in DNA. At that time there was not straightforward experimental evidence in favour of the soliton nature of these excitations. Generally speaking it is still lacking now, however modern knowledge of the DNA dynamics suggests a possibility of indirect investigation of the solitons' contribution. This can be done, for example, by in-depth analysis of the kinetics of proton exchange in an imino group in the absence of an external catalyst, see Sections 5.4, 6.4, and the Conclusion.

Impossibility of a direct comparison of the calculation data with the experiments considerably restricted the use of angular models and hindered determination of their parameters. Experimental data only enabled one to judge about the statistics of DNA openings, but they did not give any information of their dynamics. This generates a need for models which would be convenient not only for investigations of the energy transfer but also for the study of the statistical dynamics without resort to continuum approximation.

## 2.2. Introduction of a radial degree of freedom for a basepair. Peyrard-Bishop-Dauxois model

In the early 1980's apart from angular models one more mechanical aproach to investigation of DNA dynamics was developed. It was first used by E. Prohofsky's research team [149–151]. Initially the object of investigation was the behavior of an ensemble of hydrogen bonds in ammonia molecules [149] and the chains of a homopolymer DNA [150] under conditions of growing temperature. To describe an H-bond Morse potential was used for the first time [149]. Subsequently, with the use of a modified approach of self-consistent phonons it was shown that a nonlinearity of an H-bond is an important condition for accumulation of DNA stretched oscillations energy which is a necessary condition for duplex opening [151].

The works by E. Prohofsky's team provided the basis for development of a new model developed by Peyrard and Bishop [152]. That was the first mechanical approach "specialized" in the study of DNA denaturation rather than behavior of solitons in DNA. Introduction of radial variables instead of angular ones made possible investigations of the dynamics of the open states of any amplitude.

To analyze the statistical mechanics and the temperature dependence of an average distance between the chains Peyrard and Bishop used the transfer integral technique which is suitable for discrete models. This enabled the authors to give up the continuum approximation which was used in the early mechanical models and take account of the discreteness of the DNA structure. The model turned out to be analytically tractable.

In this approach two DNA chains are presented in terms of a set of variables $u_n$ and $v_n$, and H-bonds between the basepairs are described by Morse potential. Stacking interactions are taken into account via an ordinary harmonic potential. Hence the Hamiltonian of the system takes the form:

$$H = \sum_n \frac{1}{2} m\left(\dot{u}_n^2 + \dot{v}_n^2\right) + \frac{1}{2} k\left[\left(u_n - u_{n-1}\right)^2 + \left(v_n - v_{n-1}\right)^2\right] + D\left(\exp\left[-a\left(u_n - v_n\right)\right] - 1\right)^2, \quad (2.2)$$

where $m$ is the nucleotide effective mass, $k$ is the elastic constant taking account for stacking, $D$ and $a$ are, correspondingly, the depth and the width of the potential well for the interactions in a complementary nucleotide pair. Having rewritten the Hamiltonian in a simpler form and introduced new coordinates with the use of the formulae

$$x_n = \frac{\left(u_n + v_n\right)}{\sqrt{2}} \quad \text{and} \quad y_n = \frac{\left(u_n - v_n\right)}{\sqrt{2}},$$

the authors obtained

$$k\left(\left(u_n - u_{n+1}\right)^2 + \left(v_n - v_{n+1}\right)^2\right) = k\left(\left(x_n - x_{n+1}\right)^2 + \left(y_n - y_{n+1}\right)^2\right).$$

In the new coordinates the Hamiltonian takes the form

$$H=\frac{1}{2}\sum_n\left[m\dot{x}_n^2+k\left(x_n-x_{n-1}\right)^2\right]+\sum_n\left[\frac{m\dot{y}_n^2}{2}+\frac{1}{2}k\left(y_n-y_{n-1}\right)^2+D\left(\exp\left[-ay_n\sqrt{2}\right]-1\right)^2\right]. \quad (2.3)$$

This model received the name *Peyrard-Bishop model*, in what follows *PB model* [152]. The authors were not interested in the transfer of DNA as a whole, but only considered separation of the chains. Therefore to describe denaturation they used only the coordinates $y_n$.

Peyrard and Bishop calculated $<y>$ (which is an ensemble averaged stretching of H-bonds) as a function of temperature. Calculations were carried out for three values of $k$ – 0.002, 0.003 and 0.004 Å$^{-2}$. For each value of $k$, they calculated the temperature of DNA melting in the continuum approximation. They concluded that the denaturation bubble can be nucleated by energy localization due to nonlinear effects [152].

Later on Peyrard, Bishop and Dauxois determined numerically the eigenfunctions and eigenvalues for the transition integral equation [153]:

$$\int dy_{n-1}e^{-\beta\psi(y_n,y_{n-1})}\varphi_i\left(y_{n-1}\right)=e^{-\beta\varepsilon_i}\varphi_i\left(y_n\right), \quad (2.4)$$

where $\psi$ is the potential of the interaction between the basepairs $y_n$ and $y_{n-1}$, $\beta = 1/(k_BT)$ ($k_B$ – Boltzmann's constant), while $\varepsilon_i$ and $\varphi_i$ are the eigenvalue and the eigenfunction of integral equation (2.4), respectively. Besides they obtained an analytical solution of the nonlinear Schroedinger equation (NSE) in the continuum approximation.

Numerical calculations were carried out without the use of approximations. The authors made use of diagonalizing the transition integral matrix and replacing the integral by the sums of discrete contributions. Summation formulae of different orders were used. In paper [153] it was shown that the role of localized excitations in DNA denaturation grows as the temperature approaches $T_m$. Though the PB model is very simple as compared to a real duplex it adequately describes the transition observed in the experiments (see fig.1.3).

The PB model made possible a straightforward comparison of the calculation results with the experimental data. Moreover modeling with the use of a Nose–Hoover thermostat provided insight into the time-dependent dynamics of denaturation bubbles. Nucleation, motion and merging of denaturated regions lead, finally, to complete denaturation of a duplex. However numerical check of the analytical solution of NSE along with the great influence of discreteness demonstrated imperfection of the model in itself. Numerical calculations yielded great values of $T_m$ which are approximately 150 K higher than actual experimental data [153]. Besides, the transition *per se* appeared to be too smooth as compared to both the analytical solution and the experimental results.

This situation was the reason to introduce phenomenologically an anharmonic potential to describe stacking interactions more properly. As a result the model Hamiltonian took on the form [154]:

$$H=\sum_n\left[\frac{1}{2}m\dot{y}_n^2+V\left(y_n\right)+W\left(y_n,y_{n-1}\right)\right], \quad (2.5)$$

where

$$V\left(y_n\right)=D\left(e^{-ay_n}-1\right)^2$$

and

$$W\left(y_n,y_{n-1}\right)=\frac{k}{2}\left(1+\rho e^{-\chi(y_n+y_{n-1})}\right)\left(y_n-y_{n-1}\right)^2. \quad (2.6)$$

In expression (2.6) $\chi$ is decay constant for stacking interactions, while $\rho$ is a dimensionless parameter that phenomenologically takes account of denaturation cooperativity.

The new approach received the name Peyrard-Bishop-Dauxois model, in what follows PBD model [154]. The authors investigated statistical mechanics of the model by the transition integral method determining the eigenfunctions and eigenvalues of equation (2.4) with the new potential:

$$f\left(y_n, y_{n-1}\right) = W\left(y_n, y_{n-1}\right) + \frac{1}{2}\left[V\left(y_n\right) + V\left(y_{n-1}\right)\right]. \tag{2.7}$$

In view of anharmonicity of the stacking potential an analytical investigation of the PBD moled was impossible. Therefore the author calculated the dependence of $<y>$ on $T$ numerically and with the use of thermostat (stochastically), as it was done in [153]. The results obtained by these methods were in good agreement between themselves. Moreover, they appeared to be much closer to the experimental data: the melting point was 361.5 K and the transition *per se* was rather sharp.

In paper [154] it was shown that it is anharmonicity of stacking that the reason of such a sharp transition in the PBD model. The role of anharmonic stacking potential in the description of the sharp phase transition was also confirmed by modeling the DNA chains by random walks [155].

Later on the physical nature of the PBD model behavior was investigated thoroughly and the key role of the enthropy increase during opening of H-bonds and stacking was proved strictly [156]. The authors carried out a comparative investigation of the PB and PBD models and demonstrated fundamental differences in their behavior in the course of temperature increasing. In the PBD model the reason of a sharp separation of the DNA chains at moderate temperatures is the cooperativity of melting introduced by anharmonic potential (2.6). The work [156] was also an important methodic contribution into the investigations of the phase transition properties in finite-length oligonucleotides.

Independently from Peyrard and Bishop, a model with radial degrees of freedom was studied by L. Van Zandt [157]. To describe H-bonds he used a potential similar to Lennard-Jones potential. In his model each nucleotide pair had two degrees of freedom – one per a base. This enabled him to take account of shear deformations. Using numerical modeling L. Van Zandt investigated the behavior of individual soliton-like excitations in DNA of infinite length. Further investigations in this area had no concern with denaturation either and dealt mainly with the nature of excitations and the choice of the parameters [158–160].

Introduction of a radial degree of freedom for a nucleotide pair was a logical step in the development of mechanical models of DNA. However the approaches where the displacement of the bases from their equilibrium position is described in terms of rotation angles are actual so far. Moreover, recent data of simplified molecular dymamics suggest rather a low potential barrier for the escape of individual bases from the Watson-Crick helix as well as a pronounced nonlinearity of such conformational changes [161–163]. This gives additional evidence of the great importance of angular models, the most elaborated of which is the model by L.V. Yakushevich developed in 1989 [164]. Topological solitons in DNA are studied with the use of this model in a number of works – see [165–172] and references therein.

Other important tool for the study of DNA dynamics was the group of models where the radial variables are combinated with the torsional ones. Such approaches turned out to be more complicated for the analysis, but they opened up fresh opportunities for theorists.

### 2.3. Approaches using a combination of torsional and radial degrees of freedom

Investigation of DNA models where radial degrees of freedom are combined with torsion ones is a very promising trend. Such approaches enable one to take account of the helical geometry of a DNA molecule and the role of distortions of the sugar-phosphate backbone. For the first time, the radial-torsional model was suggested by Barbi et al., who specified two degrees of freedom for each basepair [173]. Besides relative lengthening of the hydrogen bond $r_n$, the authors introduced $\varphi_n$ – a relative change in the angle between the lines which connect the points of the bases' attachement to the sugar-phosphate backbone. As we can easily see, the second term differs from the degrees of freedom of the bases in the angular models. The scheme of the approach is shown in figure 2.2.

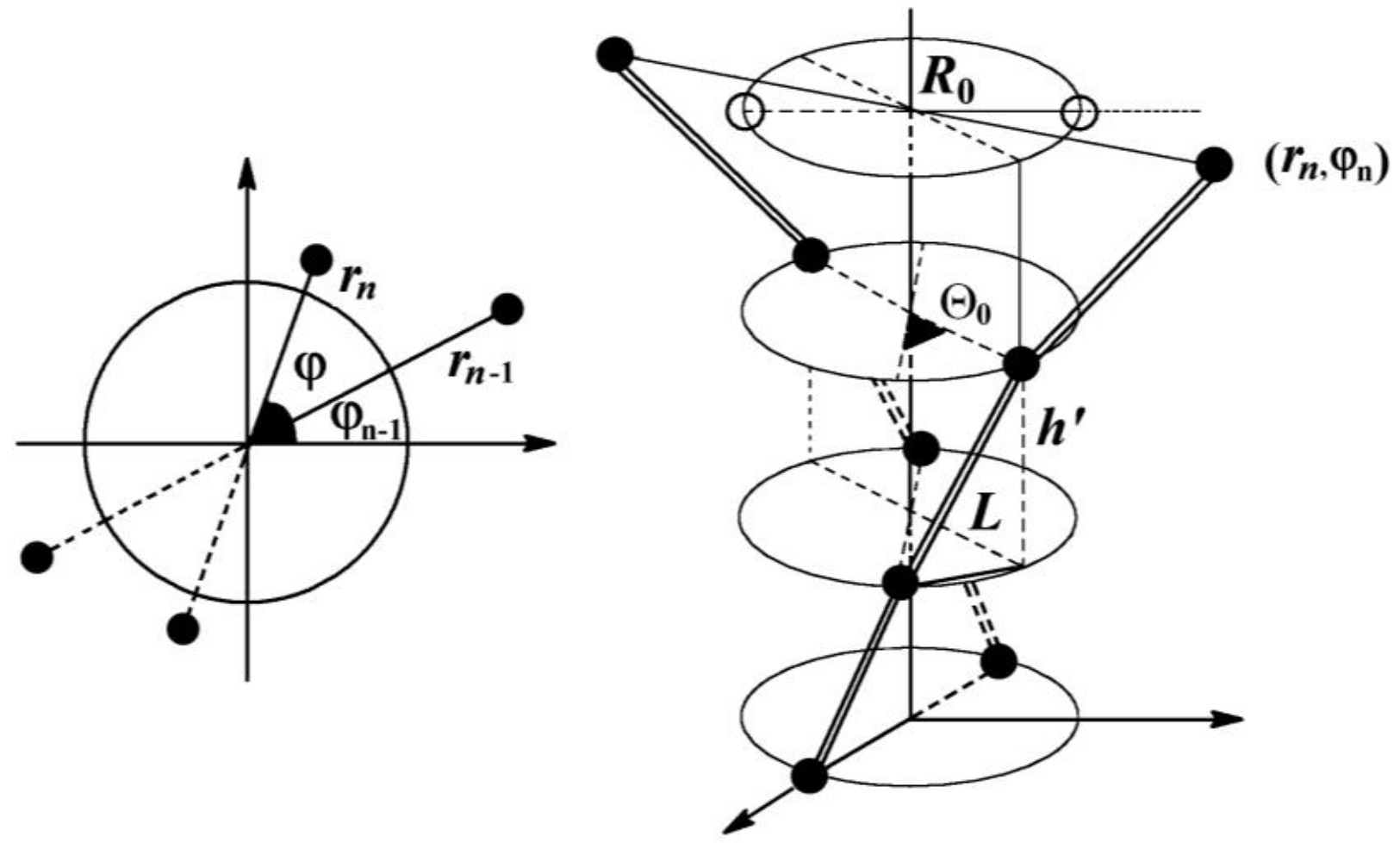

**Fig. 2.2.** Scheme of the radial-torsional model from paper [173]. $r_n$ and $\varphi_n$ are the degrees of freedom of a nucleotide pair; $R_0$ and $\Theta_0$ are their equilibrium values; $h'$ is a constant distance between the base planes. Compare figure 2.1.

The equilibrium distance $L$ between the points of attachment of the bases is

$$L=\sqrt{h^2+\left(2R_0\sin\left(0.5\Theta_0\right)\right)^2}\ ,$$

where $h'$ is a constant distance between the base planes equal to 3.4 Å. The values of $R_0$ and $\Theta_0$ are also determined from the available data on the DNA structure: $R_0$ = 10 Å, $\Theta_0$ = 36°. Lagrangian of the model has the form:

$$\begin{aligned}\Im=&\sum_n\left(m\dot{r}_n^2+mr_n^2\varphi_n^2\right)-\sum_n D\left(\exp\left[-a\left(r_n-R_0\right)\right]-1\right)^2-\\&-\sum_n C_{el}\left(\sqrt{h^2+r_{n-1}^2+r_n^2-2r_{n-1}r_n\cos\left(\varphi_n-\varphi_{n-1}\right)}-L\right)^2-\\&-\sum_n G_0\left(\varphi_{n+1}+\varphi_{n-1}-2\varphi_n\right)^2,\end{aligned}\qquad(2.8)$$

where $m$ is an effective mass, $D$ and $a$ are the depth and the width of the Morse potential well, respectively, $C_{el}$ – elastic constant and $G_0$ – is a backbone curvature constant. The first term of the Lagrangian describes the kinetic energy of the system, the second one – interaction between a pair of complementary bases. The third and the fourth terms correspond to the DNA potential energy arising as a result of changes in the DNA coiling. Subsequent modifications of the model by Barbi et al. affected only the third and the fourth terms.

Soliton solutions obtained with the use of this approach were topologically different from the solitons in the "plane" Peyrard-Bishop model, see [174]. Consideration of the helical geometry of a DNA molecule enabled investigations of the bubbles initiated by negative supercoiling of the duplex. This was done in the work by A. Campa who slightly simplified the original model [175]. A. Campa introduced a limitation of small angular deformations $\varphi_n > \varphi_{n-1}$. This enabled him to replace the fourth term in formula (2.8) by a simpler harmonical expression $\frac{b_0\left(r_{n+1}-r_n\right)^2}{2}$, where $b_0$ is an elastic constant. The bubbles induced by negative supercoiling can migrate over a long distance. Their stability is independent of the thermal noise or heterogeneity of the DNA structure. However, this property seems to stem from harmonicity of the stacking potential in A. Campa's approach [175].

An important field of application of the radial-torsional model is the study of DNA complete denaturation under the influence of an external force, i.e. micromechanical denaturation. Separation of DNA chains by mechanical "unzipping" was studied in a lot of experiments, see 3.1. Nevertheless this process cannot be described by radial models. Cocco and Monasson investigated torsion and thermal denaturation in a radial-torsion model having used the formalism of the transfer matrix [176]. Having calculated the statistical sum the authors obtained the denaturation free energy about a phase transition point.

In the approach by Cocco and Monasson the distance between the points of the bases' attachhment to the sugar-phosphate backbone was fixed and equal to $L$. Deformation of a helix was taken into account via its longitudinal extension, i.e. changes in $h'$. The part of the Hamiltonian which corresponds to the fourth term of formula (2.8) had the form:

$$E\exp\left[-\chi\left(r_n + r_{n-1} - 2R\right)\right]\left(r_n - r_{n-1}\right)^2 + V_\Gamma\left(\varphi_n - \varphi_{n-1}\right), \qquad (2.9)$$

where $\chi$ is the stacking decay constant, $E$ is its elastic constant equal to 4 eV/Å$^2$, and $V_\Gamma$ is the angular elasticity constant. This modification was also used for investigations of various vibration modes in DNA, and for interpretation of some Raman spectra and neutron scattering data [177].

Introduction of an anharmonic stacking, by analogy with Peyrard-Bishop-Dauxois Hamiltonian, improved considerably the model properties and the agreement between the calculation data and the experiments. For example, this approach enabled reproducing the force barrier at the onset of micromechanical denaturation, see Section 3.3. Besides, the authors succeeded to describe properly the macroscopic behavior of DNA in the course of its micromechanical unzipping [178]. The improved model also made posible description of the first-order phase transition observed in the course of DNA thermal denaturation [179]. Lagrangian of the "final" version of the model had the form:

$$\begin{aligned} \Im = m\sum_n\left(\dot{r}_n^2 + r_n^2\dot{\varphi}_n^2\right) - \sum_n D\left(\exp\left[-\alpha\left(r_n - R_0\right)\right] - 1\right)^2 - \\ -\sum_n K\left(\sqrt{h^2 + r_{n-1}^2 + r_n^2 - 2r_{n-1}r_n\cos\left(\varphi_n - \varphi_{n-1}\right)} - L\right)^2 \times \\ \times\exp\left[-b\left(r_n + r_{n+1} - 2R_0\right)\right], \end{aligned} \qquad (2.10)$$

where $b$ is the decay constant taking account of the finiteness of stacking interactions.

As we can easily see from the comparison of formulae (2.8) and (2.10), the anharmonic stacking is the only difference between this Hamiltonian and the original one [173]. The model by Cocco et al. is by far not an only theoretical approach taking account of the helical geometry of a DNA molecule. There are a lot of studies of various two- and three-dimensional models [180–183]. Recently this line of investigations has been in progress.

Alhough the radial-torsional models are much more complicated as to their analytical and numerical investigations they can take account of distortions of the DNA structure caused by an external torsional stresses. However most of physicochemical experiments are made on a relaxed DNA without any external mechanical stress. It can be believed that the behavior of such duplexes is described equally well by radial and radial-torsional models. Indeed, a sharp phase transition was observed both in the PBD approach [154, 156], and in the model by Barbi et al. [179].

Another characteristic of denaturation experiments is that many of them were carried out on a heterogeneous DNA. In the case of a heteropolymer duplex the most of the models lose analytical tractability. At the same time it is more convenient to perform numerical calculations for simple motion equations. In the case of a heterogeneous DNA radial-torsional models have one more serious drawback – it is rather difficult to parametrize them on the basis of experimental data. Hence simplicity of the motion equations and relative convenience of the parametrization are the main advantages of radial models in terms of theoretical investigations of the relaxed DNA dynamics.

### 2.4. Application areas of radial approaches. Problems of a heterogeneous DNA modeling

One of the main application areas of radial models is theoretical investigation of DNA denaturation. For example, the process of melting was studied with the use of a modified perturbation theory in the PB model [184]. A similar approach was used in the model where complementary H-bonds are described by a cubic potential [185]. Recent analytical approaches are more various. For example, denaturation of a homogeneous DNA in the PBD model was studied by the path integration method [186], which is not commonly used in biopolymer studies. In [187] the dependence of $T_m$ of a heterogeneous DNA on its nucleotide composition is reproduced by way of comparing the behaviour of the PB model with wetting models.

Besides, melting of a finite-length DNA in a real solution was modelled numerically. The chains were kept from final separation by an infinite distance due to modification of Hamiltonian (2.5). For example, a small positive slope was added to the plateau of the Morse potential [188]. Alternatively, a configuration space was restricted by adding an infinite wall such that $y_n$ cannot exceed a certain maximum value for all $n$ [189].

There are a lot of approaches which enable one to reproduce a sharp phase transition at a critical temperature. For example, this can be done by taking account of the finiteness of stacking interactions in an explicit form:

$$W\left(y_n, y_{n+1}\right) = \frac{\Delta H^0}{2}\left(1 - e^{-\alpha\left(y_n - y_{n+1}\right)^2}\right) + k\left(y_n - y_{n+1}\right)^2, \qquad (2.11)$$

where $\Delta H^0$ is standard stacking enthalpy, and $k$ is 2000 times smaller than that in the PBD model [190, 191]. A similar result can be obtained in the PB model by introducing an asymmetrical double Morse potential [192], or taking into account interactions between DNA and a solvent [193].

Another area where the radial models were applied is investigation of the soliton behavior in a homogeneous DNA. The peculiarity of solitons in the radial models is that they are closely related to the denaturation bubbles in a relaxed duplex. In most of the works investigations are carried out for the PB models [194–197] and the PBD models [198–200]. Modifications of the PB model are concerned with consideration of interations between the neighboring DNA coils via chains of water molecules [201, 202] and duplex distortions introduced through dipole-dipole interactions [203–205]. In some papers the behavior of solitons was studied in alternative models. Among them are the systems where the hydrogen bonds are described by potential $\varphi^4$ [206, 207], quadratic potential [208], plane zigzag model [209] etc.

A specific place among alternative approaches is held by Toda–Lennard–Jones model [210] (see [211] for review). In this model each basepair has 4 degrees of freedom (see fig. 2.3). Displacements of the bases along the duplex axis are described by Toda potentials:

$$V_{T,I}\left(\lambda'_n - d_l\right) = \frac{a''}{b''}\exp\left[-b''\left(\lambda'_n - d_l\right)\right] + a''\left(\lambda'_n - d_l\right), \qquad (2.12)$$

where $d_l$ is the equilibrium distance between the base planes ($d_l$ = 3.4 Å), parameters $a''$ and $b''$ are fitted from the experimental data, and $\lambda'_n$ is the distance between the neighboring bases of one chain (see fig. 2.3):

$$\lambda'_n = \sqrt{\left(d_l - \left(x'_n - x'_{n+1}\right)\right)^2 + \left(u'_n - u'_{n+1}\right)^2}\,.$$

The expressions for the potential of the second chain $V_{T,II}$ and $\lambda''_n$ are analogous. Complementary hydrogen bonds in the model are described by Lennard-Jones potential:

$$V_{LJ}\left(\tau_n - d_t + d_h\right) = 4\kappa\left[\left[\frac{q}{\tau_n - d_t + d_h}\right]^{12} - \left[\frac{q}{\tau_n - d_t + d_h}\right]^{6}\right], \tag{2.13}$$

where parameters κ and $q$ are fitted from the experimental data, $d_t$ is the equilibrium diameter of the double helix, and $d_h = 2^{1/6}q$ is the equilibrium length of a hydrogen bond. The quantity $\tau_n$ stands for the distance between the bases of two chains:

$$\tau_n = \sqrt{\left(d_t - v_n - u'_n\right)^2 + \left(y'_n - x'_n\right)^2}$$

hence, the length of a hydrogen bond is $\tau_n - d_t + d_h$.

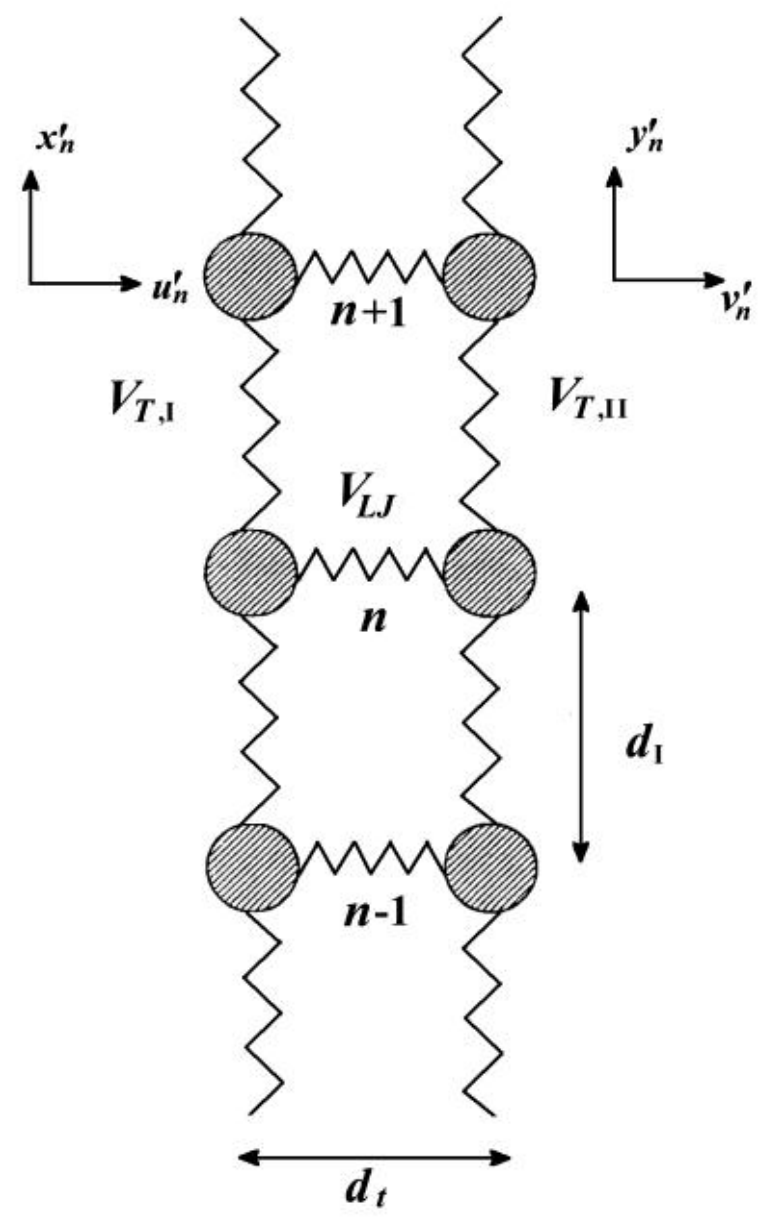

**Fig. 2.3.** Scheme of Toda-Lennard-Jones model [210].

Hamiltonian of Toda–Lennard–Jones model has the form:

$$H = \sum_{n=1}^{N}\left(\begin{array}{l}\frac{1}{2}M\left(\dot{x}'^2_n + \dot{u}'^2_n\right) + \frac{1}{2}M\left(\dot{y}'^2_n + \dot{v}'^2_n\right) + \\ +V_{LJ}\left(\tau_n - d_t + d_h\right) + V_{T,I}\left(\lambda'_n - d_l\right) + V_{T,II}\left(\lambda''_n - d_l\right)\end{array}\right). \tag{2.14}$$

Muto et al. investigated the dynamics of non-topological solitons in a homogeneous ring-shaped DNA molecule for various values of the depth of a potential well for a hydrogen bond [210]. They observed a dynamical picture of bubbles at the temperature of 310 K. The depth of the potential well was equal to 0.044 eV which is close to the value obtained by Peyrard et al. – 0.04 eV [153, 154]. One of the main results was the time scale of the DNA local denaturation: the open bubbles persisted for nearly 100 ps [210]. In the case of the PBD model, similar calculations yield not more than 4 ps even for long bubbles in a heterogeneous DNA [212]. According to experimental data, the lifetime of bubbles comes up to 1 ms [213]. Hence, the Toda-Lennard-Jones model yields the results which are closer to the experimental data by nearly two orders of magnitude. Therefore further development of this model seems to be a promising line of research. For example, modification of Toda potential will probably enable description of the conformational transitions which take place when DNA is extended along its axis (the so-called β-premeltons) [214, 215].

In our opinion, the most interesting application area of radial models is investigation of the behavior of solitons and bubbles in a heterogeneous DNA. The dependence of the opening dynamics on the nucleotide sequence is a problem which has an important applied significance in molecular biology. The pioneering studies of the energy transfer and interaction of quasi-solitons with inhomogeneities in DNA were carried out by Techera et al. in the PBD model [216]. Mass and H-bonds inhomogeneities were introduced for individual basepairs of a homogeneous duplex.

The Toda-Lennard-Jones model was the first approach applied to study the dynamics of bubbles in a duplex containing a long heterogeneous fragment [217]. This fragment had the structure [217]

……HHHHHHSWSWSWSWSWSWSWSWHHHHHHH……

The depths of potential wells for different basepairs (W – "weak", S – "strong", H – "homogeneous") had the values $\kappa_W = 0.8 \cdot \kappa$ and $\kappa_S = 1.2 \cdot \kappa$, where $\kappa = \kappa_H$ is the value for homogeneous DNA. Relation of $\kappa_W$ and $\kappa_S$ approximately corresponds to the enthalpy relations of a hydrogen bond in AT- and GC-pairs. Besides, heterogeneity was additionally taken into account via changes in the parameters of the Toda potential.

After introduction of heterogeneous fragments into the center of the duplex the total energy of the bonds per a unit of its length did not change. Nivertheless numerical calculations demonstrated that the probability of collective breaking of H-bonds grows considerably in this case. The work [217] was the first step in the theoretical investigations of the "bubble code" problem.

Certainly, there were many investigations of DNA dynamics' dependence on inhomogeneities of H-bond, mass and stacking in the 1990's. For example, Forinash et al. studied the interaction between the discrete breathers and the mass inhomogeneities in the PB model [218] and in the modification of the PB model where a cubic potential was used for the hydrogen bonds [219]. Chela-Flores and Migoni, having taken the PB model with two degrees of freedom per a nucleotide pair (see expression (2.2)), modeled a homogeneous DNA; however they introduced heterogeneity of stacking interactions [220]. In their "purine-pyrimidine" model, various values of the stacking elastic constant $k$ were specified: $k_C$ and $k_G$.

A transformational event in modelling a heterogeneous DNA became the work by Italian researchers Campa and Giansanti [221]. The authors determined individual parameters of the PBD model for AT- and GC-pairs. Having compared the theoretical melting profiles with the experimental data on the denaturation of short DNA oligomers the authors obtained $D_{AT} = 0.05$ eV, $a_{AT} = 4.2$ Å$^{-1}$ for AT-pairs and $D_{GC} = 0.075$ эВ, $a_{GC} = 6.9$ Å$^{-1}$ for GC-pairs. The calculated characteristics of stacking interactions were independent of the nucleotide sequence and were equal to $k = 0.025$ eV·Å$^{-2}$ и $\chi = 0.35$ Å$^{-1}$. It is the set of parameters' values that was used in all the subsequent works where the denaturation dynamics of a heterogeneous DNA was investigated in PBD model.

The results obtained by Campa and Giansanti enabled researchers to interpret the behavior of a heterogeneous DNA in various experiments which will be described in the following. These studies confirmed the adequacy of the parameters fitted for the PBD model whose peculiarities made it an important tool in investigations of the bubble dynamics.

In the models where only the angular degrees of freedom are present, no bubbles open. The radial-torsional approaches enable one to take account of both the radial displacements of the bases and their interaction with the torsional stress of the sugar-phosphate backbone. Such models are indispensable in the studies of denaturation of DNA fragments with nonzero supercoiling. The dynamics of such fragments cannot be described with radial models.

However the main advantage of the latter ones is that they are relatively simple. This factor becomes important when passing on to numerical calculations for the case of relaxed heterogeneous duplexes.

So, the majority of experiments on DNA denaturation are most adequately described by radial models. However, all the models of this type, except for PBD, are characterized by a harmonic intersite potential, i.e. finiteness of stacking interactions is not taken into account there [201, 202, 204–210, 217]. Being optimized and parametrized some of them would supposedly describe local unwindings of DNA better than the Peyrard-Bishop-Dauxois model does. Nevertheless for none of them such improvements have been performed. Therefore for the moment, the PBD model is the only simple mechanical approach used to investigate the denaturation of a heterogeneous DNA. This model usually enables one to reproduce the melting profiles less adequately than the nearest-neighbor models described in 1.3. However, the nature of the latter ones is phenomenological. Therefore they cannot be considered to be an adequate instrument for the study of the physical nature of denaturation. This problem is thoroughly discussed in Section 6.2.

It is known that none of simple approaches can describe all the aspects of the denaturation behavior. Nevertheless taken together these models enabled researchers to clarify the physicochemical nature of many interesting experimentally observed DNA properties. So far we have mainly dealt with investigations of the DNA thermal denaturation. However this type of experiments is by far not a solitary one. A good part of denaturation features are revealed in the experiments on micromechanical cleavage of the duplex chains. It can be said that in this field experimental and theoretical investigations of DNA denaturation complemented one another still more successfully. In what follows the material on micromechanical denaturation is given in greater detail.

## 3. INVESTIGATIONS OF MICROMECHANICAL DENATURATION OF INDIVIDUAL DUPLEXES

Experiments on individual polymer molecules are remarkable for being able to provide information about the "mechanical" properties of an ensemble of chemical bonds. This information is a unique addition to the evidence about the molecular structure. There are quite a few DNA prioperties which would not have been discovered if the techniques of micromechanical denaturation had not been invented. Theoretical models, in turn, gave an insight into the physicochemical nature of DNA properties revealed in mechanical experiments.

### 3.1. Experimental data

Though the techniques of DNA duplex mechanical unwinding are complicated and labour-consuming, experimental material on the matter is rather abundant. Investigations of this type denaturation are described in detail in some reviews [222, 223]. Smith et al. were among the first to carry out micromechanical experiments [224]. They investigated the behaviour of a duplex affected by longitudinal stretching forces of 0.1–10 pN. These forces are sufficient for the DNA stiffness resulting from its configurational entropy (caused by its numerous natural curvatures) to be overcame [225]. As a result the distance between the ends of a helix first reaches its contour length (the length of the helix axis) and then becomes a bit longer due to a small deformation of the helix *per se*.

In subsequent experiments it was shown that if the external force of 65–70 pN is applied a cooperative conformational transformation of a duplex into another form takes place. In this case, its length increases 1.7-fold as compared to the native B-DNA. The new conformation received the name "S-DNA" [215]. As the external force decreased, S-DNA became relaxed and then conformational transition into the original B-DNA took place. These conformational transitions as well as denaturation appeared to be sensitive to the ionic strength and pH of the solution [226].

Behavior of DNA in the course of twisting was studied in similar experiments carried out by Allemand et al. [227]. The authors showed that in a negatively supercoiled molecule the probability of denaturation is rather high. However in the case of positive coiling the result was surprising: DNA transformed itself into a helix with a very small pitch distance and inside-out bases. Such a structure received the name P-DNA [227].

Subsequently it was shown that DNA can denaturate in the case of longitudinal stretching of only one chain [228, 229]. In this case denaturation is also preceeded by transformation into S-ДНК. The diagram of the experiment is shown in figure 3.1.

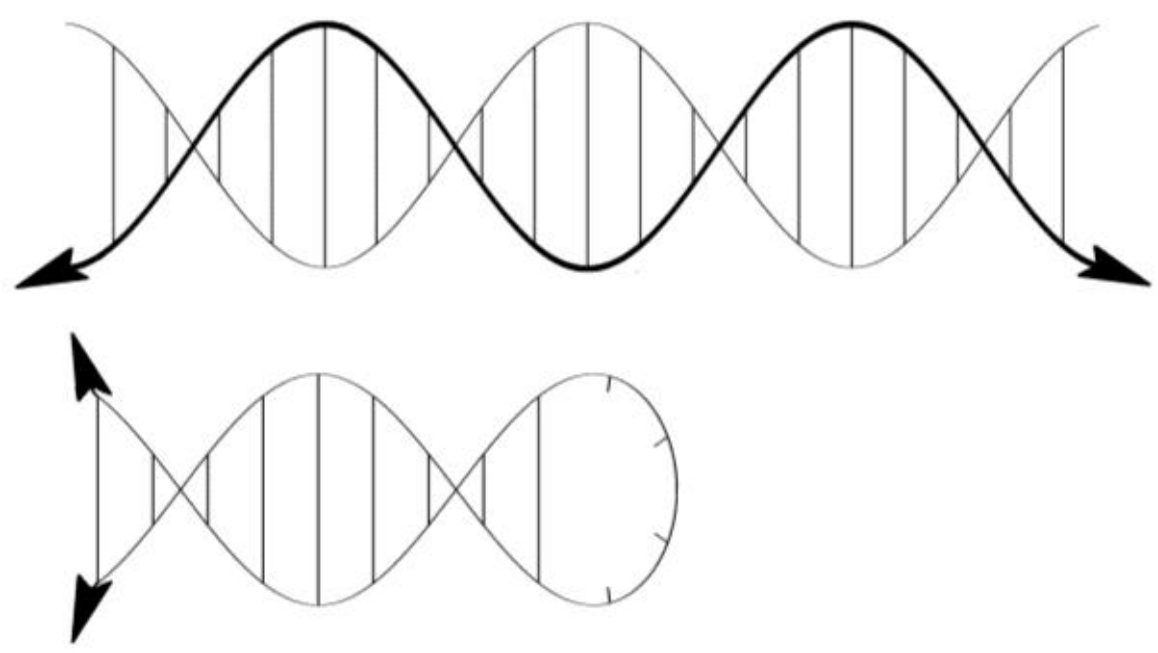

**Fig. 3.1.** Schematic representation of the experiment on longitudinal stretching DNA denaturation. At the top: denaturation of a native DNA. At the bottom: “transversal” unzipping of a hairpin that is formed in the course of relaxation of a homopolymer DNA. The diagram is adapted from theoretical work [180].

One end of a chain was attached to the microscope slide, the other – to the cantilever of the atomic force microscope. Its deflection enabled one to estimate the DNA resisting force [229]. If a regular heteropolymer was stretched, relaxation was accompanied by folding of (self-complementary) chains into hairpins. To break an adenine-thymine hairpin a force of 9 ± 3 pN was required, to break a guanine-cytosine one – a force of 20 ± 3 pN [229].

It was also shown in vitro that DNA of λ bacteriophage after B–S transition at 65 pN dissociates under the action of an external force of over 150 pN [229]. Moreover, it turned out that as the velocity of stretching decreases, the force required for denaturation also decreases: at very small stretching velocities the force of 70–80 pN was sufficient [228]. This peculiarity, as will be seen in what follows, appeared to be important for understanding the DNA denaturation behavior.

Experiments on folding and stretching of a two-chain DNA provided important information on its mechanical properties and characteristic conformational transitions. However, investigations were mainly focused on the attempts of DNA “mechanical sequencing”. They were based on the idea that a nucleotide sequence can be revealed from DNA resisting force which the duplex demonstrates when its chains are pulled by the ends. Early experiments on DNA denaturation by way of its “transversal” stretching were carried out by Bockelmann et al. in 1997 [230]. The diagram of the experiment is shown in figure 3.2.

An end of one DNA chain is attached to a glass microscope slide via a ds-DNA linker arm, the other chain is attached to a polystirol bead connected with a glass microneedle. The stiffness of the needle was very small – 1.7 pN/μm. The tip on the microneedle attached to the bead serves as a force lever. Force is measured by the deflection of the tip from its equilibrium position.

Bockelmann et al. showed that mechanical opening of DNA occurs via a series of pronounced stick-slip cycles, by analogy with a macroscopic solid friction. A force required to break one basepair varies in the range of 12–15 pN.

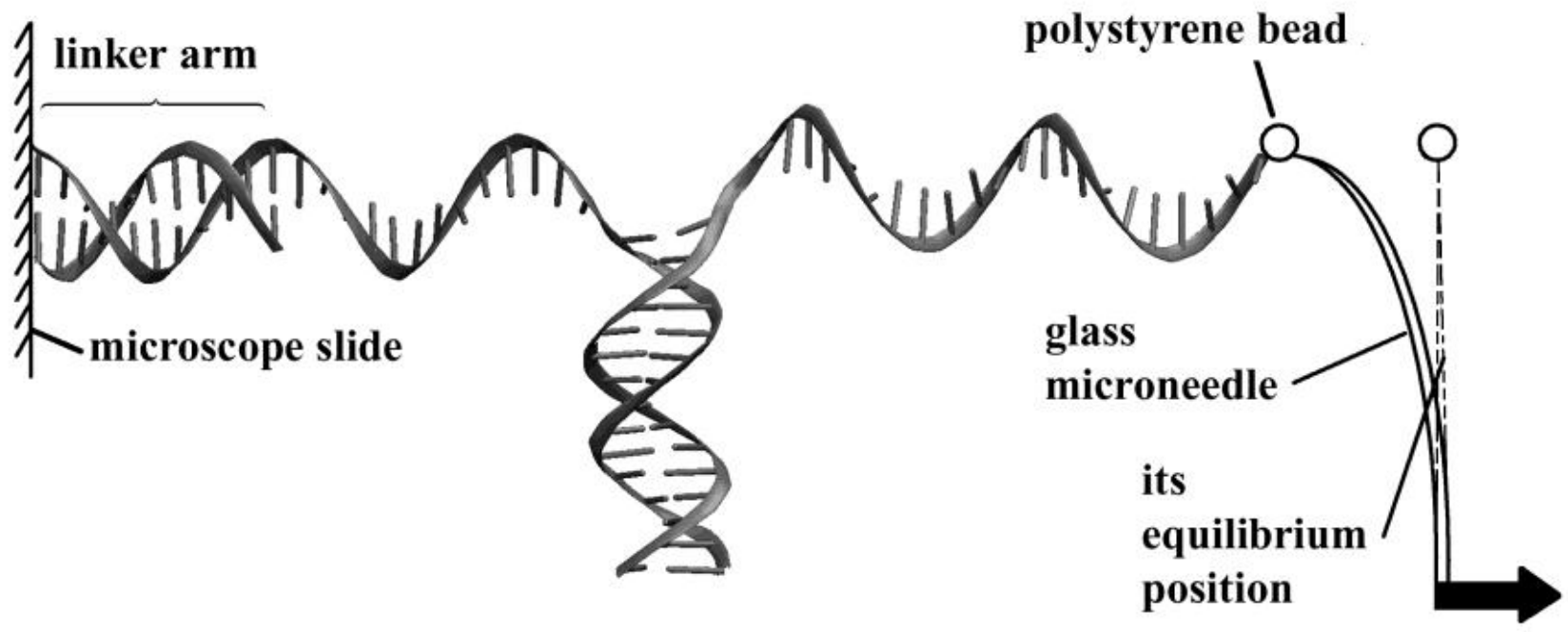

**Fig. 3.2.** Schematic representation of the experiment on the “transversal” micromechanical denaturation of DNA [230]. Explanations are presented in the text.

The same authors revealed a strong correlation (with a sequence resolution of the order of several hundreds of basepairs) between the nucleotide composition of local DNA fragments and a force required for their denaturation [231]. The denaturation profiles, i.e. the “cordinate-force” curves appeared to be absolutely symmetrical when DNA was pulled by the ends [231]. Besides, when the velocity of breaking increased from 20 nm/s to 800 nm/s they changed only slightly. The authors interpreted their results on the basis of equilibrium calculations. On the scale of hundreds or thousands of basepairs this approach is quite reasonable since transversal stretching proceeds extremely slow as compared to molecular motions.

Using a more precise tool – optical tweezers – the force profile resolution was improved to the scale of tens of basepairs [232]. The high resolution made local differences between the signals to be evident even for the opening velocity of 20 nm/s. According to the authors, these results suggest that the thermal equilibrium is not always observed in the course of measurements even when DNA unzipping proceeds at minimal velocity. At high resolution the observed signal flips between discrete values. Authors explained this behavior by transitions of the system between different minima in the complex energy landscape [232]. Therefore “mechanical sequencing” cannot be observed by methods of that time. Anyway a solution of this problem was found later: a model of “physical” sequencing by dragging a duplex through a nanopore is described [233]. The delay time in this case is determined by the temperature, the character of a nucleotide pair and the potentials’ difference.

### 3.2. Theoretical investigations of micromechanical denaturation. Analysis of equilibrium properties of the models

The models used to describe the conformational transition of a duplex caused by its longitudinal stretching are usually rather simple. DNA is presented as a chain of elements each of which can be in either of two states with the lengthes $l_1$ and $l_2$ – see [215] and references therein.

DNA denaturation under the action of a torsional stress and the influence of supercoiling on the melting process are investigated in a lot of works. This problem was studied both in Ising-like approaches [33, 35, 234–239], and in the radial-torsion models [175, 176, 240].

DNA supercoiling can fluctuate even in the absence of an external force, however these fluctuations are small. In this case the total torsional stress of a duplex is close to zero, i.e. DNA is relaxed. Stress-induced duplex destabilization caused by an external force or an interaction with specific enzymes can lead to more pronounced changes in the bubbles dynamics. Nevertheless, a detailed consideration of this case is outside the scope of this paper for the reasons given below.

First, while *in vitro* experiments on relaxed molecules provide information about the properties of the denaturation bubbles (however it may be indirect), there are no any similar techniques for investigation of a supercoiled stressed DNA.

Second, the supercoiled state is typical mainly for a living cell. Transport of a torsional stress along a duplex can be hindered not only by DNA-associated proteins but also by natural curvatures of DNA [241]. This can lead to a nonuniform distribution of supercoiling. As a result, its value on some or other fragment becomes unknown and cannot be determined. On the other hand, there is some evidence that the location of the fragments with maximum probability of unwinding depends on the extent of supercoiling [236, 242]. Besides, a considerable part of DNA in a eukaryotic cell is wound around histones. These facts make investigations of the bubble dynamics *in vivo* rather difficult.

Third, as was already mentioned in the Introduction, relaxed duplexes are better suited to the role of conductors in nanoelectronic devices. The bubble dynamics in such DNAs is determined entirely by the nucleotide sequence.

Notwithstanding the complexity of a supercoiled DNA, modeling of its dynamics gives an insight into the behavior of "torsional" bubbles. In particular, it turned out to be useful for prediction of the most probable places of the bubble openings *in vivo*. It was revealed that they often coincide with the places of a specific DNA-protein interaction [235, 236, 243, 244]. Moreover, recently some works on the coarse-grained molecular dynamics have been published where the behavior of bubbles in a negatively supercoiled plasmid is studied in the microsecond timescale [245].

The external force-induced DNA unzipping was studied with the use of the same models as the DNA melting. In the studies of micromechanical denaturation PS-type models were the most popular Ising-like approaches. Their main advantage is "fundamentality" of calculations of the changes in the DNA configurational enthropy which take place when the length of the denaturated fragment increases. Investigations of denaturation with the models of this group and comparison of the theoretical results with the experiments are presented in review by Kumar and Li [246].

PS-type models enabled elucidation of many physical bases of denaturation caused by a combined action of an external force and high temperature. The behavior of the models is studied for a wide range of the values of these quantities [247–252]. The cases when an external force is applied to complementary chains in the middle of the molecule (the so-called "eye phase") [248–251] and to their 5'-ends [252] are considered in detail. In the latter case the force causes denaturation as a result of shear displacement.

Success in the investigations of denaturation in the PS-type models was additionally contributed by the development of new methods for consideration of exluded volume interactions. For example, the use of mutually-attracting-self-avoiding walks (MASAW) enabled one to take account of the possibility of the formation of hairpin-like structures in the case of micromechanical denaturation [250]. The occurrence of complementary statistical segments in one and the same cell ("noncrossing" walks) was also excluded in the walks of this type.

The problem of phase transitions in quasi-one-dimensional systems is one of the most important in modern theoretical physics. PS-like models occupy a unique niche among the theoretical tools for its investigation. However, the dynamics of the transversal micromechanical denaturation has some peculiarities which are more conveniently studied in mechanical models. An example is provided by the slip-and-stick opening of the duplex described in the previous section. One of the well-known theoretical approaches where the nature of this phenomenon was studied is the model by Lubensky and Nelson [253, 254].

Lubensky and Nelson studied transversal unwinding of DNA by the methods of equilibrium statistical mechanics. The caharacteristic of the denaturation extent was $\mu$ – the number of open basepairs. In the absence of an external force the potential energy of the system depended only on μ: the positions of minima and maxima of this function were determined only by the nucleotide sequence. In the numerical investigations of the model it is shown that the applied force produces a "slope" of the potential energy surface [254]. This contributes toward irreversible stepwise transitions of the system between the minima whose depth grows as μ increases. It should be noted that this explanation is simplified. Having investigated equilibrium properties of the model, Lubensky and Nelson also demonstrated considerable differences in the cooperativities of the micromechanical unwinding of a homogeneous DNA and a heterogeneous DNA [253, 254].

Nevertheless, the value of the experiments carried out on individual molecules lies in the fact that they provide information on the dynamical properties of a duplex. An example is provided by fast oscillations of the signal observed in the course of micromechanical denaturation at high resolution [232]. It is also revealed that the reproducibility of the renaturation profiles obtained as the external force gradually decreased is considerably lower than the reproducibility of ordinary "force" unwinding profiles [232]. Another indirect proof of the nonlinear dynamics contribution is a considerable scatter in the values of the external force required for unwinding of homogeneius hairpins [229].

These phenomena are caused by inhomogeneity of energy distribution of the basepair stretched oscillations in a duplex, its "dynamical heterogeneity". Therefore their nature cannot be studied by equilibrium statistical mechanics. There is a need for theoretical investigations of the energy transfer and nonlinear localization in DNA duplex.

### 3.3. Role of energy transfer and localization in the dynamics of force-induced DNA unzipping

Impossibility of DNA "mechanical sequencing" had been shown theoretically several years before the relevant experiments were carried out [255, 256]. According to early works, a nucleotide sequence cannot be determined precisely because of thermal fluctuations of the single-stranded fragments which connect a duplex with a microneedle.

The contribution of the duplex dynamics *per se* was investigated later in a modified PBD model [257]. In the modified version the "standard" Hamiltonian contained an additional term describing the external force:

$$H=\sum_n\left[\frac{1}{2}m\dot{y}_n^2+D_n\left(e^{-a_ny_n}-1\right)^2+\frac{k}{2}\left(1+\rho e^{-\alpha(y_n+y_{n-1})}\right)\left(y_n-y_{n-1}\right)^2\right]-\frac{c_0}{2}\left(y_0-y_1\right)^2, \quad (3.1)$$

where $c_0$ is stiffness of the force lever, $y_0 = vt$ is its displacement whose velocity $v$ is known and constant.

M. Peyrard in [257] has shown for the first time that fluctuations would have appeared in the profiles of micromechanical denaturation even in the case of a homopolymer DNA. The reason of such behavior is "dynamical heterogeneity" caused by inhomogeneous localization of the stretched oscillations energy in a duplex. Interaction of an external force with the dynamics of nonlinear oscillations of nucleotide pairs is shown in figure 3.3 [257]. Along the *x*-axis are the coordinates of the basepairs, along the *y*-axis is the time in picoseconds. The dynamics of 512 basepairs during 1.276 ns is presented. The state of a pairs is shown in gray scale: the black colour corresponds to separation of the bases by more than 1.5 A, the white colour – to closed pairs.

It is evident from figure 3.3 that when the front of the mechanical break reaches the fragment where closed pairs prevail, a denaturation is delayed till the moment when the force-induced strain becomes large enough for it to be able to go on. Recommencement of the break leads to a release of the strain and a decrease of the force measured. A similar relaxation occurs in the case when the break front meets an open region – the process of breaking accelerates. In the case of a homogeneous DNA the slip-and-stick character of micromechanical unwinding cannot be described on the basis of the equilibrium theory.

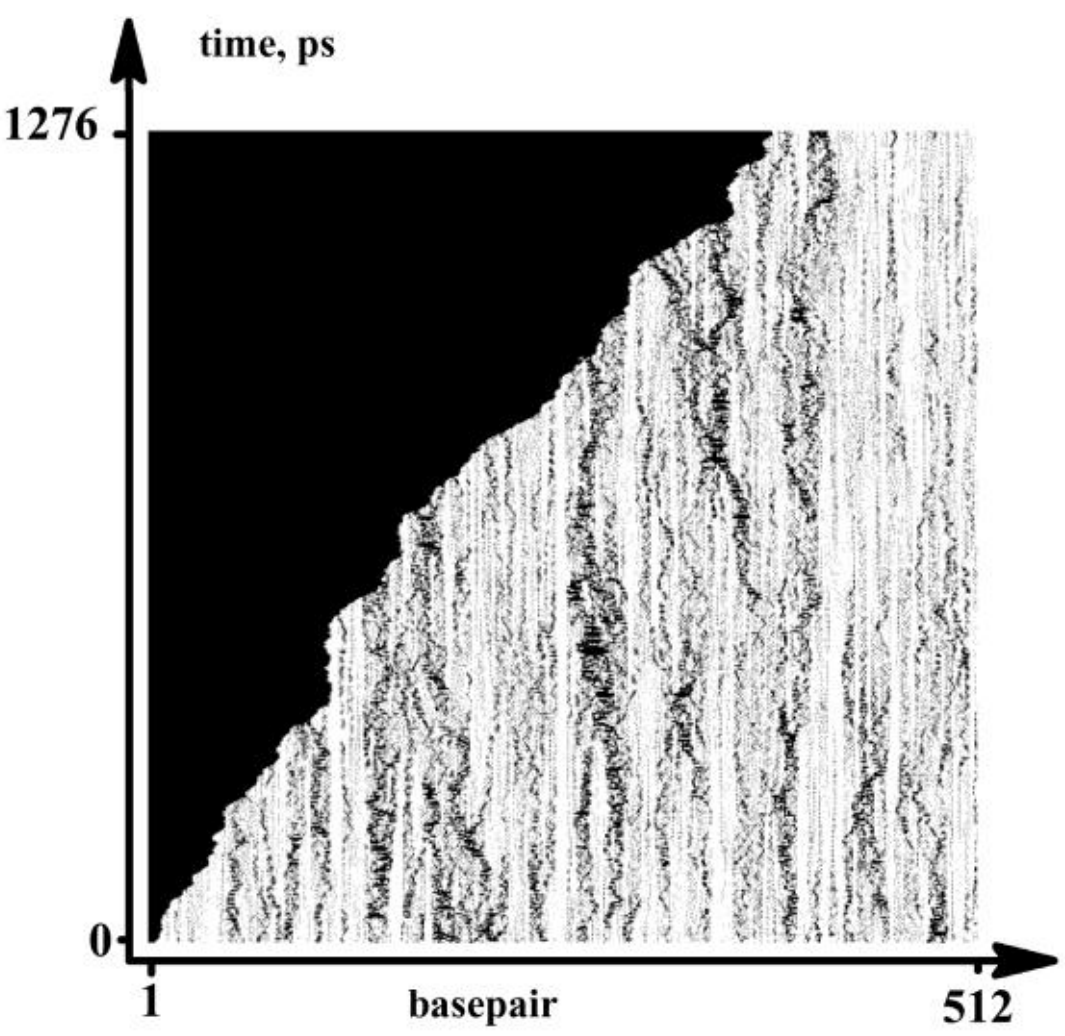

**Fig. 3.3.** Unzipping of a duplex consisting of 512 basepairs under the action of an external force in the PBD model with a term of external force. See text for details.

Later on the kinetics of DNA force unzipping in PBD model was investigated in greater detail in [13]. M. Peyrard carried out a mathematical analysis of this process. In particular, he studied how the effects of a force unzipping velocity on the resisting stress of a duplex. It should be noted that for reasons of computational resource economy use was made of very high denaturation velocities [13, 257] – 7–10 orders of magnitude higher than the experimental values. However, even under these conditions rather representative results were obtained. For the velocities of $2.5 \cdot 10^9$ – $2.5 \cdot 10^{11}$ nm/s, the calculation value of the external force was 64 pN. However, for $2.5 \cdot 10^8$ nm/s, the average force for breaking a duplex decreased to 16 pN, approaching the experimental data (nearly 13 pN) [13]. According to M. Peyrard's conclusion, it is quite a predictable result for the situation when the unzipping timescale goes down under the timescale of nonlinear localization of energy in duplex.

If we now turn back to works by Rief et al. [228, 229], see Section 3.1, it is obvious that a similar mechanism operates in the case when a micromechanical denaturation takes place as a result of stretching of one of the chains. It seems that if an external force is not large it merely stabilizes the local conformational changes arising due to thermal fluctuations. In this case S-DNA denaturates as a result of accumulation of a "critical concentration" of these changes. The external force just reduces the probability of the reverse transitions of the duplex's local fragments into more stable conformations.

Another important property of the duplex revealed in the experiments on individual molecules is the availability of a force barrier which make it difficult to onset a force unzipping [258]. Its existence was predicted by Cocco et al., who used a semi-macroscopic variant of the combined model [178]. In the authors' opinion, the existence of the barrier can be easily explained thermodinamically, since on a small fragment where the transversal stretching of the duplex takes place at a particular moment, stacking interactions survive. The scheme of this process is shown in figure 3.4. Subsequent decay of stacking partly compensates the free energy costs for breaking H-bonds (owing to enthropy growth). Obviously, this effect takes place throughout the denaturation process, excepting its onset.

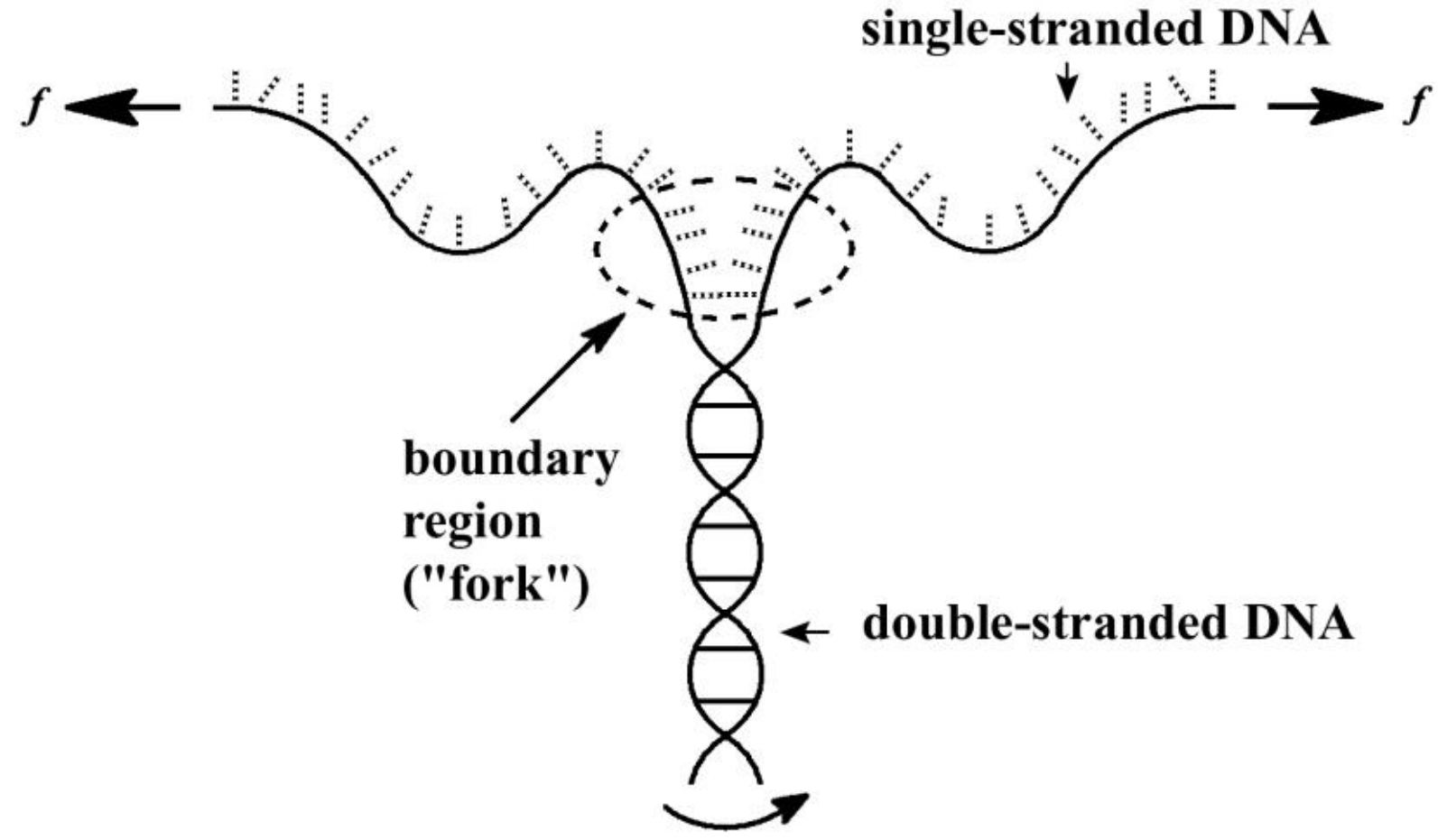

**Fig. 3.4.** A schematic view of a "transversal" force unzipping of a DNA molecule. The boundary region of an unwinding duplex is responsible for the occurrence of the force barrier. The boundary region moves along the duplex in the course of DNA micromechanical denaturation which proceeds under the influence of an external force $f$ [178].

On the microscopic scale the barrier arises as a result of superposition of the coordinate derivatives of the Morse potential and anharmonic stacking [178]. Later the force barrier was reproduced in the PBD model [259]. In paper [259] N. Singh and Y. Singh thoroughly investigated how the value of the barrier $F_c$ depends on the temperature, energy of hydrogen bonds in the basepairs and the parameters of the anharmonic stacking potential. It was revealed that the force barrier is determined only by the model parameters and independent, for example, of the duplex fragment to which the external force is applied. The values of $F_c$ calculated for the cases of breaking at an end and breaking in the middle of the chain were in fact similar. The dependence of the external force on the distance traveled by the force probe is presented in figure 3.5.

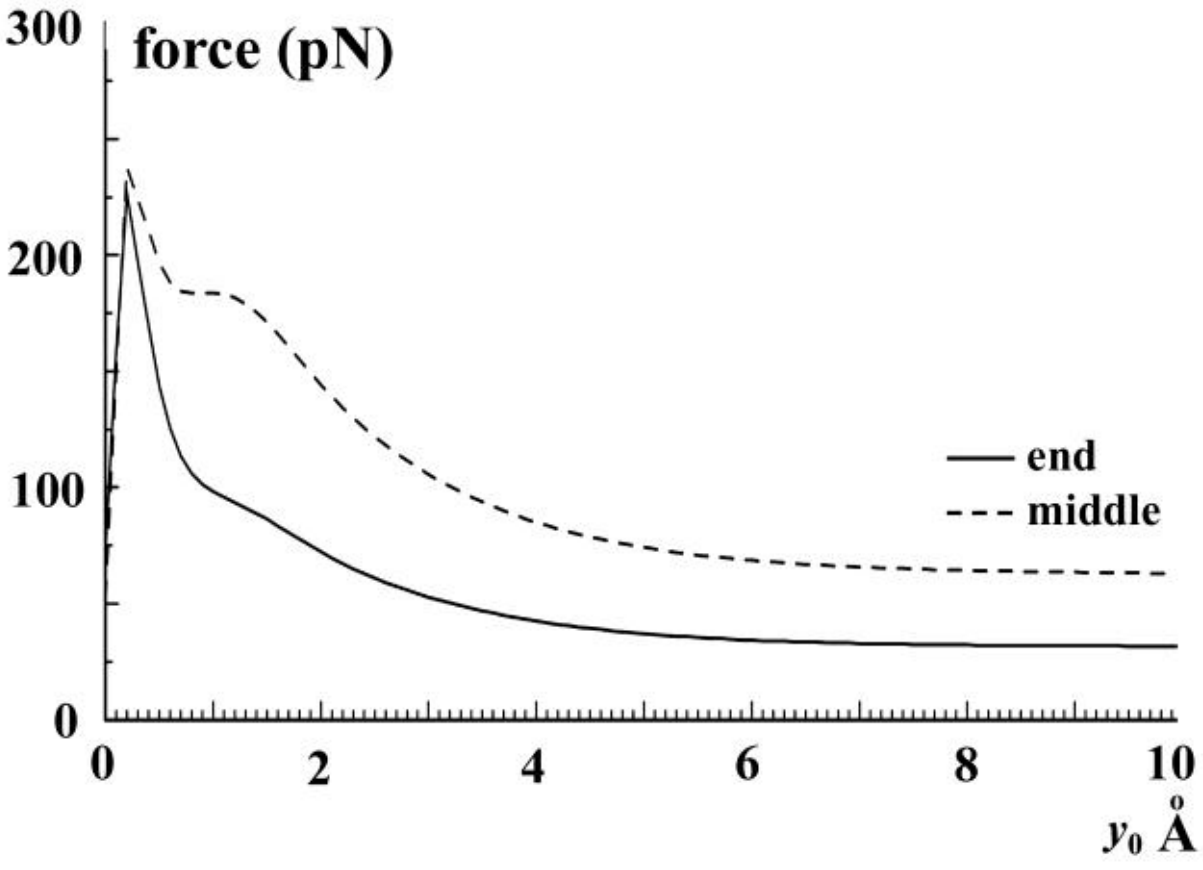

**Fig. 3.5.** Dependence of the averaged external force on the probe displacement in the PBD model [259]. Solid line indicates the graph for the case of breaking at an end of a duplex, dashed line – for the case of breaking in the middle of a duplex.

Stiffness of the probe $c_0$ (see expression (3.1)) that moves at a constant velocity turned out to be a one more factor determining the value of the force barrier. This dependence was investigated by Monte Carlo method in the work by Voulgarakis et al. [260]. A decrease of $c_0$ led to a gradual decrease of the barrier which completely disappeared for $c_0 \leq 16$ pN/nm. According to the authors this fact has a simple explanation. When a probe possessing small stiffness moves, a second energy minimum corresponding to an open state of the first nucleotide pair gradually emerges. This minimum gradually deepens, and the barrier separating two potential wells smoothes out. When the difference $y_0 - y_1$ reaches a critical value, nucleotide pairs at the ends break and a force-induced micromechanical denaturation starts. The value of 16 pN/nm corresponds to approximately 0.001 eV·Å$^{-2}$ which is, for example, 25 times less than the elastic constant of stacking, see [221]. In the other limit case, i.e. when the probe stiffness is very large the second minimum is not formed. For $c_0 \geq 10$ eV·Å$^{-2}$ the value of the force barrier is determined only by the maximum coordinate derivative of the total potential.

Theoretical investigations of the slip-and-stick micromechanical breaking, force barrier at its onset, conformational transitions of a duplex and of other peculiarities of the DNA denaturation behavior have made a considerable contribution into the physics of biopolymers. They allowed refining a lot of peculiarities of a micromechanical denaturation which were not revealed by experimenters and to clarify their physical nature. In turn, experiments on the force denaturation of individual duplexes enabled one to test out different models of DNA and improve them.

At present we witness active investigations of the denaturation dynamics of DNA which require joint efforts of theoricians and experimenters. One of the most actual problems in this field is to study the bubble dynamics in a heterogeneous DNA at temperatures much lower than $T_m$. This problem is discussed in what follows.

## 4. EXPERIMENTAL AND THEORETICAL INVESTIGATIONS OF DENATURATION BUBBLES AT VARIOUS TEMPERATURES

The main subject of the previous Chapters was theoretical and experimental investigations of a complete denaturation – a process resulting in DNA dissociation into 2 strands. Even Chapter 2 devoted to mechanical models is focused mainly on radial and radial-torsion approaches in which a complete denaturation is possible. Experimental investigations of DNA melting allowes improvement of the models and their parametrization. In turn, the improved and parametrized models gave insight into the physical nature of cooperative DNA melting. However, the use of theoretical approaches is not limited by investigation of a complete denaturation. Mathematical modeling has played a great role in the studies of the open states arising at temperatures considerably lower than melting temperature $T_m$.

### 4.1. Behavior of bubbles in short DNA. Self-complementary oligomers quenching technique

As was already mentioned in Section 1.3, DNA fragments with different concentrations of AT-pairs denaturate at different temperatures. This property of the heteroduplex appeared to be useful in the studies of denaturation with the use of photometry in the wavelength range from 260 to 268 nm. Among theoretical investigations in this field an important role belongs to the attempts to understand the role of energy transfer and localization in the bubble dynamics with the use of the PBD model.

Short DNA consisting of two or three fragments which greatly differ in the concentration of AT-pairs is the most convenient object for mechanical modeling. In view of a short length of the bubbles arising in AT-rich domains, the effects of excluded volume in the course of denaturation are insignificant in such DNA, see Section 1.2. Therefore the calculated melting profiles of the oligomers are siutable for comparison with the experimental results.

The main disadvantage of short duplexes is the finite size effects. The normalized photometer signal is approximately equal to $1 - \theta_{ext} \cdot \theta_{int}$, see expression (1.10). If a molecule's length is short, the portion of nondissociated duplexes $\theta_{ext}$ starts decreasing at approximately the same temperature as the signal *per se* apprears. In other words one cannot determine what contribution into the reduction of hypochromism is made by the denaturation bubbles and what contribution – by completely dissociated molecules.

The finite size problem was successfully solved by G. Zocchi's research team in 2003 [54, 55]. Their method received the name "quenching technique". It is based on the capacity of single strands of self-complementary oligomers to form hairpin structures .

When a DNA solution (heated to a certain temperature) is sharply cooled down the initial duplex is formed only by the molecules which failed to denaturate completely. In the course of cooling single chains have no time to find a partner for the formation of a double helix. Being self-complementary, they form hairpins and no longer can form the initial duplex. The scheme of the method is shown in figure 4.1.

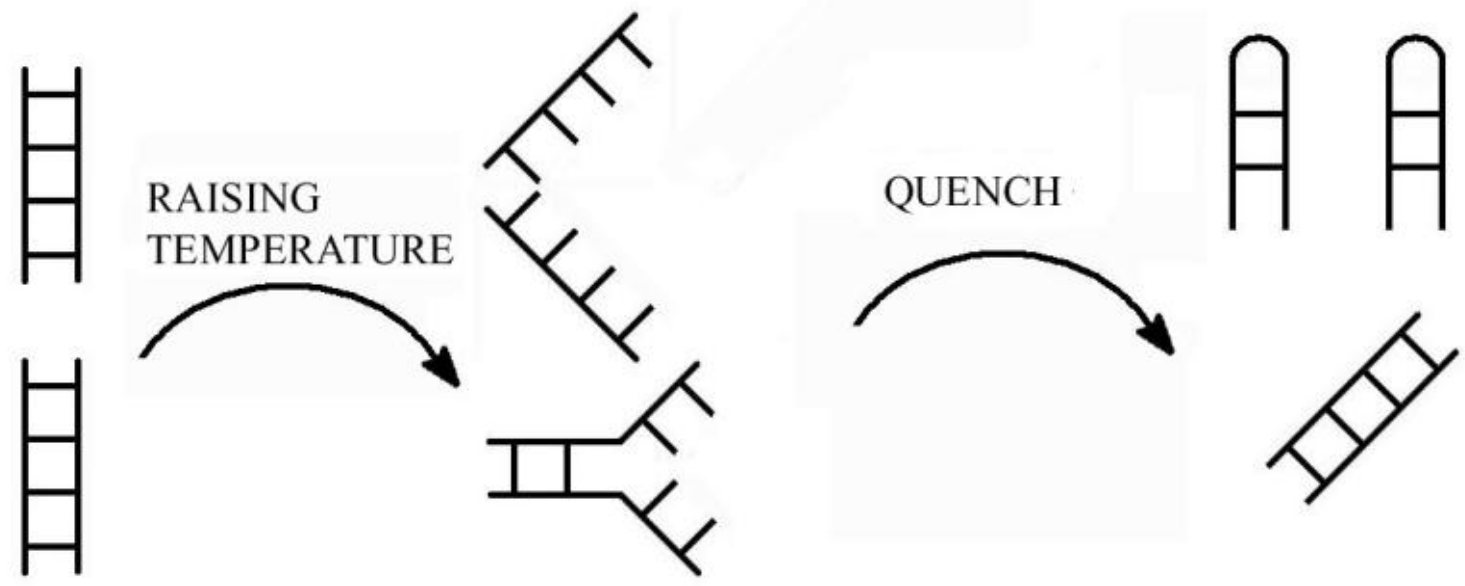

**Fig. 4.1.** Scheme of self-complementary oligomers' quenching technique. Explanations are in the text.

Since the molecular mass of the hairpins is twice less than the mass of the duplexes, they can be separated by electrophoresis in agarose gel. Having dyed the gels with ethydium bromide one can easily obtain a fluorescent profile – a curve for the fluorescence dependence on a linear coordinate. A relative concentration of hairpins and duplexes is determined from the ratio of the the areas under the peaks on the profile.

Hence, combining photometer measurements with quenching technique one can calculate not only the portion of open basepairs $f(T)$, but also the concentration of completely dissociated molecules $p(T)$. Using these quantities one can easily find $<l>(T)$ – the averaged (over the subset of the partially open molecules) fractional length of the bubble [55]:

$$<l>(T)=\frac{f(T)-p(T)}{1-f(T)}. \qquad (4.1)$$

Montrichok et al. studied two sequences which they conventionally denoted as $L_{48}AS$ and $L_{42}V_1$ [54]. The hairpin structures formed by these oligomers are indicated in figure 4.2.

Oligomer $L_{42}V_1$ has AT-rich domain in the center. The length of GC-rich domains at the ends of the duplex is small, only 12 basepairs. Therefore completely dissociated molecules start appearing in the solution at relatively low temperatures. At 60°C their fraction attains the value 0.1. Denaturation behavior of $L_{42}V_1$ can suggest the role of large fluctuations in the AT-rich domains: these fluctuations "break" GC-rich domains at the ends [54].

$L_{48}AS$ has the GC-rich domain at an end and its length is 27 basepairs. Therefore $p(T) \approx 0$ even at 65 °C, though $f(T)$ starts increasing at about 45 °C – at the same temperature as it does in $L_{42}V_1$. However on the interval 75–80 °C there is a sharp spike in the fraction of dissociated duplexes – from 0.15 to 0.97. At the same time the temperature for which $f(T) = 0.5$, is about 67 °C. In other words, the contribution of completely dissociated molecules into the total photometry signal at $T_m$ is very small. Gradual increase of $f(T)$ for $L_{48}AS$ suggests low cooperativity of AT-rich domains denaturation. Flattening out of $<l>(T)$ (nearly 0.7) at 75 °C and a subsequent sharp spike of $p(T)$ suggest that in GC-rich domains the cooperativity of denaturation is, on the contrary, very high.

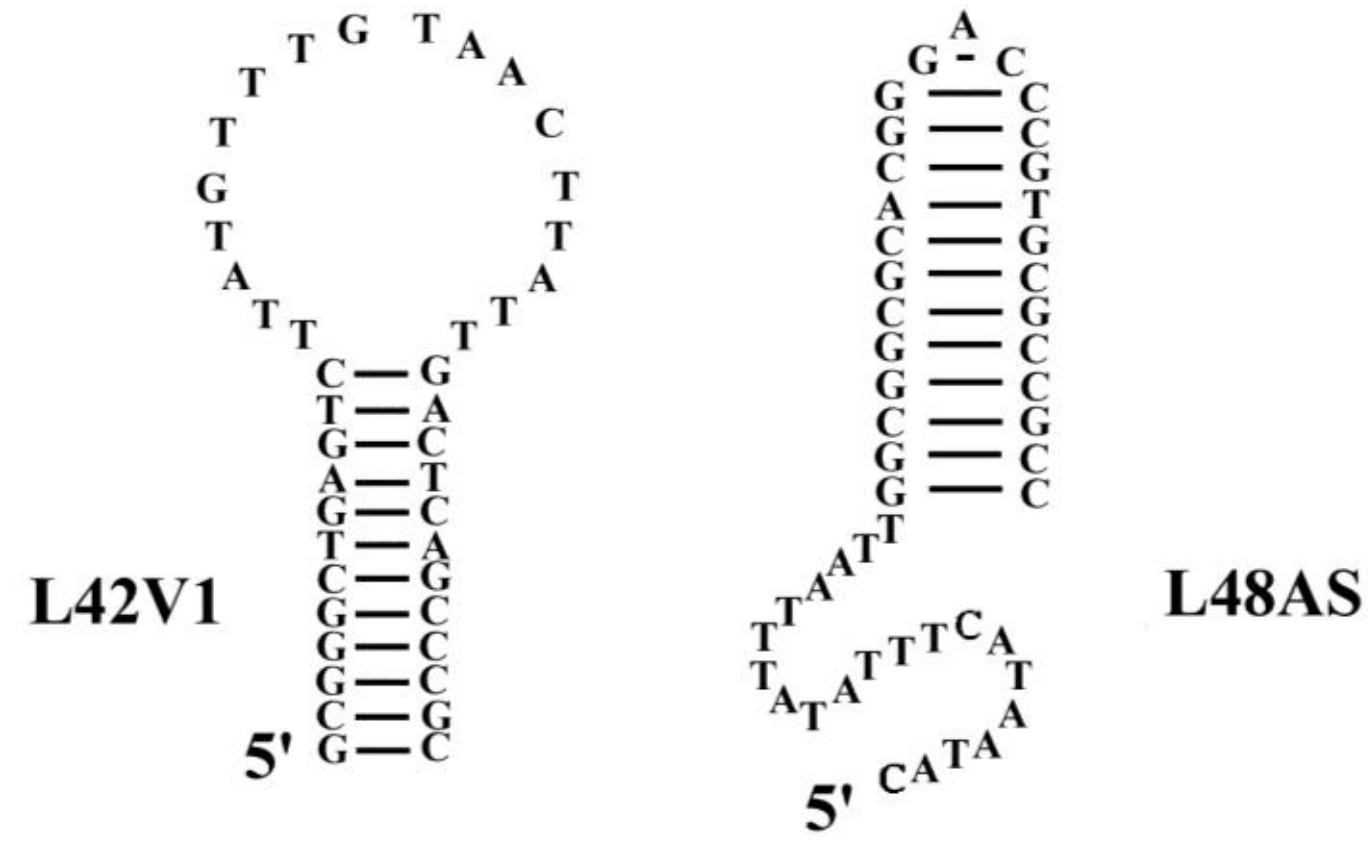

**Fig. 4.2.** Hairpin structures formed as a result of quenching of $L_{48}AS$ and $L_{42}V_1$ oligomers [54]. The length of $L_{48}AS$ and $L_{42}V_1$ is equal to 48 and 42 basepairs respectively.

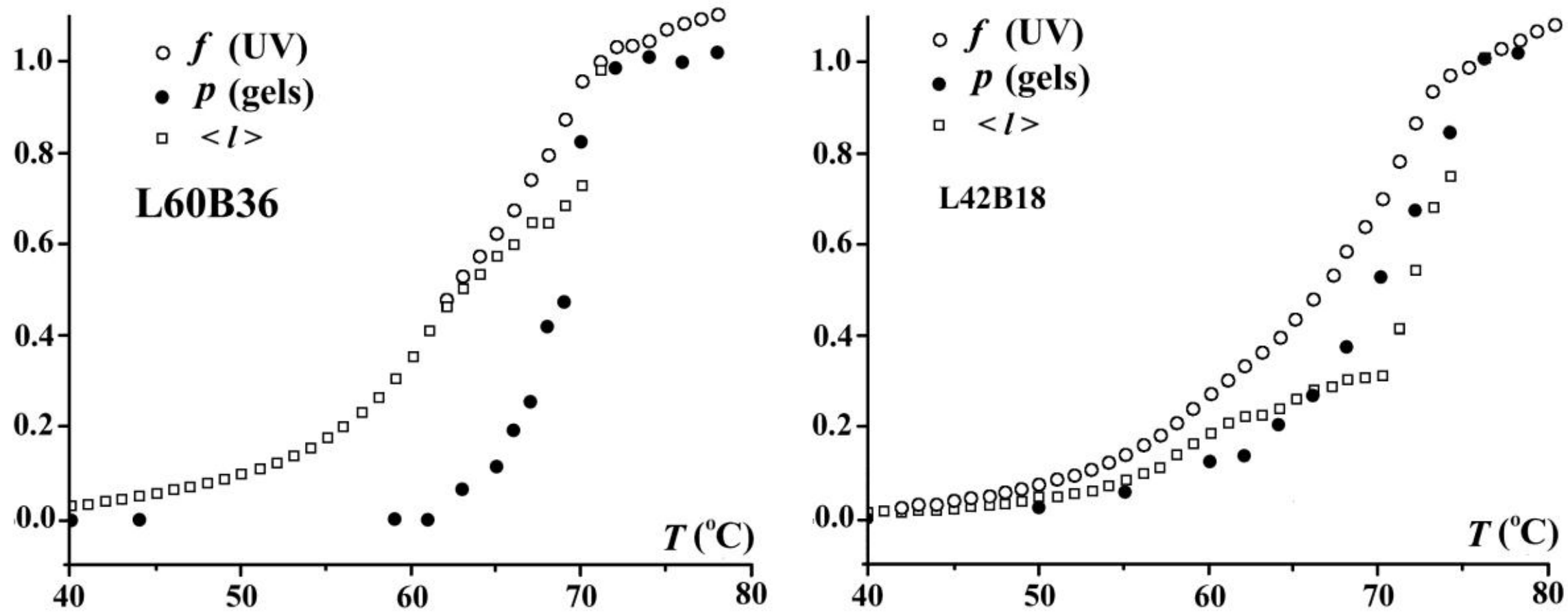

**Fig. 4.3.** Melting profiles of $L_{60}B_{36}$ and $L_{42}B_{18}$ oligomers [55]. Solid circles – *p*(*T*), empty circles – *f*(*T*), squares – <*l*>(*T*).

Subsequently the technique was improved and the properties of oligomers with AT-rich domains in the midle of the molecule were investigated more thoroughly [55, 261]. In paper [55] two oligomers – $L_{60}B_{36}$ and $L_{42}B_{18}$ were studied. The first one consisted of 60 basepairs and included AT-rich domain of 36 pairs length, for the second one these values were 42 and 18, respectively.

It is evident from the melting profiles shown in figure 4.3, that the length of a bubble <*l*>(*T*) in both the oligomers flattens out due to melting of AT-rich domains, but as the temperature increases further it demonstrates a sharp spike. This supports the data on the high melting cooperativity GC-rich domains in DNA.

For $L_{60}B_{36}$, the value of <*l*>(*T*) on the plateau is approximately equal to 0.6. This value coincides with the relative length of the AT-rich domain (36/60 = 0.6). At the same time the plateau *per se* is poorly marked. For an "opposite" characteristic of melting cooperativity, the quantity $\sigma(T) = f(T) - p(T)$ was introduced. This quantity reflects the amount of basepairs involved in the denaturation bubbles. For $L_{60}B_{36}$, the value of $\sigma(T)$ at 65 °C was 0.5, i.e. approximately 0.5/0.6 = 0.83 molecules were in the state of a duplex with an open central part. Further melting of GC-rich end domains proceeds with high cooperativity, as is seen from figure 4.3.

The denaturation behavior of L42B18 was different. The plateau achieved by <*l*>(*T*) on the interval from 40 to 70 °C, was well-marked. The value of <*l*>(*T*) on the plateau was approximately 0.3, while the ratio between the length of the AT-rich domain and the total length of the duplex was equal to $18/42 \approx 0.43$. The maximum value of $\sigma(T)$ was approximately 0.2. These differences suggest a higher cooperativity of the denaturation of $L_{42}B_{18}$ as compared to $L_{60}B_{36}$, though the increase of $<l>(T)$ occurs equally sharply in both the oligomers when heated to a temperature higher than 70 °C. As the oligomer length grows down the maximum $\sigma(T)$ further decreases. In oligonucleotide $L_{33}B_{9}$, which was studied in the next work of the research team, $\sigma(T)$ did not exceed 0.1 which is approximately 3 times less than 9/33 [261].

For the duplexes with the AT-rich domain surrounded by the GC-rich end domains of equal lengths, a "critical length" was obtained by way of extrapolation which was found to be approximately 20 basepairs. A shorter DNA possessing a similar structure dissociates "at once" without forming an intermediate with a bubble in the middle [261]. Oligomers having the AT-rich domain in the end demonstrated quite a different dependence of the denatutation behavior on the oligomer length. For these sequences, extrapolation to zero value of $\sigma(T)$ yields a critical length close to zero [261].

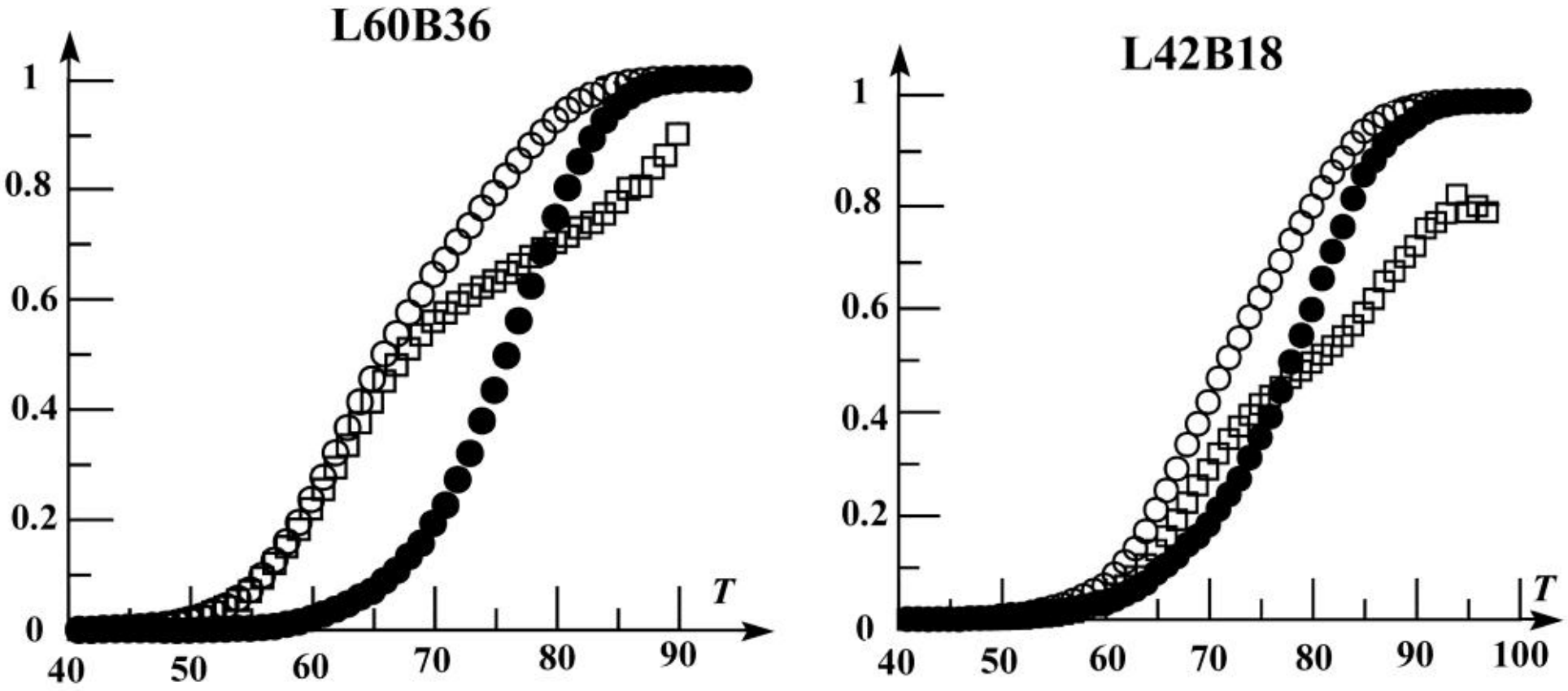

**Fig. 4.4.** Calculated denaturation profiles of short self-complementary oligonucleotides $L_{60}B_{36}$ and $L_{42}B_{18}$ from paper [55], obtained by Ares et al. [262]. Solid circles – $p(T)$, empty circles – $f(T)$, squares – $<l>(T)$.

To describe the results of the experiments G. Zocchi's research team used the nearest-neighbor approach [62, 261]. Besides, these results were interpreted by Ares et al. in the PBD model [262]. Using the standard Metropolis algorithm [263], Ares et al. investigated the PBD model by Monte-Carlo method having performed a number of short realizations. Figure 4.4 illustrates calculated denaturation profiles of oligonucleotides $L_{60}B_{36}$ and $L_{42}B_{18}$.

As is seen from comparison of figure 4.3 and figure 4.4, the data obtained for $L_{60}B_{36}$ are in agreement with the experimental results though the plateau is marked not so well. For $L_{42}B_{18}$ the dynamics of $<l>(T)$ is reproduced less adequately, though at the qualitative level the main features of this dependence coincide with the experimental data.

Ares et al. also investigated how the character of denaturation of short DNA depends on their length and composition [262]. For the characteristic of the statistical significance of the bubbles, the authors introduced a special quantity $\sigma_{av}$. Its value reflects the maximum portion of basepairs occuring in the bubbles rather than in individual chains. Strictly speaking, $\sigma_{av}$ is not a maximum difference $f(T) - p(T)$, but represents a difference of the areas restricted by the curves $f(T)$ and $p(T)$ divided into the half-width of the temperature interval.

Figure 4.5 shows the calculated dependence of $\sigma_{av}$ on the DNA length for oligomers with the AT-rich domain in the middle of the duplex and for the molecules with the AT-rich domain at an end. For the former group, extrapolation yields a critical length equal to approximately 22 basepairs. For the latter group this length in close to zero. It is evident from figure 4.5 that the probability of the chain separation without the formation of an intermediate is actually coincides with the experiment for both the sequences.

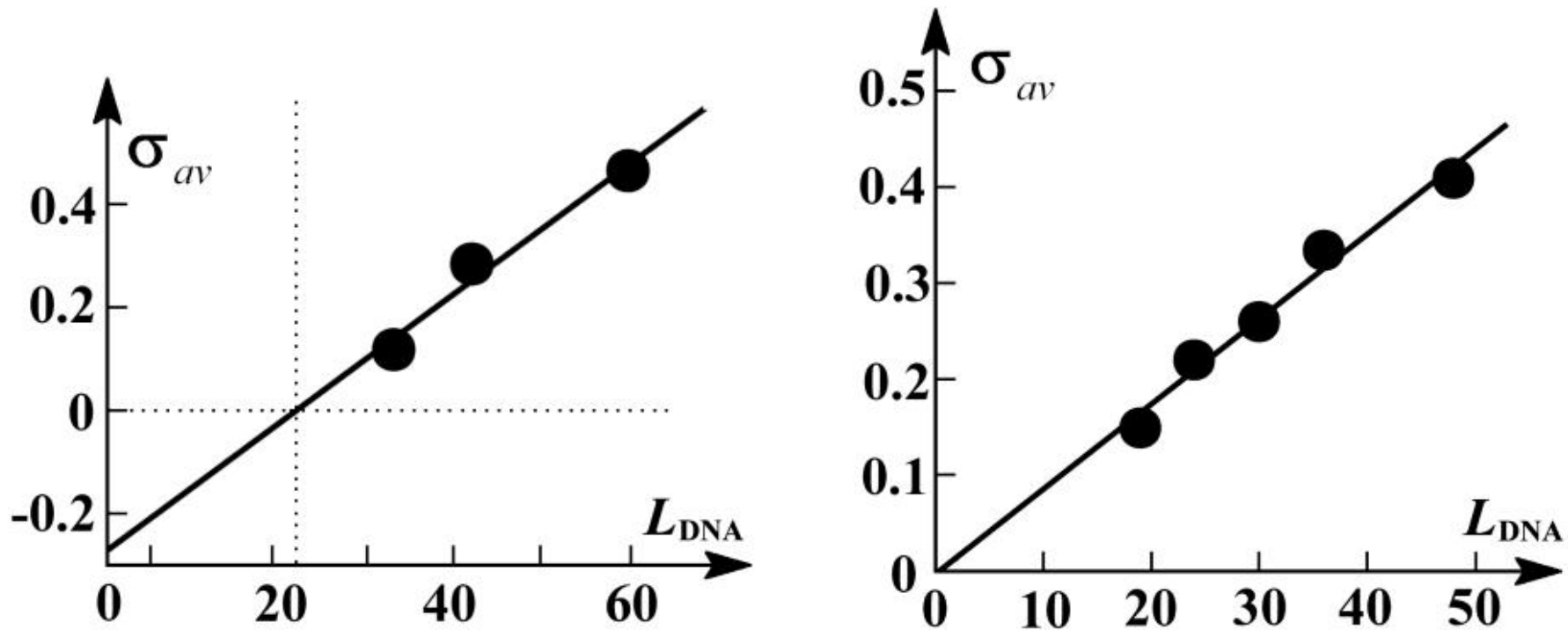

**Fig. 4.5.** Simulated dependence of the open basepairs' maximum fraction $\sigma_{av}$ on the oligomer length $L_{\mathrm{DNA}}$ for molecules with the AT-rich domain at the center (on the left) and at an end (on the right) [262].

Satisfactory agreement between the simulated and experimental profiles enabled concluding that the PBD model has no need for an additional fitting of the parameters available or introduction of new ones [262]. However soon van Erp et al. investigated the model by a different method which received the name "direct integration method" [264], see also [265]. Introduction of a displacement potential which prevents entire separation of the chains enabled them to study the duplex behavior over long periods of time. The calculated profiles were inconsistent with the experiment. Van Erp et al. made a conclusion that for a short DNA, the PBD model needs an additional optimization, though the adequacy of the approach *per se* is in no doubt [264]. In particular, they proposed that heterogeneity of stacking interactions should be introduced.

On the other hand, qualitative agreement of the calculations performed by Ares et al. [262] with the experimental data is an essential result which illustrates the possibilities of the PBD approach. In particular, one of the key achievements is the calculation of the critical length. Apparently it could not have been calculated on the basis of the nearest-neighbor models. Besides, the difference in the melting cooperativities of AT- and GC-rich domains is described at the qualitative level, however the influence of strong fluctuations of denaturated AT-rich domains is not reproduced [262].

Hence, a combination of UV-photometry with the hairpin quenching technique enabled the discovery of some important characteristics of a heterogeneous DNA unwinding. First, this is the influence of open fragments on the remote closed ones. Second, it is a strong dependence of DNA fragments' melting cooperativity on their base composition. Theoretical calculations, in turn, demonstrated that it is very important to consider the transfer and linear localization of the energy when modeling the denaturation behavior. The role of these processes in the duplex dynamics considerably grows when passing on to temperatures well below $T_m$. This was clearly demonstrated in S1 nuclease cleavage assay experiments which will be described in what follows.

## 4.2. Investigation of denaturation bubbles by S1 nuclease cleavage assay. Roles of energy transfer and localization

A direct comparison of the simulated data with the experiments is one of the key problems of DNA modeling. In the investigations of the duplex dynamics at low temperatures it was partly solved by enzymatic hydrolysis of "low-temperature" denaturation bubbles [29]. Hydrolysis was performed with endonuclease S1. This enzyme effectively cleaves a single-stranded molecule however demonstrates little activity towards a duplex [266]. Hydrolysis of a double-stranded DNA by endonuclease S1 proceeds 75000 times slower than the cleavage of a single-stranded molecule [266]. It is widely accepted that this reaction is caused by hydrolysis of denaturated (single-stranded) fragments only. Its bulkiness (molecular weight is about 31 kDa) hinders interaction with small denaturation bubbles, but not with the large ones.

Endonuclease S1 cleavage assay in combination with investigation of the PB model for the same sequences was first used in the work by Choi et al. [29]. Relatively low heat stability of DNA promotor regions is a well known fact, however such investigations were performed only for temperatures close to $T_m$ [128]. Hence Choi et al. were the first to investigate opening of a heterogeneous DNA at a moderate temperature of 28 °C. The authors assumed that initiation of the transcription occurs as a result of the enzyme contact with the transiently denaturated fragment of DNA.

To test the hypothesis the authors used stochastic dynamics (Langevin equation) in the PB model. In the course of stochastic calculations an event of a bubble emergence was registered every time when a group of 10 or more basepairs was transiently separated by a distance longer than the threshold one, which was 2.1 A. In this case the coordinate of a bubble was believed to be the base occuring at its center. In the course of a computational experiment they performed 100 separate PB model realizations per a sample with a running time of 1 ns each. Having averaged the calculations over realizations Choi et al. obtained the instability profiles which represented the dependence of the bubble formation probability on the coordinate (nucleotide pair) [29].

The simulated instabiliy profiles were compared with relevant profiles obtained in the experiment. After end labeling of the encoding chain of each sample by phosphorus isotope $^{32}$P they were incubated with S1-endonuclease at 28 °C. After that the hydrolysis products were separated by electrophoresis. The cleavage points were determined from the mass of the products obtained as a result of hydrolysis. The relation of cleavage probabilities at different point of DNA basepair sequence was assessed by relation of areas on the experimental cleavage density profile.

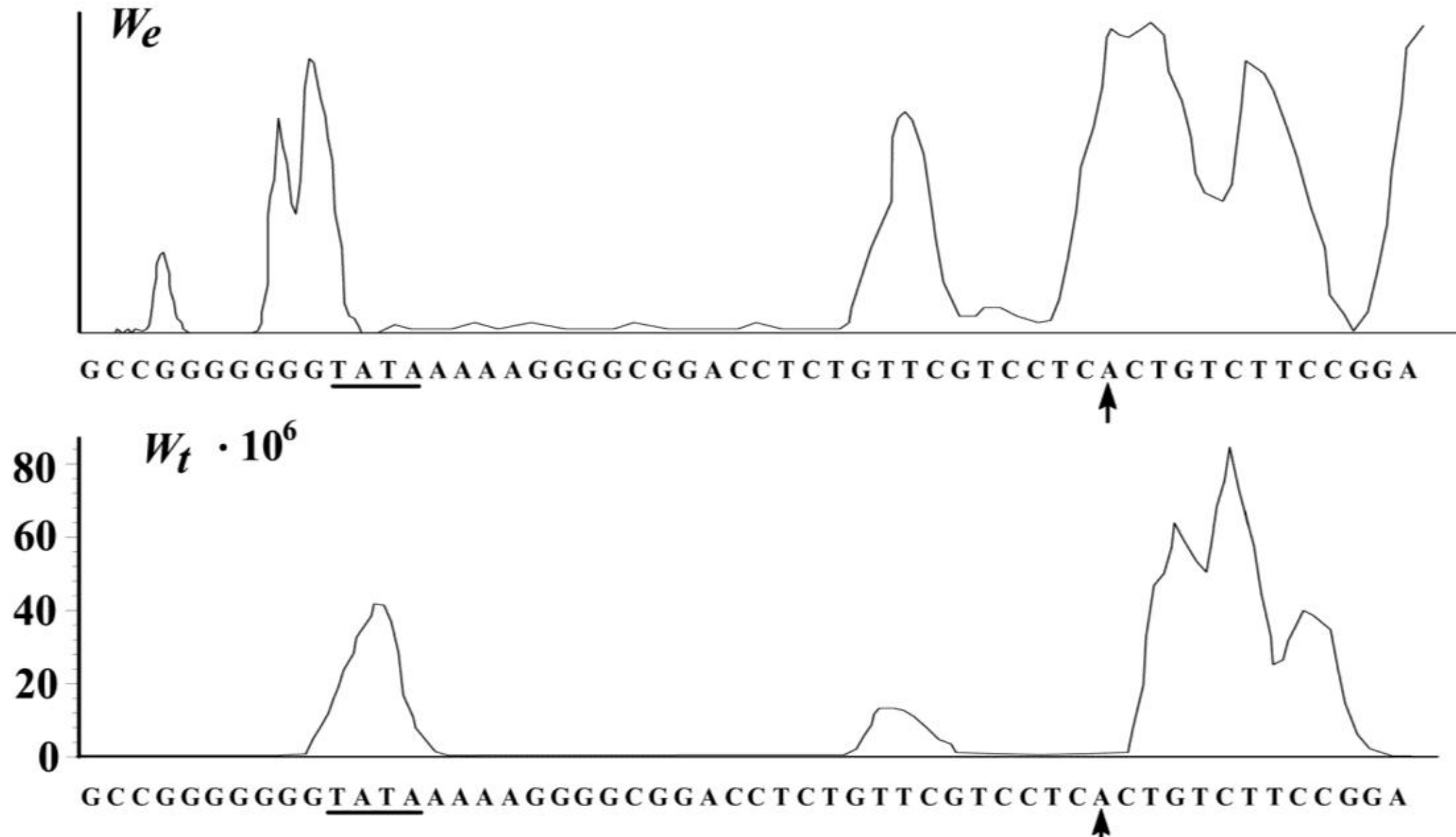

**Fig. 4.6.** Comparison of the experimental (at the top) and calculated (at the bottom) instability profiles for the second sample [29]. The transcription start site is indicated by the arrow, the TATA-box is underlined. $W_t$ is the calculated probability of the occurrence of a given basepair in a bubble consisting of 10 or more basepairs; $W_e$ is a relative value of this probability obtained experimentally.

In the work by Choi et al. [29] four DNA samples with known primary structures were investigated. The first sample was a fragment of a human gene consisting of 62 pase pairs. It did not contain any regulatory regions and was used as a control. It was shown in the experiment and in the calculations that in all the fragments of this sample the probabilities of a denaturation bubble formation are approximately similar and very small.

The second sample was a fragment of adenovirus DNA consisting of 86 basepairs containing AdMLP promoter. The fragment included the transcription start site (TSS) and the site of regular binding of TBP-protein (TATA-box). The primary structure of the central region of the second sample which included these sites is given as the absciss of the instability profiles shown in fig. 4.6 and 4.10.

In figure 4.6 the results of modeling are compared with the experimental instability profiles obtained for the second sample. There is a good agreement between these results. In both the profiles one can see clear peaks in the TSS and TATA-box and also a small peak in the vicinity of the 9-th pair from TSS to the 5'-end. The positions of the calculated peaks are shifted by 2–3 basepairs towards the 3'-end, but the width and the relative intensity in fact coincide.

The third and the fourth samples are a fragment of a virus DNA with AAV P5 promoter and its mutant variant in whose TSS A and T bases are replaced by G and C respectively. Apart from TSS, the fragments include the binding site for YY1 regulatory protein.

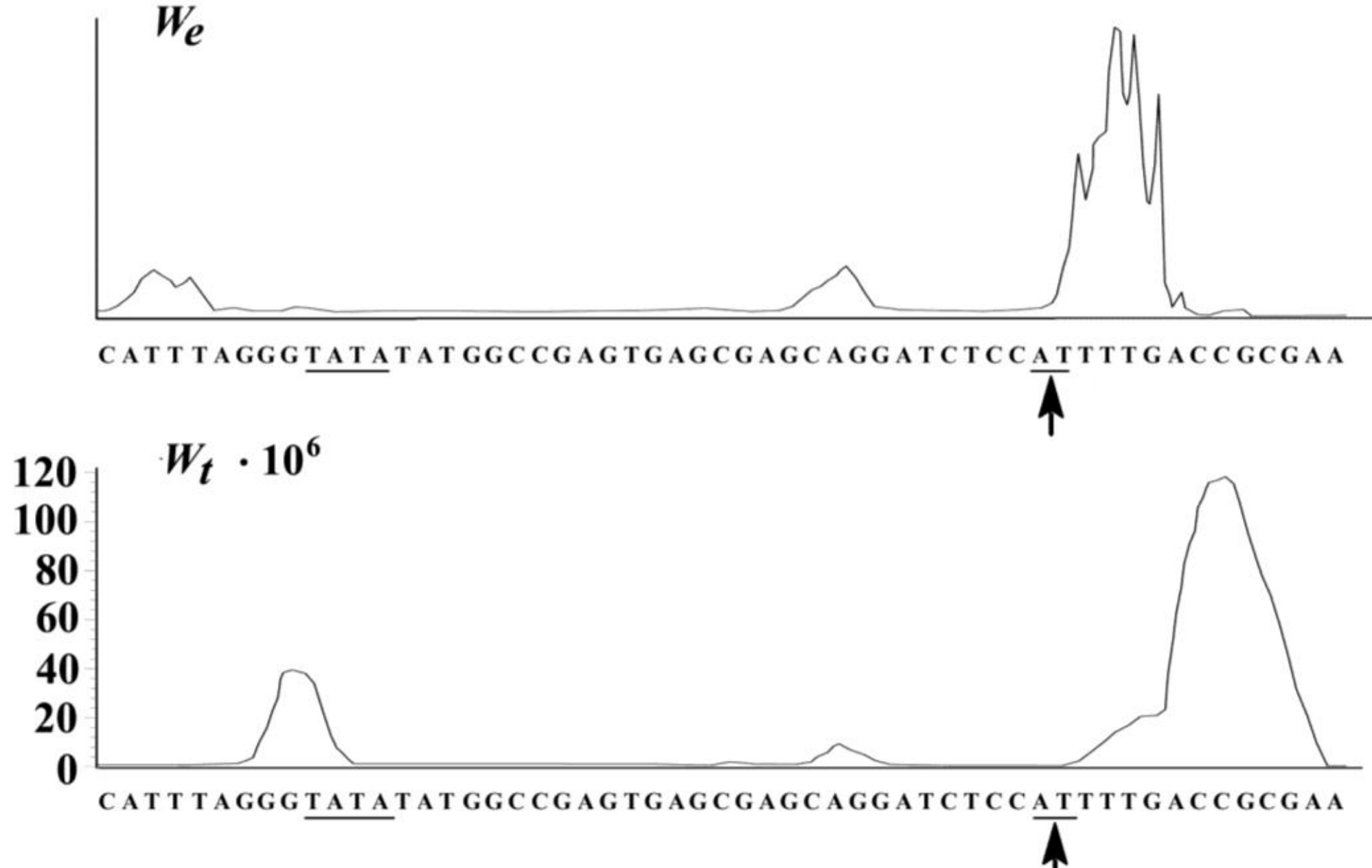

**Fig. 4.7.** Comparison of the experimental (at the top) and calculated (at the bottom) instability profiles for the third sample. The site of YY1 protein binding is underlined. TSS is underlined and marked by an arrow. $W_t$ is the calculated probability of a given basepair occurrence in a bubble consisting of 10 or more basepairs; $W_e$ is a relative value of this probability obtained experimentally.

The sequence of the central part of the oligomers including these sites serves as an absciss for the profiles shown in figures 4.7, 4.8 and 4.9. Figure 4.9 illustrates the calculated profiles of both DNA.

Figures 4.7 and 4.8 also demonstrate good agreement between the calculated and experimental results, except for the position of the peak in the site of YY1 binding. Both in the model and in the experiment the mutation led to disappearance of the peak in TSS, however the stability in the vicinity of the YY1 binding site sharply decreased. The authors explain this result by reduced competition of various promoter sites for available thermal energy [29].

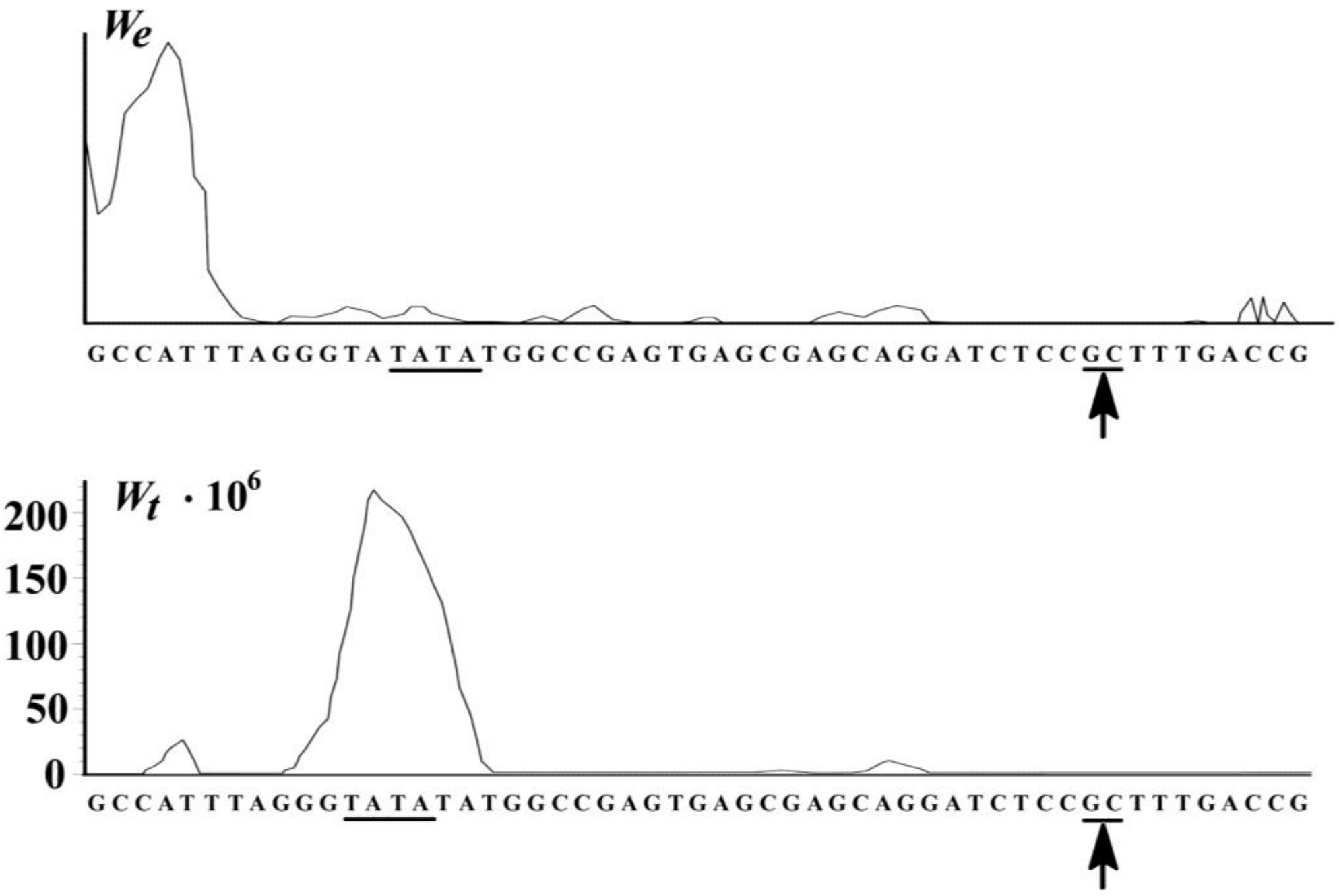

**Fig. 4.8.** Comparison of the experimental (at the top) and calculated (at the bottom) instability profiles for the fourth sample. The site of YY1 protein binding is underlined. TSS is underlined and marked by an arrow. $W_t$ is the calculated probability of a given basepair occurrence in a bubble consisting of 10 or more basepairs; $W_e$ is a relative value of this probability obtained experimentally.

The work by Choi et al. [29] suggests some conclusions. First of all they demonstrated that the sites if preferential DNA opening coincide with the regions important for initiation of its transcription. According to the authors, the enhanced probability of openings is inherent in the nucleotide sequence *per se*. In other words, there may be a certain "bubble code". Besides, it is said that the PBD model is suitable for successful prediction of the promoter properties of some or other DNA fragments and does not need any optimization [29].

After a while some additional results concerned with the problem were published [30]. First, a promoter from T7 bacteriophage was added to the promoter regions AdMLP and AAV P5 for which a good agreement of the model and experimental data was also obtained. Second, in [30] the investigation statistics was improved ten-fold: 1000 realizations were performed for each sample. Finally, it was demonstrated that it is just the nonlinear nature of stacking interactions in the PBD Hamiltonian ($\rho \neq 0$) that plays the key role in the agreement between the calculated and experimental data. Linearizing the stacking totally eliminates the successful comparison with the experimentally observed transcription [30].

Though early experiments on mechanical modeling of the dynamics of DNA promoter regions were convincing [29, 30], their conclusions raised doubts of other researchers. Van Erp et al. investigated the statistics of a local denaturation in the PBD model by direct integration method [264]. This method was mentioned in 4.1. It enabled investigations of the bubble statistics with little computational burden as compared to Langevin dynamics. Therefore denaturated regions having great length and, as a consequence, negligible statistical weight could be the object of investigation. In the work by van Erp et al. the statistics of a denaturation bubble consisting of 1–50 basepairs was studied [264].

The results obtained by van Erp et al. had some essential differences from both the experimental data and the results of Langevin dynamics [265, 264]. First of all, though the mutation reduced the peak of the instability profile in TSS, it did not influence the bubble statistics in the vicinity of the YY1 binding site. Strictly local influence of the nucleotide replacement in TSS is shown in figure 4.9, where the calculated profiles of the third and the fourth samples consisting of 10 basepairs are compared.

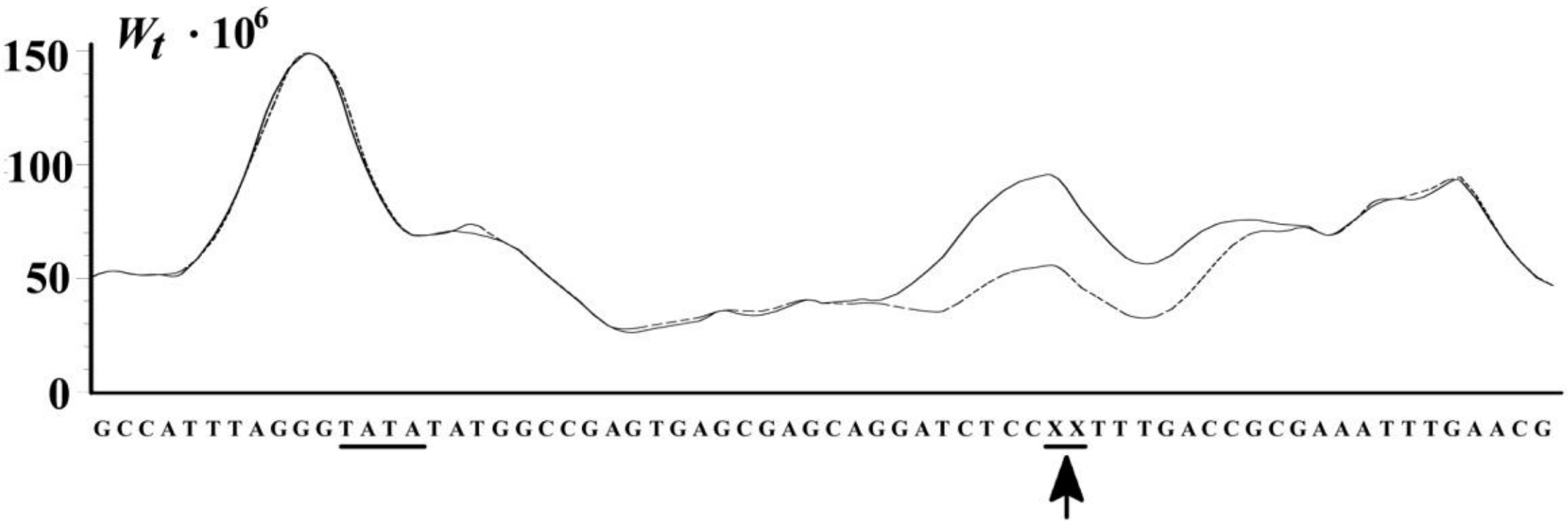

**Fig. 4.9.** Comparison of the computational results by van Erp et al. for the third and the fourth samples [265]. $W_t$ is the calculated probability of the occurrence of a certain basepair in a bubble consisting of 10 or more basepairs. The solid line represents the instability profile of the third sample, the dashed line – the relevant profile of the fourth sample. The YY1 protein binding site is underlined, TSS are underlined and marked by arrows: XX corresponds to AT in the third sample and GC – in the forth sample.

Besides, it is evident from figure 4.9 that in the calculated curves the stability of the TSS and YY1 binding site differs from that of the other duplex areas far less than in the experiment, see [29]. Noticeable peaks of the instability profile were also observed in the control sample: their value was not less than in the fragments containing promoter regions.

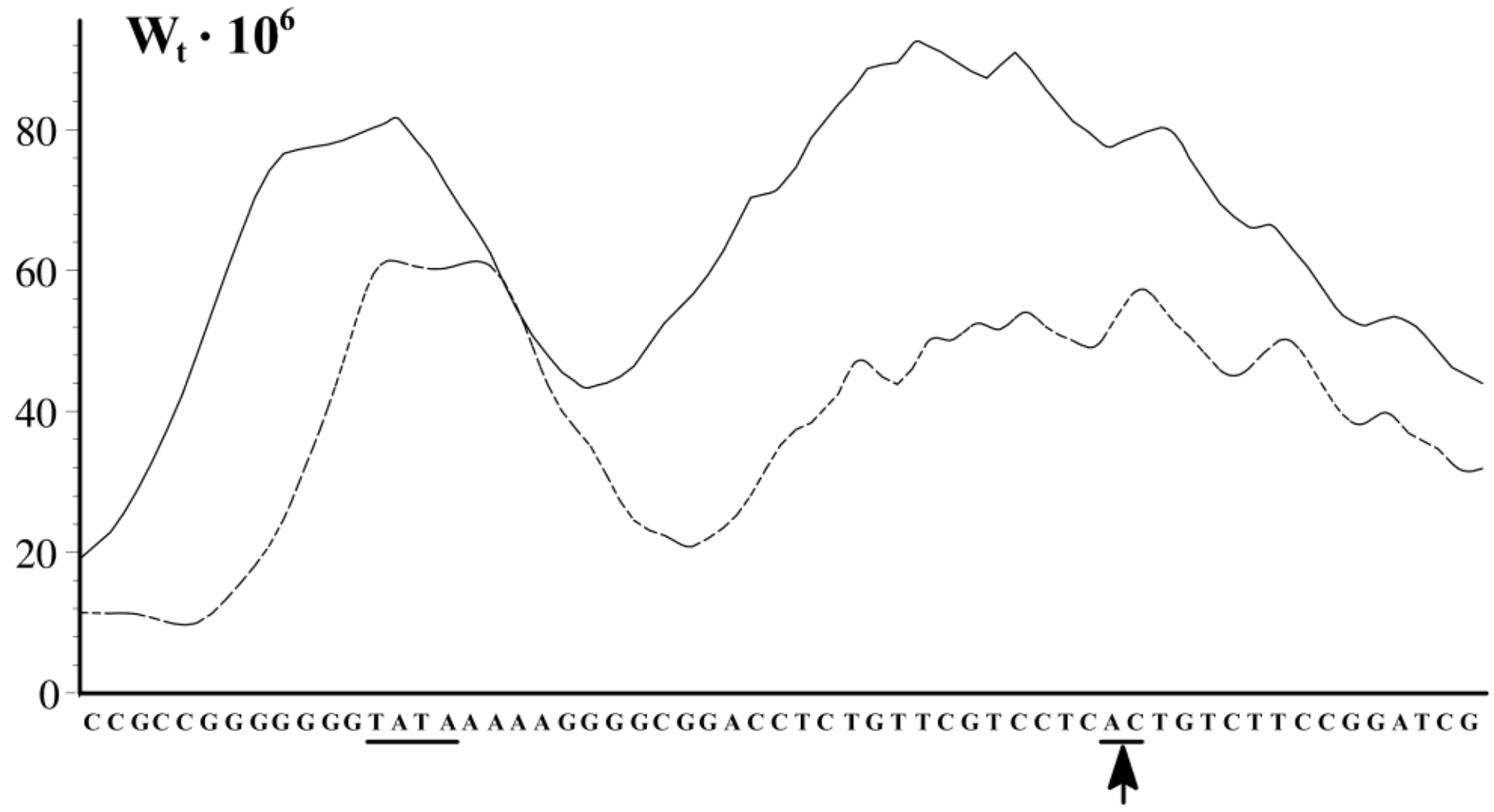

**Fig. 4.10.** Comparison of the instability profiles of the second sample obtained in [265] (dashed line) and [267] (solid line). $W_t$ is the simulated probability of the occurrence of a certain basepair in a bubble consisting of 10 or more basepairs. The TBP protein binding site is underlined. TSS is underlined and marked by an arrow.

Figure 4.10 illustrates the calculated profiles of the 10-basepair bubbles in the second sample which were obtained in works [265] and [267]. If we compare figure 4.10 with figure 4.6 we can see that the semianalytical approach by Rapti et al. as well as the direct integration by van Erp et al. yield the results which are far different from the experiment.

As in the paper by van Erp et al., the difference between the profiles of the third and the fourth sample was concerned only with the peak of instability in TSS, that is the influence of the mutation was strictly local [267]. The peaks *per se* are expressed poorly as compared to the experiment or calculations by Choi et al. [29].

On the other hand approaches by van Erp et al. and Rapti et al. enabled to obtain instability profiles with little expenditure of computing time as compared to Langevin dynamics. Subsequently Rapti et al. developed an even more efficient method for investigation of equilibrium thermodynamic properties of the PBD model for a heterogeneous DNA [268]. It was based on the effective density approximation: AT-pairs were considered as defects in a homogeneous chain of GC-pairs. The positions and heights of the peaks on the profiles were determined only by the size of the bubbles under investigation and the local concentration of AT-pairs.

This method enabled a lot of promoters to be studied [268, 269]. It is characteristic that the calculated peaks of the instability profiles by no means always coincided with the fragments responsible for primary interaction with enzymes. This suggested a conclusion that the PBD model should be improved so that it be able to describe a local denaturation of a heterogeneous DNA more adequately [269]. This conclusion is valid for, at least, the case when the instability profiles are obtained from the calculation of the equilibrium thermodynamic properties of the model.

Noteworthy is the problem of DNA modeling at low temperatures. In this temperature range theorists deal with two different problems.

The first problem is the development of an effective algorithm for searching the promoter sites of the genome on the basis of the DNA primary structure. The difficulty of its solution consists in the fact that the transctiption initiation is just associated with “collective” openings of the basepairs. Therefore the algorithms not always can be based on the nearest-neighbor models (NNM), for example. There DNA stability in NNM depends on the nucleotide sequence only on a lengthscale of 1–2 basepairs. In this connection the instability profiles are usually obtained by averaging over the areas containing a given number of basepairs [270, 271].

The only exclusion is the model which was recently developed by Kantorovitz et al. [272]. This model operates with the probability of simultaneous opening of several pairs. It enables direct computations of the probability profiles for a simultaneous opening of several adjacent basepairs. The approach by Kantorovitz et al. allows computations for DNA of several tens of kilobases in length. Along with this effective tool for genome analysis, and some other approaches have played a great role in solving the problem of the search for promoters. Another examples are provided by the models by C. Benham [33, 35, 234–236, 243, 244], E. Yeramian [135, 136] etc.

The second problem is the role of the dynamics of DNA openings in the genome functioning. It is not a mere coincidence that the work by Choi et al. is entitled “DNA dynamically directs its own transcription initiation” [29]. This problem is urgent for the development of the molecular biology of gene. Neither the Ising-like approaches nor the investigation of the equilibriun properties in the PBD model are suitable for its solution since they do not take account of the bubble dynamics.

Only investigation of the PBD model with the use of Langevin dynamics enabled researchers to reveal well-marked differences between the denaturation behavior of the promoter sites and that of the rest of the macromolecule. This method enabled Alexandrov et al. to obtain two types of instability profiles for each of the samples considered in the paper by Choi et al. [212].

The first type corresponded to the probability of opening of a bubble of a certain length: the results obtained in fact coincided with the data found by Choi et al. [29]. The reason of the seeming difference was in the minor part: in paper [212] the origin of a bubble was considered as its coordinate, whereas in [29] it was its center.

The second type of the profiles in [212] reflected the lifetime of the bubbles which open in different fragments of the duplex. The lifetimes of large bubbles which open at TSS and the sites of binding of regulatory proteins turned out to be much higher than that of the bubbles opening in other places. Besides, the lifetime of the bubbles which opened in the vicinity of YY1 binding site increased greately as a result of mutation, i.e. in the fourth sample as compared to the third one.

A correlation between the probability of the bubble opening, its lifetime, the amplitude of the chains divergence was thoroughly investigated in the next fork by Alexandrov et al. [273]. Having studied the dynamics of eight eucariotic promoters the authors made unexpected conclusions. First, the points of the most long-living openings by no means always coincide with the maxima of the instability profile. Second, a large amplitude of the denaturation bubble by no means always implies the increased value of its lifetime [273]. Sunsequently Alexandrov et al. modified the PBD model having included there a heterogeneity of the stacking potential $k$ (see expressions (2.5) and (2.6)) [274]. The stacking parameters were fitted for 10 combinations of adjacent basepairs on the basis of a comparison of the simulated data with the melting profiles of regulary structure oligomers. Investigations of the new model with the use of Langevin dynamics confirmed the results obtained in the previous works [212, 273]. Moreover, due to consideration of the stacking heterogeneity, the dynamical distinctions of the binding sites of specific proteins became much more apparent [274].

The modified PBD model played a great role in the investigations of a correlation between the opening dynamics of various DNA regions and their biological functions [275, 276]. In recent years the antecedence of a local duplex opening at the onset of transcription has been proved in vitro [18, 19]. This stresses the relevance of mechanical modeling of a heterogeneous duplex.

Nevertheless, in low-temperature DNA investigations, a shortcoming of the PBD model such as very little lifetime of the bubbles ($\leq 4$ ps) becomes apparent. For comparison: the shortest time of an open state obtained *in vitro* is 0.92 ns [277]. Moreover, this lifetime characterizes neither a bubble, nor even an opening of a single nucleotide pair but a short-time escape of a single base from the Watson-Crick helix (for further details see Chapter 5). Interaction of DNA with a nonspecific enzyme like endonuclease S1 obviously requires rather long-living bubbles. It is not impossible that for adequate description of their behavior, models of higher level are needed.

Besides, the lifetimes of the bubbles obtained in the model depended on their size only slightly. In experiments these times can differ by three orders of magnitude or greater depending on the bubble size [213]. Investigation of the kinetics of denaturation bubbles became possible relatively recently due to a minor modification of a well-known method – fluorescence correlation spectroscopy. In what follows we will describe this technique and relevant investigations in greater detail.

### 4.3. Range of denaturation bubbles lifetimes. Fluorescent correlation spectroscopy investigations

Fluorescence correlation spectroscopy (FCS) is widely used in chemical and biological physics. The basic idea of this method is registration of fluorescence from a very small volume of a solution as a time function [278]. At every instant of time the signal value is proportional to the number of fluorescent molecules in the sample volume which fluctuates about a certain mean value. Since the volume is small relative fluctuations of the signal reach a considerable value. The main causes of the signal fluctuations are chemical transitions of particles into a nonradiating state and back as well as diffusion of molecules through the sample volume. Having constructed autocorrelation functions one can find a time scale of these processes. Hence FCS enables one to get information on the kinetics of the chemical reactions in a sample volume, diffusion coefficients and equilibrium concentrations of molecules [278].

In order to obtain an autocorrelation function one registers the values of fluorescence fluctuations $\delta F(t)$ at different equidistant instants of time $t_1$, $t_2$, $t_3$, ..., $t_i$, ..., $t_P$ and fluctuations at “delayed” moments $\delta F(t + \tau)$. The products $\delta F(t_i)\cdot\delta F(t_i + \tau)$ obtained are divided by $P$, i.e. simply averaged over time. Varying $\tau$ one gets the autocorrelation function $G_{fluor}(\tau)$ [278]:

$$G_{fluor}(\tau) = \frac{\langle \delta F(t) \cdot \delta F(t+\tau) \rangle}{\langle F(t) \rangle^2}, \tag{4.2}$$

where $F(t)$ is the absolute value of fluorescence and the angle brackets $\langle ... \rangle$ imply averaging over time. Hence, $\delta F(t)$ is a difference between $F(t)$ and $\langle F(t) \rangle$. Sometimes it is convenient to present formula (4.2) not in terms of fluctuation values but in terms of absolute values of the signal:

$$G_{fluor}(\tau) = \frac{\langle F(t) \cdot F(t+\tau) \rangle - \langle F(t) \rangle^2}{\langle F(t) \rangle^2}. \tag{4.3}$$

Early investigations of the DNA denaturation behavior by FCS were carried out by G. Bonnet et al. [279]. The authors studied the denaturation kinetics of specific hairpin structures with the sequence formula 5'-CCCAA-$(B)_n$-TTGGG-3', where B corresponded to A or T, and $n$ varied from 12 to 30. Fluorescent molecules were obtained by way of covalent binding (tagging) of a fluorophore to 5'-end and a quencher – to 3'-end. As a result structures called "molecular beacons" were constructed [279]. In a closed "beacon" a fluorophore and a quencher are close to each other and can interact directly: as a result, a fluorescence cannot take place. When the beacon's ends are far apart enough, a fluorophore is removed from a quencher and the molecule becomes fluorescent.

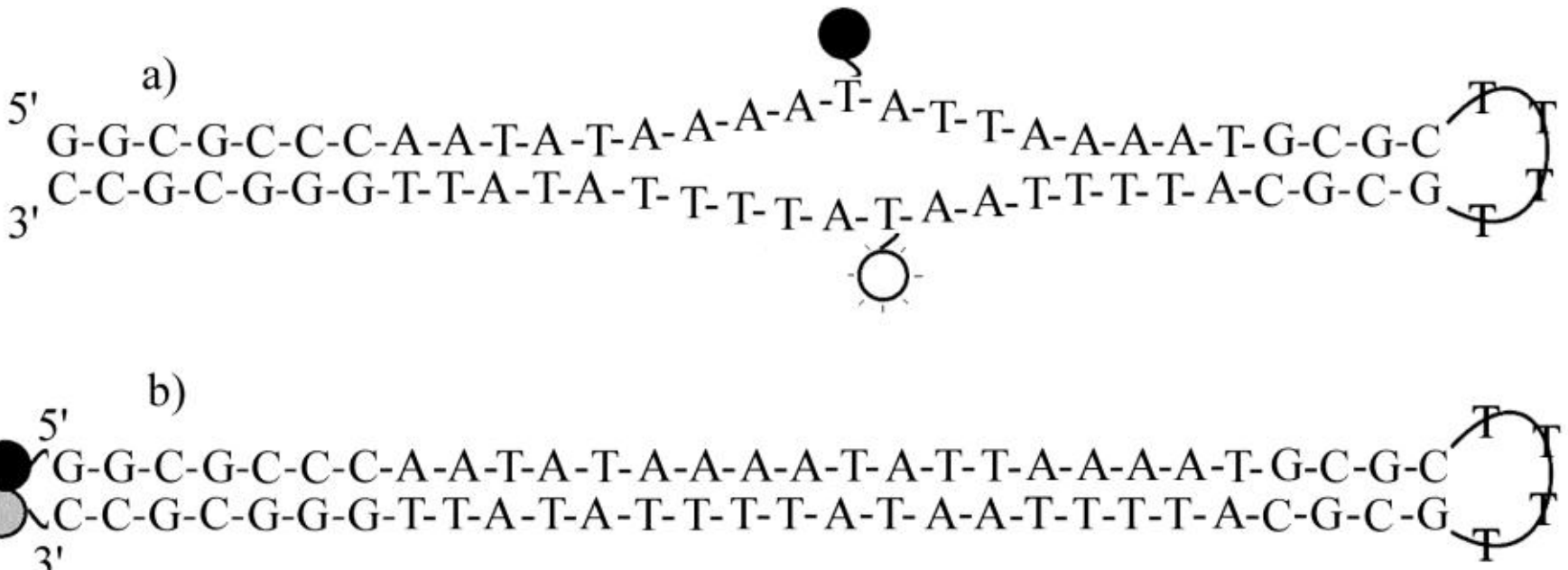

**Fig. 4.11.** Beacons obtained from oligonucleotide $M_{18}$; the solid circle indicates a quencher (DABCYL) [213]. **a)** scheme of beacon $M_{18}$ with an internal tagging: the empty circle indicates the excited state of a fluorophore (6-carboxyrhodamine). b) scheme of beacon $M_{18}$ with an end tagging: the gray circle corresponds to a fluorophore in contact with the quencher.

Subsequently a similar technique was used to study the kinetics of local openings in AT-rich domain of self-complementary oligomers [213]. For this purpose the bases in the center of an AT-rich domain were tagged by the fluorophore and the quencher. The authors investigated beacons obtained on the basis of three oligomers which they called $M_{18}$, $A_{18}$ and $(AT)_9$. In $M_{18}$ oligomer, the AT-rich domain consisted of randomly alternating adenine and thymine bases. Therefore even in the case of full denaturation single strands could not form any secondary structures which would stabilize the fluorescent form. The Figure 4.11 illustrates two types of beacons synthesized on the basis of $M_{18}$ and a simplified scheme of the possible states of a beacon.

It is evident from the figure that fluctuations of the fluorescent signal are caused not only by diffusion but also by opening and closing of the tagged fragment of duplex.

Beacons with internal and end tagging were also obtained for oligomers $A_{18}$ and $(AT)_9$. The tagging points are underlined and unpaired fragments are given in bold on the following schemes of their primary structure. Beacon $A_{18}$ had the structure: 5'–GGCGC CCAAA AAAAA ATAAA AAAAA GCGC**T TTT**GC GCTTT TTTTT ATTTT TTTTT GGGCG

CC–3'. Its fluorescent form can form different shifts of 5'–$A_{18}$–3' strand with respect to 3'–$T_{18}$–5'. These shifts exclude a contact between a fluorophore and a quencher. As a result, the lifetime of this form should be much longer than that of $M_{18}$. The fluorescent forms of the beacon $(AT)_9$: 5'–GGCGC CCATA TATAT ATATA TATAT GCGC**T TTT**GC GCATA TATAT ATATA TATAT GGGCG CC–3' were thought to live still longer: its bubbles can self-hybridize to form cruciforms due to self-complementarity of the single chains of the AT-rich domain.

Measurements of the total fluorescence of the beacons as a function of temperature $I(T)$ enabled G. Bonnet et al. to find the equilibrium constants $K(T)$ in the DNA opening reaction. The fraction of open beacons $p_B(T)$ was calculated by the formula [279]:

$$p_B(T) = \frac{I(T) - I_c}{I_o - I_c}, \tag{4.4}$$

where the fluorescence of open beacons $I_o$ corresponds to the maximum signal at 363 K and higher, $I_o = I(363)$, and the fluorescence of closed beacons $I_c$ is obtained by extrapolation of the low-temperature signal to absolute zero $I_c = I(0)$ [279]. With the knowledge of $I(T)$ the authors calculated the equilibrium constant as a function of temperature:

$$K = \frac{k_-}{k_+} = \frac{I(T) - I_c}{I_o - I_c} \Bigg/ \frac{I_o - I(T)}{I_o - I_c} = \frac{p(T)}{1 - p(T)}, \tag{4.5}$$

where $k_-$ is the rate constant for the beacon opening in the AT-domain, $k_+$ is the rate constant for the an open state relaxation (closing) [213].

The autocorrelation function of a beacon $G_{beacon}$ is a product of the diffusion term $G_{diff}$ and the kinetic term $G_{chem}$:

$$\begin{aligned} G_{beacon} &= \frac{\langle I(t_0) I(t_0 + \tau)\rangle - \langle I(t_0)\rangle^2}{\langle I(t_0)\rangle^2} = G_{diff} G_{chem} = \\ &= \frac{I}{B} \frac{1}{1 + \dfrac{t}{\tau_{diff}}} \left(1 + \frac{1 - p_B}{p_B} \exp\left[-t(k_- + k_+)\right]\right), \end{aligned} \tag{4.6}$$

where the angle brackets $\langle ... \rangle$ denote averaging over all the momenta $t_0$, $B$ is the mean number of the beacons in the sample volume, $\tau_{diff}$ is the characteristic time of the beacon diffusion through this volume. The value of $p_B$ corresponds to $p_B(T)$, that is $(1 - p_B)/p_B = 1/K$, $I$ is a fluorescent signal, $k_- + k_+ = \tau_{reaction}^{-1}$ is the summary rate (inverse time) of the reaction. In order to obtain this parameter G. Bonnet et al. calculated an autocorrelation function for a special beacon-control tagged only with fluorophore [279]:

$$G_{diff} = \frac{I}{B} \frac{1}{1 + \dfrac{t}{\tau_{diff}}}. \tag{4.7}$$

Solving the system of equations

$$K = \frac{k_-}{k_+} \quad ; \quad k_+ + k_- = \frac{1}{\tau_{reaction}}, \tag{4.8}$$

it is easy to obtain the expressions for the rate constants [279]:

$$k_- = \frac{1}{\tau_{reaction}} \frac{K}{1+K} \quad , \quad k_+ = \frac{1}{\tau_{reaction}} \frac{1}{1+K}. \qquad (4.9)$$

If there is the only fluorescent form of a beacon, then the rate constants $k_-$ and $k_+$ can be determined uniquely. Such results were obtained for the beacons with a short complementary region capable of being either open or closed [279].

If there are many fluorescent forms, the analysis of the autocorrelation functions can only give information about the range of the relaxation rates $k_+$. For the beacons obtained from $M_{18}$, $A_{18}$ and $(AT)_9$, these functions were not monoexponential. The value of $k_+$ strongly depended on the length of the open fragment, i.e. on the number of open bases [213]. An example of a normalized auto-correlation function characterizing the relaxation kinetics of the beacon $M_{18}$ at 33 °C is presented in figure 4.12.

It is evident from the figure that the lifetimes of the states with a bubble in the tagging point are in the range of $10^{-6}$–$10^{-3}$ s. By the standards of Langevin dynamics these are huge lifetimes – 6–9 orders of magnitude higher than the timescale typical for the PBD model. These times suggest a high activation barrier preventing quick relaxation of the open states.

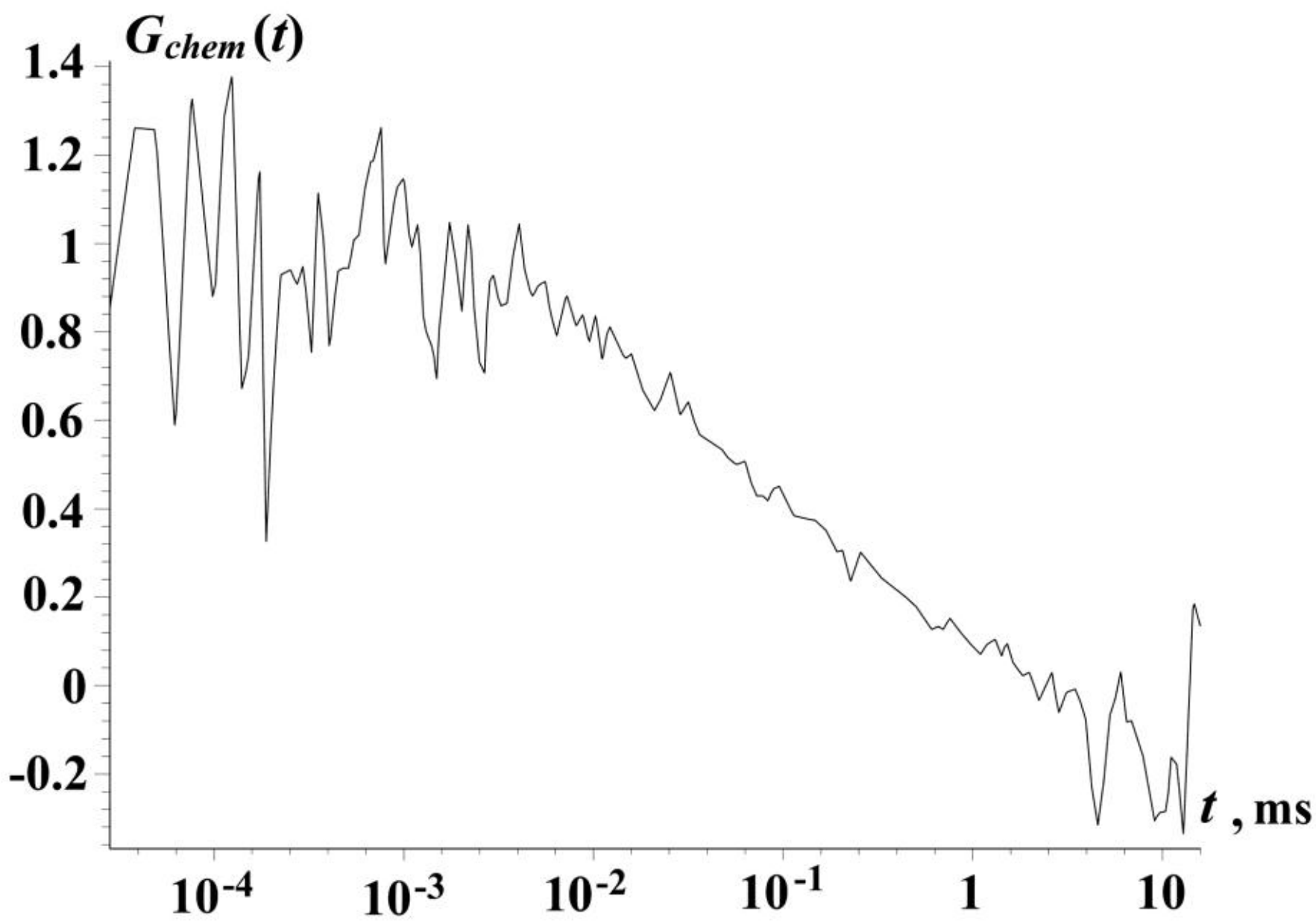

**Fig. 4.12.** Autocorrelation function $G_{chem}$ of the molecular beacon $M_{18}$ with internal tagging at 33 ºC [213].

Notwithstanding the possibility of an additional stabilization of fluorescent forms in $A_{18}$ and $(AT)_9$ (see above), the ranges of $k_+$ practically coincided in all the beacons. Moreover, the investigations of their temperature dependence in the range of 20–50 °C demonstrated that for different beacons, the values of the thermodynamic parameters of relaxation are very close. All these facts suggest a similar mechanism of stabilization of the open states.

According to the authors, the bubbles in DNA may stabilize mainly by re-stacking of bases in the single-stranded domains. Just these interactions prevent formation of any alternative variants of the secondary structure in $A_{18}$ and $(AT)_9$ [213]. As we will show in Section 5.3, in general this explanation is correct, though some improvements are needed.

Hence, Altan-Bonnet et al. were the first to assess directly the lifetimes of denaturation bubbles [213]. The wide range and great values of this quantity are an important and interesting result. However, the primary structure of the beacons used in the work was rather specific: their AT-rich domains consisting of 18 basepairs did not include GC-basepairs at all. As a result, the equilibrium constants for the opening at the tagging points appeared to be very

high – from 0.005 to 0.02 at 30 °C. This is 3–4 orders of magnitude higher than the constants for individual nucleotide pairs obtained by calculations [129, 280] and in the experiments, see Section 5.2. This raises apparent contradictions between the results of FCS and other data on the kinetics of DNA opening. For example the mutual independence of the bases "flipping out" (single basepair openings) kinetics in neighboring pairs is an established fact [281–286]. This would be impossible if collective openings prevailed over single ones, i.e. if the latter would have much lower equilibrium constants.

The main problem of the next Chapter is to compare the denaturation bubbles and single basepair openings in terms of kinetics and thermodynamics. For this purpose we will consider the investigations of an escape of individual bases from the Watson-Crick helix performed by the proton exchange methods. We will explain seeming kinetic contradictions between the data of this method and FCS and consider the problem of bases' re-stacking upon opening of several neighboring basepairs.

## 5. INVESTIGATIONS OF DNA OPENINGS AT LOW TEMPERATURES

As have been shown in Chapter 2, three degrees of freedom are sufficient for a simplified description of DNA denaturation. They are a radial separation of the chains, a local change of supercoiling and an angular displacement of a base around the axis of the sugar-phosphate backbone. All the three degrees take part in the formation of a denaturation bubble, but in the case of a single basepair opening the first two degrees play an insignificant role. This case and the scheme of the degrees of freedom are shown in figure 5.1.

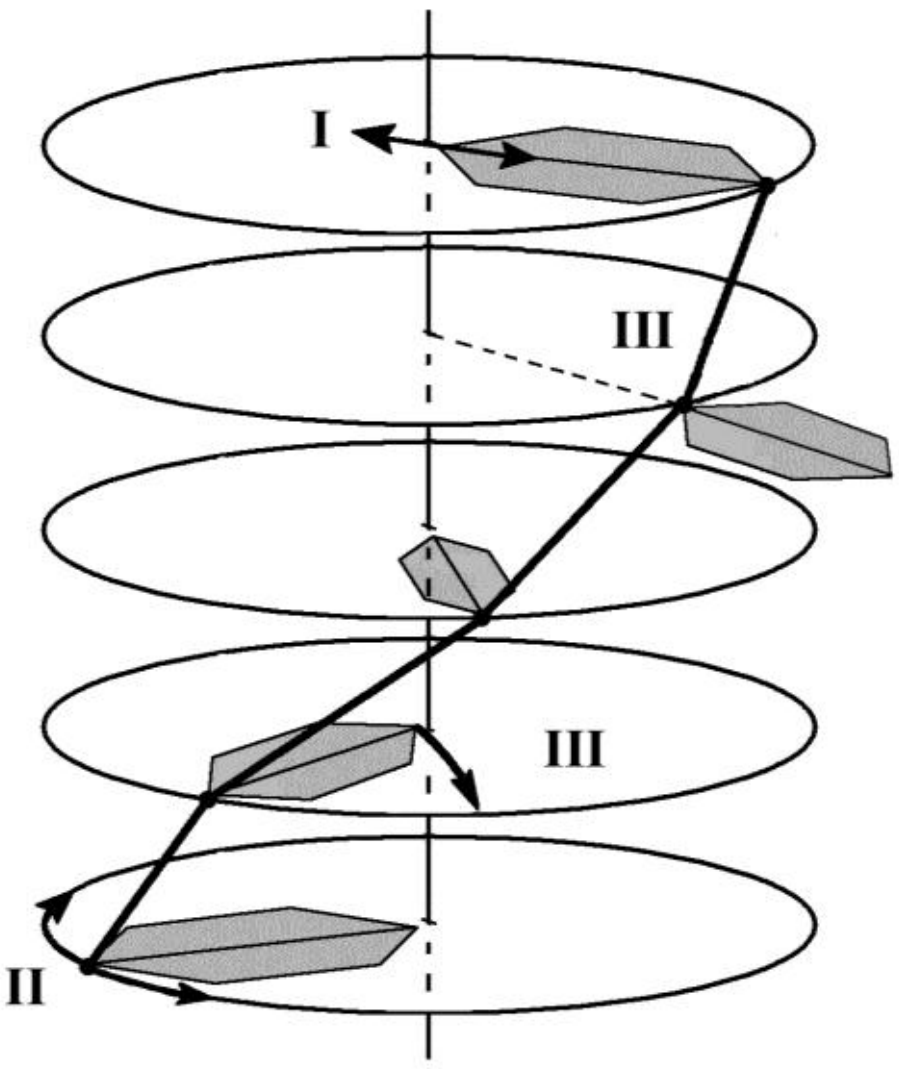

**Fig. 5.1.** Scheme of the main degrees of freedom of nucleotide pairs in simple DNA models: only one DNA chain is shown. I. Radial separation. II. Local change in supercoiling. III. Angular escape of a base from the stack – a "flip-out", which depend almost not at all on the other two degrees of freedom.

The main way of independent opening of a nucleotide pair is an angular displacement of its bases. Therefore in the English-language literature the term "flip-out" is often used as a synonym of a single basepair opening. We will also use this term.

### 5.1. Investigations of bases flipping out by $^1$H NMR method

Opening of individual basepairs plays a key role in many specific DNA-protein interactions. In this case a base falls into an enzyme's active center or another specific binding site after escaping from a Watson-Crick helix [287, 288]. This mechanism is general for initial interaction of DNA with many methyltransferases [289–291] and glycosylases [292–294], as well as in the case of RNA replication [295, 296] and transcription [18, 19, 296].

The contribution of DNA dynamics into its primary interaction with enzymes is one of the most important problems in molecular biology. These interactions take place at temperatures much lower than $T_m$. Under these conditions the probability of the bases flipping out considerably prevails over the probability of a bubble formation since the latter process is usually characterized by a far greater activation enthalpy.

During all the time of a flip-out interactions of a flipped base with some atoms of DNA duplex partially preserve [161]. Even having escaped from the Watson-Crick helix, it usually maintains bonds with one of the DNA grooves.

To study the flip-out kinetics at temperatures lower than 35 °C use is made of the method based on the capacity of imino groups of guanine and thimine to exchange hydrogen atoms with molecules of the solution. The exchange is possible only from the open states and proceeds with the participation of a catalyst – any molecule or ion capable of a reversible capture of $H^+$. The replacement of imino protons is registered by $^1$H-NMR methods. This enables one to measure the reaction rate of individual bases flip-out by varying the catalyst concentration. We will dwell upon the process of the catalytic exchange of protons.

First we will consider the situation in a free nucleoside. Let nuH be a nucleoside capable of imino-proton exchange, and acc – a proton acceptor. As a result of a collision, a complex nuH··acc is formed capable of transition to $nu^-{\cdot\cdot}H^+acc$ form and conversely. Further exchange dynamics can be conveniently presented in terms of nucleosides ionization. pK for proton acceptor is the negative common logarithm of equilibrium constant of its dissociation reaction:

$$K_A = \frac{[A]\left[H^+\right]}{\left[AH^+\right]}. \tag{5.1}$$

Value $pK_{acc}$ is numerically equal to pH of half-ionization of acceptor molecules. The concentration ratio of nuH··acc and $nu^-{\cdot\cdot}H^+acc$ is equal to relation of $K_A$ of acceptor and nucleoside [297]:

$$\frac{[\mathrm{nuH\cdot\cdot acc}]}{\left[\mathrm{nu^-\cdot\cdot H^+acc}\right]} = \frac{K_{acc}}{K_{nu}} = \frac{10^{pK_{nu}}}{10^{pK_{acc}}}. \tag{5.2}$$

Usually pH of the solution is far smaller than $pK_{nu}$ and the probability of the reaction of a proton capture from the solvent

$$\begin{aligned} &\mathrm{nu^-\cdot\cdot H^+acc + H^+ \rightarrow nuH\cdot\cdot H^+acc} \\ &\mathrm{nuH\cdot\cdot H^+acc \rightarrow nuH\cdot\cdot acc{+}H^+} \end{aligned}, \tag{5.3}$$

for the complex $nu^-{\cdot\cdot}H^+acc$ is several orders of magnitude higher than the probability of the transition back to nuH··acc through returning of its “own” proton. Therefore it is the deprotonation of the nucleoside that is the limiting step of the process. Then the rate of the proton exchange $k_{ex}$ may be taken to be equal to the rate of its reversible transfer to the acceptor $k_{tr}$. This quantity is determined by the expression:

$$k_{tr} = k_{coll}[\mathrm{acc}]\eta \quad \left[\frac{l}{mol\cdot s}\cdot\frac{mol}{l} = s^{-1}\right], \tag{5.4}$$

where $k_{coll}$ is the frequency of successful collisions expressed through the rate constant of the second order, η is the fraction of the complexes capable of proton exchange, and [acc] is the concentration of acceptor molecules. The dimensions of the quantities are given for the sake of clarity. The typical value of $k_{coll}$ is approximately $10^9$ l · mole$^{-1}$· s$^{-1}$ for most acceptors and an order of magnitude greater for $OH^-$ ions [297].

Having expressed η in terms of pK for $pH < pK_{nu}$ (see (5.2))

$$\eta = \frac{\left[\mathrm{nu}^- \cdot\cdot \mathrm{H}^+\mathrm{acc}\right]}{\left[\mathrm{nu}^- \cdot\cdot \mathrm{H}^+\mathrm{acc}\right] + \left[\mathrm{nuH} \cdot\cdot \mathrm{acc}\right]} = \left(1 + \frac{\left[\mathrm{nuH} \cdot\cdot \mathrm{acc}\right]}{\left[\mathrm{nu}^- \cdot\cdot \mathrm{H}^+\mathrm{acc}\right]}\right)^{-1} = \left(1 + 10^{\mathrm{pK_{nu}} - \mathrm{pK_{acc}}}\right)^{-1},$$

we finally express the rate of proton exchange in a free nucleoside as:

$$k_{tr} = \frac{1}{\tau_i} = \frac{k_{coll} \cdot [\mathrm{acc}]}{\left(1 + 10^{\mathrm{pK_{nu}} - \mathrm{pK_{acc}}}\right)}, \tag{5.5}$$

where $\tau_i$ is the exchange time. Indeed, this expression is a little more complicated since usually pK of an acceptor is higher than pH of a medium [297]. For example, for ammonia $pK = 9.3$. Using formula (5.1) it is easy to calculate the concentration of its uncharged molecules for $pH = 7$:

$$[\mathrm{NH_3}] = [\mathrm{NH_3}]_{total} \left(1 + \frac{\left[\mathrm{NH_4^+}\right]}{[\mathrm{NH_3}]}\right)^{-1} = \frac{[\mathrm{NH_3}]_{total}}{1 + 10^{9.3-7}} \cong 0.05[\mathrm{NH_3}]_{total}, \tag{5.6}$$

where $[\mathrm{NH_3}]_{total}$ is the total concentration of ammonia. The general form of the expression for the catalyst active concentration is:

$$[\mathrm{acc}] = [\mathrm{acc}]_{total} \left(1 + 10^{\mathrm{pK_{acc}} - \mathrm{pH}}\right)^{-1}. \tag{5.7}$$

Now, it is meaningful to consider the proton exchange kinetics for a nucleotide in a duplex. Since this reaction requires an open state, the exchange time $\tau_{ex}$ is determined by the equilibrium constant $K_d$. It is equal to the ratio between the averaged time of an open state and the averaged time of a closed one.

It should be stressed that in literature these times are called “closing time” and “opening time”. Therefore they are denoted as $\tau_{cl}$ and $\tau_{op}$, respectively. The dependence of the exchange time on these quantities has the form:

$$\tau_{ex} = \frac{\tau_i}{\alpha}\left(\frac{\tau_{cl}}{\tau_{cl} + \tau_{op}}\right)^{-1} = \frac{\tau_i}{\alpha}\left(1 + \frac{1}{K_d}\right),$$

where $\alpha < 1$ is the accessibility factor taking account of the limited access of the catalyst to the proton of an imino group. It is believed that this parameter is close to 1 for $NH_3$ and imidazole; for tris (tris-oxymethylaminomethane) and other amines it is about 0.3–0.5 [282–284].

Since $\tau_{cl}$ is several orders of magnitude less than $\tau_{op}$, the proton exchange time may be expressed as:

$$\tau_{ex} \approx \frac{\tau_i}{\alpha K_d}. \tag{5.8}$$

However expression (5.8) is valid only for small concentrations of the external acceptor of $^1$H. If this concentration is high a considerable part of the exchange events will occur upon the first opening. In the limit of an infinite catalyst concentration $\tau_i \to 0$. Nevertheless, $\tau_{ex} = \tau_{op}$ since the exchange cannot take place until a base flips out of the duplex. Hence the expression for $\tau_{ex}$ takes the form:

$$\tau_{ex} = \tau_{op} + \tau_i \left(\alpha K_d\right)^{-1} = \tau_{op} + \frac{1 + 10^{\mathrm{pK_{nu}} - \mathrm{pK_{acc}}}}{k_{coll} [\mathrm{acc}] \alpha K_d}. \tag{5.9}$$

At high concentrations of the catalyst $\tau_{ex}$ becomes linearly dependent on $[acc]^{-1}$. The angle of slope of this curve is inversely proportional to $\alpha K_d$, and $\tau_{ex}(0) = \tau_{op}$.

Different imino protons yield clearly distinguishable peaks in NMR spectra. This enables one to calculate $\tau_{ex}([acc]^{-1})$ for individual guanine and thymine bases. One of the methods for registering a flip-out is a catalytic exchange of a proton for a deuteron: at the moment of the exchange of $^1$H for $^2$H a relaxation occurs. In this case a relative widening of a resonance band at half-height is related to $\tau_{ex}$ as [298]:

$$\left(\tau_{ex}\right)^{-1} = \pi\left(\Delta\nu^{b} - \Delta\nu^{aac}\right), \tag{5.10}$$

where $\Delta\nu^b$ is is the width of the resonance band for a certain catalyst concentration, and $\Delta\nu^{aac}$ is the width of in its absence.

The value of $\tau_{ex}$ can also be calculated from the time of longitudinal relaxation $T_1$ which is measured by the inversion recovery times [282–284, 286, 299]. The method is based on a sselective spin inversion $^1$H followed by measurement of their recovery times. The time $T_1$ is related to $\tau_{ex}$ by a simple expression:

$$T_1^{-1} = \tau_{ex}{}^{-1} + T_{10}{}^{-1},$$

where $T_1$ and $T_{10}$ are the times of relaxation with and without a catalyst, respectively.

The value of $T_{10}{}^{-1}$ is a sum of the contributions of proton-proton dipolar interactions and an exchange independent of an external catalyst. In the course of the latter protons can transfer to $OH^-$ ions and to nitrogen atoms of complementary bases – $N_1$ adenine or $N_3$ cytosine. The second type of the transfer was called "internal catalysis". Its mechanism will be considered in detail in Section 5.4.

The catalytic contribution of $OH^-$ in a weakly alkaline DNA solution is negligible. This can be easily verified by setting pH = 9 and substituting the typical values of the other parameters $k_{coll} = 10^{10}$, $K_d = 10^{-5}$–$10^{-6}$, $pK_{acc} = 15.7$, $pK_{nu} = 9.5$ into expression (5.9). Internal catalysis is not very effective too, since pK of the atoms of $N_1$ adenine and $N_3$ cytosine are very small. Therefore even low concentrations of an external catalyst can lead to a considerable accelaration of $^1$H exchange.

High concentrations of a catalyst enable one to measure $\tau_{op}$ and $K_d$ much more precisely. However at high [acc] the correlation time of DNA molecules grows since the ionic strength of the solution increases [300]. This leads to an additional acceleration of relaxation which should be taken into account so that to avoid underestimation of $\tau_{op}$ and overestimation of $K_d$ [300, 301]. An example is provided by $\tau_{op}$ of the fourth GC-pair of oligomer d(GCGCATCGCG)2 for which earlier the value of 16 ms was obtained for 15 °C [283]. A more recent assessment made with allowance for  the influence of the ionic strength was 21 ms, while without this allowance – 17 ms [301], which is close to the early result.

The influence of high catalyst concentrations was studied only in 1996–1997 [300, 301]. Nevertheless, in the main part of early investigations use was made of low [acc], not higher than 200 mmol/l [281–283, 320, 284, 285]. Therefore the influence of high [acc] was reflected on the adequacy of their conclusions only slightly.

There are two ways to solve the problem of high ionic strength. The first one implies calculation of a decrease in $T_{10}$ from $(\Delta\nu^b - \Delta\nu^{aac})$ of non-exchangeable protons [300], or from the dependence of $\Delta\nu^{aac}$ of imino protons on $[H^+acc]$ at low pH – when [acc] is low [299] (see expression (5.7)). A suggestion was also made to carry out all the investigations under conditions of an equal ionic strength [301]. The second way implies measurement of $T^1$ by magnetization transfer from water. This method is based on selective inversion of the $^1$H spins in $H_2O$ molecules followed by measurement of an increase in the absorption of imino group protons [302–305]. Magnetization can be transferred only in the DNA open state (as in another methods), but the value of $T^{10}$ in this case is independent of the ionic strength which enables one to use arbitrary large [acc]. Therefore in more recent kinetics measurements this method was used as the main one [305–307] or in combination with other methods which were used as a control [277].

The varying parameters in many investigations of the flip-out were the length and the primary structure of oligonucleotides, the temperature, pH and the ionic strength of solutions, the exchange catalysts. Measurements of relaxation rates of imino group proton spins and their comparison with the relaxation dynamics of the niclei of other atoms enabled researchers to make some essential conclusions about DNA fluctuations at moderate temperatures. Further we will describe the features of the DNA dynamics at $T \leq 308$ K which were revealed by $^1$H NMR methods.

## 5.2. Kinetics and thermodynamics of single basepair openings

Early investigations of the nucleic acid dynamics based on proton exchange were carried out for small concentrations of $H^+$ acceptors. At first the duplex opening was studied on the basis of the exchange kinetics of hydrogen for tritium whose amount was determined by the scintillation counter [308, 309]. Later on, with the advent of $^1$H NMR techniques, proton-deuteron or proton-proton exchange came into use. A good review of the investigations performed on DNA as well as on RNA is provided in the works by Gueron, Kochoyan and Leroy [281–283]. Strictly speaking, the method described in Section 5.1 enables one to study the flip-out kinetics only for thymine and guanine bases. It is highly unlikely that the complementary bases flip out of their stack too. However, for simplicity we will believe that a flip-out in the basepairs occurs cooperatively – this fact is suggested by some literature data [162, 306].

Till the 1980's it was widely accepted that basepairs can open only collectively and the open states survive for tens and even hundreds of milliseconds [143, 144]. This time is sufficient for the exchange even if the pK difference is more than 3 and [acc] $\leq 10^{-3}$ mol/l, taking into account that usually $k_{coll} \leq 10^9$ l $\cdot$ mol$^{-1}$ $\cdot$ s$^{-1}$, see expressions (5.5) and (5.9). The values of $K_d$ calculated for small concentrations of the acceptor were approximately 0.01. This fact was in contradiction with the data of other methods: for example, the kinetics measutements of RNA binding with mercury yielded $K_d \approx 0.002$ [310]. According to hydrodynamical measurements, for DNA, $K_d \leq 10^{-4}$ [311]; the analysis of DNA interaction with formaldehyde demonstrated that $K_d \approx 10^{-5}$ ([312], cited by [282]).

However in 1985 the exchange rate in nucleic acids was shown to be sensitive to the concentration of the proton acceptor [313, 314]. For RNA, Leroy et al. obtained $\tau_{op} = 3$ ms and $K_d \approx 10^{-3}$ at 27 °C [313] which was in agreement with the data of other methods [310]. Later on, comparing the kinetics of the catalytic exchange in DNA and in free nucleosides Gueron et al. found $K_d \approx 10^{-5}$ for AT-pairs [281]. Hence the contradictions between the proton exchange data and the results of other investigations were eliminated.

At the same work [281] it was proved that in the absence of an external catalyst the main way for $^1$H exchange is an internal catalysis. Noticeable differences in $\tau_{op}$ for adjacent basepairs, as well as the difference in $K_d$ for AT- and GC-pairs by approximately an order of magnitude suggested that openings occur individually [281]. Their noncooperativity and weak interdependence were later demonstrated for short B-DNA of different lengthes and primary structures [282–286], as well as for Z-DNA [315]. Hence, flip-outs turned out to prevail over denaturation bubbles at low temperatures.

The opening kinetics of a basepair is determined not only by its chemical nature but also by the nearest neighbors. Here we can conveniently point out three factors determining the dynamics of individual basepairs.

First of all this is the oligomer length and the proximity of a basepair to the end. The rate of the catalytic exchange of a proton in the end pairs is so high that their dynamics cannot be studied by $^1$H NMR. This effect is known as "end-fraying" [316]. Nevertheless it is proved that even in an end pair both the H-bonds and the stacking with an only neighbor should be broken for a proton to exchange [317]. The end effects in DNA consisting of more than 12 basepairs are observed up to the third pair reckoned from the end. In shorter duplexes, they probably involve even the central pairs [316].

The second factor is the influence of the "primary structure context", i.e. adjacent basepairs. The data on it are somewhat contradictory. For example, though a GT-mismatch distorts the duplex and has anomalous kinetics ($K_d \approx 0.0007$), its influence does not go beyond nearest neighbors in the stack [318]. On the other hand, $\tau_{op}$ of the central thymine base of 5'-AAA[T]AGA-3' and 5'-CAA[T]AGA-3' fragments differ threefold, though their $\tau_{cl}$ are close. Here we give the values for the 8-th basepair of L oligomer and the 13-th basepair of M oligomer from paper [319]. Hence the flip-out kinetics may be sensitive to the sequence of non-adjacent bases, up to three ones on each side.

The third factor is the position of a nucleotide pair in a specific nucleotide sequences. For example this can be a connect point between the segments one of which consists only of purine bases while the other – of pyrimidine ones [282]. For the central GC-pairs of oligomer 5'-d(GGAAAGCTTTCC)2, at 15 °C $\tau_{op} = 7$ ms, and $K_d = 1.5 \cdot 10^{-6}$. For comparison: for the same pairs of the "control" sequence 5'-d(CCTTTCGAAAGG)2, these values are equal to 40 ms and $3 \cdot 10^{-7}$ [282]. Another well-known fragment with anomalous kinetics is 5'-GTGT-3': for its second AT-pair, $\tau_{op}$ is 8-fold smaller than the relevant value for the fragment 5'-GTCT-3' [28]. Moreover, the stability of GC-pairs in 5'-GTGT-3' is also reduced as compared to 5'-GTCT-3', though to a lesser extent.

A special place among specific structures is held by the so-called tracts – sequences of identical bases. A-tracts are studied especially well. These are sequences of the form of 5'-$A_n$-3', where $n \geq 4$, or 5'-$A_nT_m$-3', where $n + m \geq 4$ [320]. In the early investigations $\tau_{op}$ of the bases occuring in the center of a tract was found to be 80–120 ms at 15 °C [320]. The kinetics of the boundary AT-pairs of the tracts remained quite ordinary [320, 285].

X-ray analysis demonstrated that A-tracts are stabilized by three-centered hydrogen bonds between the chains and characterized by a great propeller twist [321–323]. Such a structure was called B'-DNA [320, 285]. Molecular dynamics method confirmed that it survives in a solution too, though the exact values of the parameters greatly depend on the protocol choice [324]. Besides the three-centered H-bonds, A-tracts are stabilized due to enhanced energy of stacking interactions [325, 326].

Having replaced thymine bases by deoxyuridine ones in oligomers containing A-tracts Warmlander et al. revealed that the main contribution into the stabilization of these structures is made by $CH_3$ groups of thymine [325]. In B'-DNA not only the duplex, but also the open states *per se* demonstrate extra stability. Their $\tau_{cl}$ achieves 0.4–1 μs, that is 1–2 orders of magnitude higher than for AT-pairs outside the tracts [299]. Besides, for A-tracts, the cases of collective flip-outs of adjacent nucleotide pairs are shown [307].

Not less interesting are G-tracts described by Dornberger et al. [277]. Abnormally fast kinetics of singular openings is observed over the whole length of these tracts: $\tau_{op} \leq 12$ ms and $\tau_{cl} \leq 6$ ns [277]. Oligomers containing G-tracts are studied by the methods of IR spectroscopy, circular dichroism and NMR in combination with molecular dynamics [327]. It is shown that G-tract is characterized by a specific structure. It retains the main features of an ordinary B-DNA and at the same time has a high tendency of transition into A-form [327].

Hence a nucleotide sequence determines the peculiarities of the duplex structure to a large extent. It is responsible not only for the strength of the bonds between each base and its neighbors but also for the possible trajectories of the base's fliping out of the duplex. In terms of thermodynamics, the context of primary structure and the size of the base determine the activation enthropy $\Delta S^{\dagger 0}$ for its flip-out. In the case of single openings this value can play a crucial role, see below.

As was already briefly mentioned, a displacement of a base in the direction perpendicular to the duplex axis occurs only at the beginning of a flip-out. Its further motion proceeds along a complicated trajectory where the base retains its hydrogen bonds with the duplex [161]. According to Bouvier and Grubmuller, even at the beginning of a flip-out none of the DNA active vibration modes has a decisive significance in this process [163].

Sensitivity of a flip-out to the chemical nature, size and the primary structure context of the base, complexity of the flip-out trajectories lead to a wide scatter in the values of its thermodynamic parameters. For example, the activation enthalpy $\Delta H^{\dagger 0}$ varies from –25 [28] to 146 kJ/mol [307]. The reasons are quite obvious. Escape of some bases requires simultaneous breaking of a number of bonds, making $\Delta H^{\dagger 0}$ high. However in this case the free energy barrier $\Delta G^{\dagger 0}$ decreases due to a great activation enthropy $\Delta S^{\dagger 0}$: when many bonds are broken, a lot of trajectories “convenient” for a flipping out appear. Other bases can flip out from the Watson-Crick helix “by trick rather than by force” without breaking a number of bonds at a time. In this case the trajectory is more complicated and $\Delta S^{\dagger 0}$ is often below zero, thus compensating the activation barrier. This fenomenon is called a compensation effect. It is common in the chemistry of complex compounds [328].

The experimental values of the flip-out activation parameters for the thymine bases obtained by different researchers from 1987 till 2005 are summarized in Table 5.1. For clarity the values of $\Delta S^{\dagger 0}$ are replaced by the enthropy contribution – the product $T \cdot \Delta S^{\dagger 0}$, where T is equal to 288 or 293 K (the differences are insignificant in our opinion) For the quantities taken from the tables given in [28, 329–331], the values of inaccuracy are given. If these values are absent in the original tables we do not give them either.

When the original tables listed only two activation parameters we calculated the rest one using the expression $\Delta G^{\dagger 0} = \Delta H^{\dagger 0} - T\Delta S^{\dagger 0}$. The relevant values are marked by the symbol «ǂ»: their inaccuracy is not given. If the inaccuracy does not exceed 200 J/mol it is not given either.

In the other cases $\Delta H^{\dagger 0}$ was calculated with the use of the expression [28]:

$$\ln\left(\frac{1}{T\tau_{op}}\right) = \ln\left(\frac{k}{h}\right) + \frac{\Delta S^{\dagger 0}}{R} - \frac{\Delta H^{\dagger 0}}{RT}, \qquad (5.11)$$

where $k$ is Boltzmann's constant, $h$ is Planck’s constant. We used the values of $\tau_{op}$ for different temperatures which we took from the original tables. With no tables we got the data by digitizing the graphs from papers. The values of $\Delta G^{\dagger 0}$ and $T\Delta S^{\dagger 0}$ were calculated using the same data. The only exclusion is the paper by Moe et al. [318] where the values of $\Delta G^{\dagger 0}$ are presented in tables. For the values of $\Delta H^{\dagger 0}$ calculated (by digitizing) from graphs by work [318], inaccuracy is not given, but for $T\Delta S^{\dagger 0}$ it is taken equal to the inaccuracy of $\Delta G^{\dagger 0}$. The data obtained by digitizing are marked by the symbol “◊” in the column “reference”.

The data listed in the table are arrangesd in such a way that the thymine bases standing side by side are as similar as possible in the primary structure context. This enables one to compare thermodynamic parameters of the bases with a similar environment. The pairs investigated are given in the brackets. The figures in the parentheses at the beginning and at the end of a nucleotide sequence fragment indicate the number of the pairs to 5'- and 3'-end of the oligomer, respectively. If a base is close to one of the ends, 5'- or 3'-end *per se* is labeled. The bases involved in A-tracts are underlined. The asterisk «*» marks the nucleotides of the guanine-thymine pair investigated in [318].

In the table 5.2 the values of the same quantities GC-pairs flipping out are presented. Both the means of calculation and the parameters’ denotations are the same as in table 5.1. Out of 32 AT-pairs whose activation parameters are given in table 5.1, 16 pairs are characterized by $T\Delta S^{\dagger 0} \leq 0$. In six more pairs high positive values of the activation enthropy can be explained by their positions within A-tracts or closeness to an end of the duplex. Actually, the influence of the end effects is observed up to the third pair [316]. Therefore for the bases in the vicinity of the ends an escape from Watson-Crick helix by rupture of a significant number of interactions is more probable: some of them have likely been opened as yet due to the end-fraying effect.

**Table 5.1.** Activation parameters for AT-pairs in different primary structure contexts. See the text for explanations

| Primary structure context | $\Delta H^{\dagger 0}$, kJ/mol | $T\Delta S^{\dagger 0}$, kJ/mol ($T$ = 288–293 K) | $\Delta G^{\dagger 0}$, kJ/mol | Ref. |
|---|---|---|---|---|
| 5'-CCT[T]TCG-(5) | 52 | –6.4 | 58.4 | [282]$^{\Diamond}$ |
| (5)-GCT[T]TCC-3' | 57 | –1.4 | 58.4 | [282]$^{\Diamond}$ |
| (4)-AAT[T]T*GC-(1) | 82 | 3 ± 8 | 79 ± 8 | [318]$^{\Diamond}$ |
| (4)-AAT[T]TGC-(1) | 83 ± 8 | 19 ± 7 | 64$^{\ddagger}$ | [330] |
| (10)-TAT[T]TGC-3' | 118 ± 3 | 55$^{\ddagger}$ | 63 | [331] |
| ( 6)-TAT[T]TAT-(4) | 122 ± 3 | 58$^{\ddagger}$ | 64 | [331] |
| (2)-CTT[T]TAT-(3) | 146.8 | 84.2 | 62.6 | [307]$^{\Diamond}$ |
| (7)-ATT[T]ATT-(3) | 100 ± 5 | 38$^{\ddagger}$ | 62 | [331] |
| (5)-ATT[T]GCG-3' | 46 ± 17 | –12 ± 14 | 58 $^{\ddagger}$ | [330] |
| (11)-TAT[T]GC-3' | 75 ± 12 | 17$^{\ddagger}$ | 58 | [331] |
| (11)-ATT[T]GC-3' | 121 ± 4 | 62$^{\ddagger}$ | 59 | [331] |
| (1)-CTT[T]CGA-(4) | 65 | 7.7 | 57.3 | [282]$^{\Diamond}$ |
| (4)-AAT[T]CGC-(1) | 88 | 13 ± 12.5 | 75 ± 12.5 | [318]$^{\Diamond}$ |
| (6)-CTT[T]CC-3' | 46 | –8.9 | 54.9 | [282]$^{\Diamond}$ |
| ( 6)-TGT[T]CTA-(4) | 33 ± 4 | –26$^{\ddagger}$ | 59 | [331] |
| (3)-ATC[T]ATT-(7) | 88 ± 4 | 24$^{\ddagger}$ | 64 | [331] |
| 5'-CC[T]TTC-(6) | 43 | –14 | 57 | [282]$^{\Diamond}$ |
| (4)- AGC[T]TTC-(1) | 55 | –3.5 | 58.5 | [282]$^{\Diamond}$ |
| (3)- GAA[T]TCG-(2) | 76 | –3 ± 4 | 79 ± 4 | [318]$^{\Diamond}$ |
| ( 3)-GAA[T]TT*G-(2) | 79 ± 4 | –1 ± 4 | 80 ± 4 | [318]$^{\Diamond}$ |
| (3)- AAA[T]TTG-(2) | 71 ± 13 | 7 ± 13 | 64$^{\ddagger}$ | [330] |
| (3)- GAA[T]TCG-(2) | 47.6 | –10.9 | 58.5 | [299]$^{\Diamond}$ |
| (2)- CGA[T]CGC-(1) | 65 | 7.5 | 57.5 | [281]$^{\Diamond}$ |
| (3)- AGA[T]CTG-(2) | 29 ± 10 | –31 ± 10 | 60$^{\ddagger}$ | [28] |
| (2)- AGA[T]CAC-(1) | 71 ± 7.5 | 14 ± 7.5 | 57 | [329] |
| (2)- AAA[T]AAA-(8) | 90 ± 2.5 | 27$^{\ddagger}$ | 63 | [331] |
| (6)- AAA[T]AGA-(4) | 54 ± 4 | –7$^{\ddagger}$ | 61 | [331] |
| (3)- ACA[T]GTG-(2) | 25 ± 13 | –31 ± 13 | 56$^{\ddagger}$ | [28] |
| (10-AGA[T]GCG-3' | 54 ± 12 | –7$^{\ddagger}$ | 61 | [331] |
| (5)-ATG[T]GCG-3' | –25 ± 25 | –81 ± 24 | 56$^{\ddagger}$ | [28] |
| (5)-CTG[T]TCT-(5) | 71 ± 4 | 11 ± 4$^{\ddagger}$ | 60 | [331] |
| (5)-ATC[T]GCG-3' | 33 ± 8 | –23 ± 11 | 56$^{\ddagger}$ | [28] |

An analogous situation is observed in A-tracts where the duplex structure is additionally stabilized by three-center H-bonds and the stacking reinforced by propeller twist [285, 321–323]. This stabilization is the cause of the high $\Delta H^{\dagger 0}$ values for the bases in the tracts.

The positive values of $T \cdot \Delta S^{\dagger 0}$ in the rest seven AT-pairs are on the average 2–14 kJ/mol which is almost commensurable with the standard error of thermodynamic measurements, see tables 5.1. and 5.2. The only exclusion is the $AT_7$ pair of oligomer L from the work by Coman and Russu, for which $T\Delta S^{\dagger 0}$ = 23.8 kJ/mol [331]. In most cases the enthropy component either enhances the activation barrier of the thymine base flip-out or reduces it slightly.

**Table 5.2.** Activation parameters for GC-pairs in different primary structure contexts. See the text above for explanations

| Primary structure context | $\Delta H^{\dagger 0}$, kJ/mol | $T\Delta S^{\dagger 0}$, kJ/mol (T = 288–293 K) | $\Delta G^{\dagger 0}$, kJ/mol | Ref. |
|---|---|---|---|---|
| (1)-GCA[G]ATC-(4) | 42 ± 13 | –22 ± 14 | 64ǂ | [28] |
| 5'-GTA[G]ATC-(3) | 67 ± 3 | 6 ± 3 | 61 | [329] |
| (3)-ATA[G]AAC-(7) | 92 ± 8 | 28ǂ | 64 | [331] |
| (3)-TTC[G]AAA-(2) | 65 | 2.3 | 62.7 | [282]◊ |
| 5'-CGC[G]ATC-(3) | 46 | –14.4 | 60.4 | [283]◊ |
| 5'-CGC[G]AAT-(5) | – | – | 7 ± 16 | [318]◊ |
| 5'- GC[G]ATC-(11) | 104 ± 8 | 41ǂ | 63 | [331] |
| 5'- GC[G]ATC-(11) | 109 ± 3 | 45ǂ | 64 | [331] |
| (2)-AAA[G]CTT-(3) | 30 | –28 | 58 | [282]◊ |
| (6)-T*TT[G]CG-3' | – | – | 79 ± 25 | [318]◊ |
| (6)-TTT[G]CG-3' | 92 ± 21 | 31 ± 19 | 61ǂ | [330] |
| (6)-TGT[G]CG-3' | 67 ± 8 | 6 ± 9 | 61ǂ | [28] |
| (6)-TCT[G]CG-3' | 29 ± 16 | –31 ± 17 | 60ǂ | [28] |
| (3)-ATA[G]AAC-(7) | 92 ± 8 | 28ǂ | 64 | [331] |
| 5'-AGT[G]ATC-(3) | 77 ± 6 | 16 ± 6 | 61 | [329] |
| (4)-ATC[G]CG-3' | 47 | –12.3 | 59.3 | [283]◊ |
| (6)-TTC[G]CG-3' | – | – | 88 ± 21 | [318]◊ |
| (4)-CAT[G]TGC-(1) | 62 ± 8 | 1 ± 7 | 61ǂ | [28] |

The data on the guanine bases flipping out are more scarce. A complete set of activation parameters is presented in table 5.2 only for 15 GC-pairs. Only 5 of them are characterized by $T\Delta S^{\dagger 0} < 0$. This can be explained not only by more strong bonds in the GC-pair but also by a greater size of the guanine base as compared to the thymine one.

In the GC-pairs where $T\Delta S^{\dagger 0} > 0$, the consistent patterns are practically the same as in AT-pairs. The greatest $\Delta H^{\dagger 0}$ and $T\Delta S^{\dagger 0}$ are observed in GC-pairs placed near the duplex ends. As for the rest of the pairs, the greatest $T\Delta S^{\dagger 0}$ are characteristic for guanine bases placed between two adenines. What is interesting is that in the case of $GC_4$ pair of oligomer d(AGTGATCTAC):(GTAGATCACT) investigated in [329], the end effects and the occurrence between the adenine bases are probably annihilated. This leads to moderate values of $\Delta H^{\dagger 0}$ and $T\Delta S^{\dagger 0}$ – 67 and 6.1 kJ/mol. The values of $T\Delta S^{\dagger 0}$ for all the rest GC-pairs do not exceed 16 kJ/mol.

Generalizing the data of tables 5.1 and 5.2 we can say that the ratio between the enthalpy and enthropy contributions into the flip-out activation barrier is determined by a set of factors. The main of them are the base size, the strength of complementary H-bonds, the energy of stacking interactions and the primary structure context – up to three bases from each side.

The smaller is the number of interactions which are disrupted during a base flipping out and the more complicated is the trajectory of its escape, the lower are $\Delta H^{\dagger 0}$ and $\Delta S^{\dagger 0}$ of this process. Therefore opening of several neighboring bases should be characterized by a great $\Delta H^{\dagger 0}$ compensated by high $T\Delta S^{\dagger 0}$. A striking example is provided by high $\Delta H^{\dagger 0}$ of a concerted flipping out reaction of two adjacent bases in A-tract obtained in [307]. Its value is approximately 147 kJ/mol, see table 5.1.

The process of a denaturation bubble nucleation involving a number of neighboring bases should be characterized by still higher $\Delta H^{\dagger 0}$ and $\Delta S^{\dagger 0}$. This assumption can be confirmed by the analysis of the results of FCS of molecular beacons. In what follows we will assess the activation parameters of the bubble nucleation on the basis of the temperature dependencies of their kinetics obtained by Altan-Bonnet et al. [213].

## 5.3. Thermodynamical differences between flip-outs and denaturation bubbles. Explanation of disagreement between FCS results and data from other technuques

Fluorescent correlation spectroscopy of molecular beacons [213, 279, 332] is the only technique which enables direct investigations of the denaturation bubbles kinetics in DNA. The principle of this method is briefly described in Section 4.3. In the work by Altan-Bonnet et al. kinetic and thermodynamic parameters of the bubble relaxation in the beacons of different primary structures are studied [213].

Hereafter we are interested in bubble nucleation rather than in its relaxation. As will be shown in what follows, this process implies simultaneous opening of $N$ adjacent basepairs. Assesments of the minimum $N$ and the arguments that $N > 1$ are given in Section 6.1. The most probable value of $N$ is 3–6 basepairs. The bubble nucleation corresponds to the first reversible reaction in the scheme:

$$closed \quad DNA \leftrightarrow bubble_N \leftrightarrow bubble_{N+1} \leftrightarrow bubble_{N+2} \leftrightarrow bubble_{N+3} \leftrightarrow \ldots, \qquad (5.12)$$

where the subscript stands for the length of the open fragment in the nucleotide pairs.

A bubble of 4–6 pairs in length is sufficient for a fluorophore and a quencher to separate from each other, see Section 5.3.1. Therefore the $bubble_N$ is a fluorescent form with a high probability.

The bubbles consisting of $N$ and more basepairs have relatively long tifetimes and obviously make the main contribution into the total fluorescence. Therefore the contribution of the short-living open states involving less than $N$ pairs can be neglected. Hence, we can admit that the lifetimes of the open state $\tau_{cl,bub}$, obtained by Altan-Bonnet et al. [213] characterize only the bubbles.

In this approximation the averaged time of the closed state $\tau_{op,bub}$ can be easily calculated from the expression

$$\tau_{op,bub} = \frac{\overline{\tau}_{cl,bub}}{K_{d,bub}}$$

where $\overline{\tau}_{cl,bub}$ is the averaged time of an open state, and $K_{d,bub}$ is an equilibrium constant for the beacon opening reaction in the tagging point. Accordingly, the activation parameters of the bubble nucleation may be obtained by a simple addition of the activation characteristics for relaxation and the standard thermodynamic values.

The standard thermodynamic parameters were estimated for the premelting temperature interval [326, 333]. The upper temperature limit of this interval for each beacon is chosen as $T$ of the growth onset of $\frac{d(\ln[I(T)])}{dT}$, where $I(T)$ is the normalized value of the total fluorescence. Strictly speaking, the equality $K_{d,bub} = \frac{I(T)}{1 - I(T)}$ is not proved. However a close relation of $K_{d,bub}$ with $I(T)$ suggests that an increase in $\frac{d(\ln[I(T)])}{dT}$ coincides with the growth of $\frac{d\left(\ln\left[K_{d,bub}\right]\right)}{dT}$ which corresponds to a transition to the melting phase.

The value of $\left|\frac{K_{d,bub} - I(T)}{1 - I(T)}\right|$ for the pre-melting interval was considered to be small as compared to the inaccuracy of $I(T)$. Therefore it was taken to be equal to zero. This approximation is quite appropriate since our aim is, mainly, to estimate qualitatively the thermodynamic characteristics of bubbles.

The lower limit of the interval of the standard parameters estimation was set as temperature at which the ratio between $I(T)$ and the normalized inaccuracy of its measurement reduces to 8. This enabled the inaccuracy associated with the signal logarithmation to be neglected. Besides, as $T$ falls beyond the lower limit of the estimation interval, the value of $\frac{d(\ln[I(T)])}{dT}$ of the beacons starts increasing. The nature of this effect will be considered in detail in Section 6.5.

The estimation intervals for the thermodynamic parameters were: 38–52 °C for $M_{18}$, 30–54 °C for $A_{18}$ and 31–44 °C for $(AT)_9$. The temperature dependencies of the times $\tau_{cl,bub}$ were calculated from the relaxation kinetics data. The data were kindly provided by O. Krichevsky under whose leadership the FCS experiments were carried out [213]. The activation parameters of the bubble closing were found with the use of formula (5.11). To calculate the standard thermodynamic quantities from temperature dependences of $K_{d,bub}$ use was made of the expression

$$\ln\left[K_{d,bub}\right] = -\frac{\Delta H^0}{RT} + \frac{\Delta S^0}{R}, \tag{5.13}$$

The data on the temperature dependence of fluorescence $I(T)$ for the beacon $M_{18}$ are furnished by O. Krichevsky. Points $I(T)$ of the other two beacons were obtained by digitizing of the curves in figure 2 from paper [213]. The digitizing error was not taken into account in the calculations of the thermodynamic quantities. The normalized error of measurement $I(T)$ of $A_{18}$ and $(AT)_9$ beacons was taken equal to the error for $M_{18}$ which is about 0.005.

The values of the activation parameters of the bubble opening are presented in table 5.3. As in the case of tables 5.1 and 5.2, the activation enthropy changes $\Delta S^{\dagger 0}$ are replaced, for clarity, by the enthropy contributions $T\Delta S^{\dagger 0}$.

The values of $\tau_{op,bub}$ corresponding to the activation barriers given in table 5.3 fall in the range of 0.4–2.3 ms. These times are well within the range obtained by extrapolation of the activation thermodynamic parameters of the flip-out into 38 °C which is 0.25–7 ms, see expression (5.11) and table 5.1.

As one would expect, the activation enthropy appeared to be positive and rather high in all the cases. However, for $A_{18}$ the values of $\Delta H^{\dagger 0}$ and $T{\cdot}\Delta S^{\dagger 0}$ are considerably reduced as compared to other beacons. This is due to specific properties of the center of $A_{18}$ which is essentially an A-tract, i.e. B'-DNA, see Section 5.2. Possible mechanisms of a reduction of $\Delta H^{\dagger 0}$ and $T\Delta S^{\dagger 0}$ in such structures are described below in Section 5.3.3. It should be noted that $\Delta H^{\dagger 0}$ of base flipping is relatively high in A-tracts (see table 5.1), on the contrary to $\Delta H^{\dagger 0}$ of bubble formation. Hence the behavior of $A_{18}$ is a perfect illustration of a principal difference between these types of the open state.

**Table 5.3.** Activation thermodynamic parameters of the bubble nucleation in molecular beacons, kJ/mol. $T = 38$ °C. The tagging points are underlined

| Beacon | Structure of AT-rich region, from 5'- to 3'-end | $\Delta H^{\dagger 0}$ | $T\Delta S^{\dagger 0}$ | $\Delta G^{\dagger 0}$ |
|---|---|---|---|---|
| $M_{18}$ | AATATAAAA**T**ATTAAAAT | 120 ± 23 | 61 ± 25 | 59 ± 1.5 |
| $(AT)_9$ | ATATATATA**T**ATATATAT | 111 ± 25 | 53 ± 26 | 58 ± 1 |
| $A_{18}$ | AAAAAAAAA**T**AAAAAAAA | 84 ± 22 | 27 ± 23.5 | 57 ± 1.5 |

The main result of the FCS of molecular beacons is a wide range of $\tau_{cl,bub}$ which comprises $10^{-6}$–$10^{-3}$ s [213]. However since the characteristic $\tau_{cl}$ is usually within $10^{-7}$ s, the similarity of $\tau_{op}$ and $\tau_{op,bub}$ suggests that $K_d$ differs from $K_{d,bub}$ by 1–4 orders of magnitude. $K_{d,bub}$ obtained by FCS differs drastically from that obtained by other methods [311, 312]. Besides, for such enormous $K_{d,bub}$ the vast majority of the imino protons should exchange from the bubbles. This is in conflict not only with the data on interindependence of flip-outs but also with the whole kinetics of $^{1}$H exchange described in Section 5.1.

The simplest way to resolve this contradiction is to believe that the results obtained by FCS of the beacons are unreliable, erroneous. For example, according to Peyrard et al., a fluorescent label can have a great influence on the duplex dynamics [334]. However if we bring together some facts which are not interrelated at first sight, the reliability of the FCS data becomes obvious.

First, let us consider a similar order of magnitude of the times in all the beacons. As was already mentioned, $A_{18}$ could form shifted structures and $(AT)_9$ was additionally form cruciforms [213]. However actually the times of the beacon relaxation differed moderately. The reason is not retaining of the stack in the open states of a duplex but its fast formation anew. Formation and decay of the stacking was shown to occur in tens-to-hundreds nanoseconds timescale in the investigations performed on a one-chain polycytidylic acid [335].

The existence of stacking in one-chain nucleic acids is confirmed for molecules of different lengthes: from dimers [74, 336, 337] to polynucleotides [39, 338–340]. According to the data of differential scanning calorimetry, when short chains reassociate into a duplex at low temperatures, most of them have the form of a single helix [341, 342]. This fact is also confirmed by a combination of microcalorimetry investigations and neutron scattering [343]. The absolute value of oligomer dissociation enthalpy at melting temperature is almost twice as high as the absolute value of its renaturation enthalpy [341]. It seems clear that energy release as a result of renaturation is fewer than energy cost of denaturation by virtue of reassociation of single DNA strands with rebuilt stack [ibid.].

By various estimates, the equilibrium constant for the stacking rupture in a one-chain DNA at temperatures below 40 °C falls in the range of 0.05–0.5 and depends greatly on the nucleotide sequence, see for example [340, 341]. Besides, it has long been known that a one-chain poly(A) is one of the most stable forms in an aqueous solution [338, 339]. A DNA fragment consisting of adenine nucleotides has a considerable persistent length as compared to one-chain fragments of a different primary structure [344]. It was shown with the use of molecular beacon spectroscopy that high rigidity of these fragments is caused by considerable stacking enthalpy [332].

Formation of stacking interactions in unwound fragments should considerably reduce $\Delta H^0$ of a bubble opening. This effect seems to be especially pronounced in the beacon $A_{18}$ containing a polyadenylic fragment. Moreover $A_{18}$ has the lowest $\tau_{cl,bub}$. This fact is in good agreement with the data on a possibility of individual DNA chains reassociation. In such chains stacking interactions are partly retained, see above. Reduced values of $\Delta H^0$ for a bubble opening and $\Delta H^{\dagger 0}$ for its closing suggest a low activation enthalpy of its nucleation. This confirms a possibility of a radial separation of the chains without decay of the stacking.

An important role of stacking interactions is implicitly pointed to by the results of Raman spectroscopy. According to these data the main part of stacking interactions in polynucleotide poly(A):poly(T) is retained even after a wide-scale opening of H-bonds at temperatures above 65 °C [333]. In another DNA investigated in that work – poly(A-T):poly(A-T) – the stacking disrupt simultaneously with opening of H-bonds. The primary structures of the DNA investigated coincide with the middle areas of the beacons $A_{18}$ and $(AT)_9$, respectively, while their lengths are several thousands basepairs. However none of these DNA formed any alternative secondary structures (shifted structures or cruciforms) in the course of investigations notwithstanding repetitive heating to 85 °C and higher [333]. This enables us to hold that the contribution from the formation of such structures in FCS is small to negligible.

Now let us turn to the problem of the discrepancy between the data from $^{1}$H NMR and FCS. To understand its nature we should address to the material of the previous chapters of our review. To explain the contradictions between the results of FCS of the beacons and the data of other methods we can point out four mutually supportive factors.

The first factor will be conditionally called sequential because it results from DNA nucleotide sequence.

### 5.3.1. Sequential factor. Role of DNA primary structure

It is well known that stability of a DNA fragment is determined by its nucleotide sequence. Typical $\Delta H^0$ for of AT- and GC-pairs' opening used in the nearest-neighbor models are 35.5 and 39.3 kJ /mol, respectively [345]. Standard entropy changes $\Delta S^0$ for nucleation of a little bubble must be the same order of magnitude for the cases of a AT- and GC-pair rows. First, suppose a simultaneous opening of all bubble's nucleotide pairs during its nucleation. Second, let us assume that $\Delta S^0$ depends on the nucleotide sequence only slightly. Therefore, proceeding from the well-known expression for the equilibrium constant $K = \exp[-\Delta G^0(RT)^{-1}]$ we can see that $K_{d,bub}$ decreases in an arbitrary fragment approximately 4.5-fold when replacing each next AT-pair by GC-pair.

Actually, in view of the compensation effect (see Section 5.2), $K_{d,bub}$ should depend on the nucleotide sequence somewhat less. However for $N$ equal to 3–6 basepairs, the differences of $K_{d,bub}$ by two orders of magnitude or more are quite possible. This fact is confirmed in vitro [29, 30]. The high temperature stability of the end fragments of the end-tagged beacons (see fig. 4.11,b from paper [213]) is also very indicative. Fluorescent melting profiles of the middle- and end-tagged beacons are shown in figure 5.2.

Separation of fluorophore from a quencher in an end-tagged construct is promoted by the possibility of a free rotation around P–O bonds. Besides, separation of the chains in this case is promoted by end-fraying effect involving three terminal basepairs [316]. Nevertheless, the ends of the beacons are far more stable than their centers. Low stability of a terminal pair is compensated in this case by the complementary H-bonds and stacking in its adjacent GC-pairs.

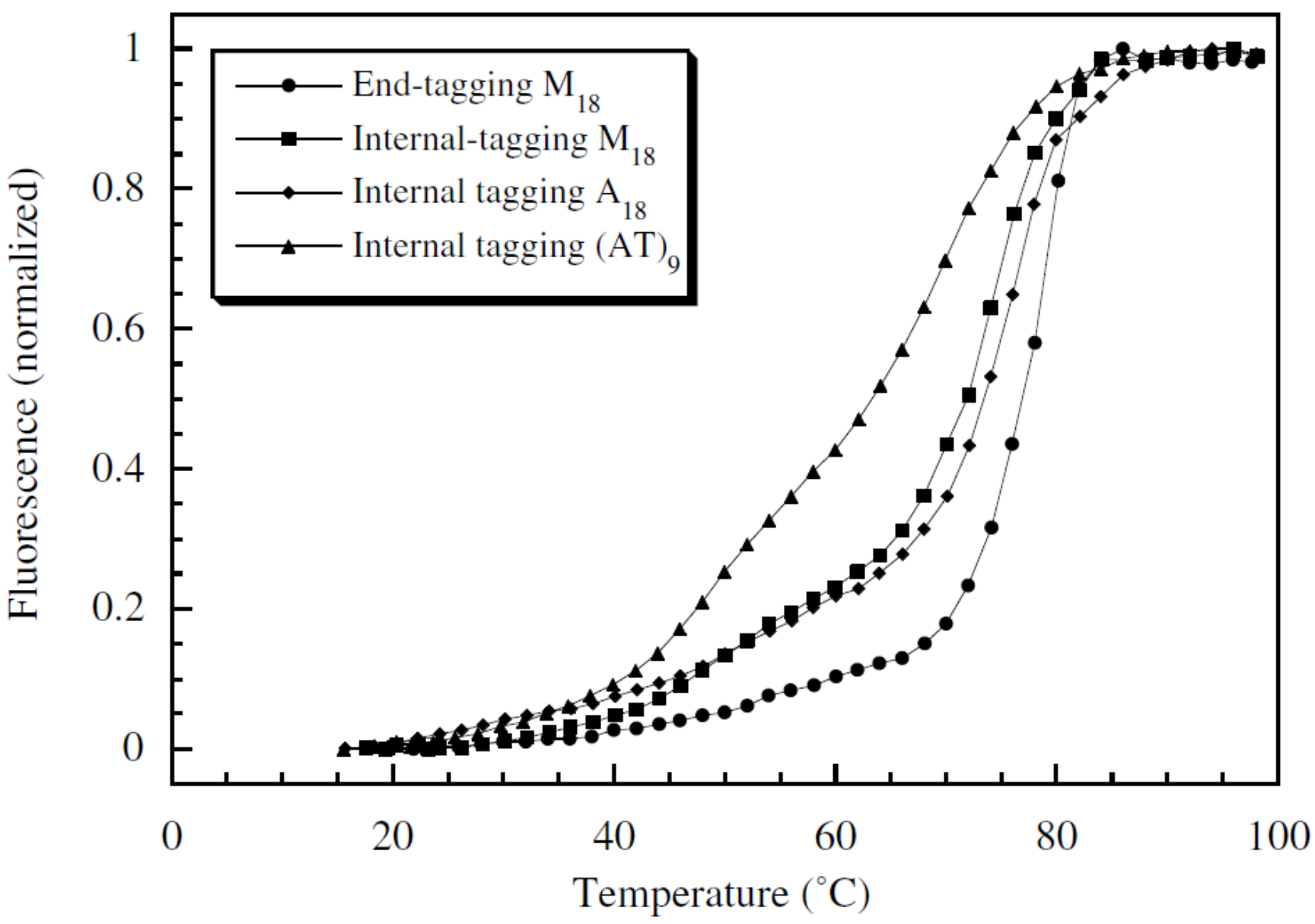

**Fig. 5.2.** Normalized fluorescent melting profiles of the middle-tagged beacons $M_{18}$ (■), $A_{18}$ (♦), $(AT)_9$ (▲), and the end-tagged beacon $M_{18}$ (●) [213], see figure 4.11. Each profile reflects the portion of open beacons calculated from formula (4.4).

Hence nucleation of a bubble involves simultaneous openings of a certain minimal number of basepairs $N$, see equation (5.12). Unfortunately $N$ cannot be determined exactly. The only thing that is highly probable is that $N$ is larger than a unity.

One of the proofs of this fact is the phenomenon of the critical length of oligomers which will be considered in the next section. This section is devoted to description of the second factor concerned with different principles of $^{1}$H NMR and FCS of molecular beacons.

### 5.3.2. Technique-dependent factor and some peculiarities of $^{1}$H NMR

We have described the phenomenon of a critical length for oligomers consisting of two terminal GC-rich domains and an AT-rich domain between them in Section 4.1. It should be noted that almost all the oligonucleotides under investigation had such a structure. If the length of such DNA is less than a certain threshold, a nucleation of denaturration bubble is prohibited. Instead the duplex dissociates as a plain chemical complex, i.e. without intermediate states. The critical length should be related to $N$ by rather a simple relation. Actually, even the smallest bubble should contain a sufficient number of bases for a short duplex to destabilize. The value of the critical length is found both in vitro [261] and by way of calculations [262] and is equal to 20–22 basepairs. The typical length of oligomers on which the flip-out kinetics was studied was almost twice less, see tables 5.1 and 5.2.

In the presence of a catalyst a dissociated oligonucleotide exchanges all the protons during tens of nanoseconds. This is several orders of magnitude less than the characteristic interval between the impulse signals in the inversion recovery of magnetization transfer techniques. A complete dissociation of the chains will also not “be seen” in the case of proton-deuteron exchange since the resonance broadening will be too great. Similar properties should be shared by the bubbles as well.

As the catalyst concentration approaches 100 mmol/l one flipping out occasion of five ones leads to proton exchange. Therefore if the time of a complete dissociation of a duplex is at least several orders of magnitude higher than $\tau_{op}$, the main role in $^{1}$H exchange belongs to just a flip-out. Moreover for high [acc] a difference in one order of magnitude is sufficient! This suggests a simple conclusion: the ratio between the contributions of bubbles and flip-outs into the proton exchange process depends not on $K_{d,bub}/K_d$ but on $\tau_{op,bub}/\tau_{op}$. This peculiarity enables one to investigate a flip-out at rather high temperatures. For example, even at 35 °C the time of dissociation of oligomer $d(CGCGATCGCG)_2$ exceeds 120 ms, and its equilibrium constant is not higher than 0.00085 [283].

Denaturation behavior of oligomers investigated by NMR is, in turn, a good illustration of the role of a sequential factor (see above). For example, the main cause of the oligomer $d(CGCGATCGCG)_2$ stability of is the abundance of GC-pairs there. For comparison we can consider a nucleotide d(CAACTTGATATTAATA):d(TATTATTATCAAGTTG) of a similar length. Its terminal AT-domain completely denaturates between 7 and 26 °C [346]. Therefore the low stability of AT-domains in the beacons is in good qualitative agreement with the denaturation data on other nucleotides.

However the role of the sequential factor can also be estimated quantitatively, though approximately. We need only compare the fluorescent data with the photometric profiles of short DNA with similar primary structures. These are self-complementary oligomers $L_{36}AS$ and $L_{60}B_{36}$ which is studied by quenching technique [55, 62, 261]. Their AT-rich domains have the lengthes of 16 and 36 basepairs, respectively and contain very few GC-pairs – 2/16 and 4/36.

To estimate the role of the primary structure it is convenient to introduce an artificial parameter $T_{st}$ – the temperature of the melting onset. Let it be a temperature for which the photometric signal reaches 1 % of the maximum value. The values of $T_{st}$ in oligomers $L_{36}AS$ and $L_{60}B_{36}$ are 40 and 39 °C, respectively. Assuming that the dependence of $T_{st}$ on the portion of GC-pairs [GC] is similar to the relevant dependence for $T_m$ and using the empirical formula by Marmur and Doty $T_m = 69.3 + 41[GC]$ [56] we get on the average $T_{st} \approx 34.6$ °C for $[GC] = 0$. This value exceeds $T_{st}$ of the beacon $M_{18}$ by as little as 4.6 °C which suggests an important role of the sequential factor.

However a direct comparison of "fluorescent" $T_{st}$ of the beacons with "photometric" $T_{st}$ of the hairpins, most probably, leads to a certain overestimation of a difference between them. Indeed, enhancement of absorption at the wavelength of 260–268 nm is caused by decay of stacking interactions, see Section 1.1. At low $T$ the fraction of disrupted stacking may be moderate even in one-chain fragments of the bubbles. As a result, $T_{st}$ for the total normalized fluorescence should be somewhat lower than a "photometric" $T_{st}$. This leads to underestimation of the role of the sequential factor. It is noteworthy that this effect should be especially pronounced for the beacon $A_{18}$, since stacking in polyadenylic single strands is the most stable, see above.

Besides, the influence of the primary structure can depend on the length of the AT-rich domain. In recent investigations the stability of a small DNA fragment was shown to depend not only by its nucleotide sequence but also by the stability of the fragments on the neighboring coils of the helix. The influence of steric effects was clearly demonstrated with the use of ultraviolet laser photolysis of guanine [347]. According to these, large premelting fluctuations in the AT-rich domain strongly affect the stability of the fragments placed 10–11 basepairs away from this domain, i.e. after a full turn of the double helix.

Hence in a fragment consisting of 18 AT-pairs and forming nearly 2 coils, fluctuations in the neighboring regions of the central domain should mutually enhance. In view of the foregoing it becomes clear that the stability of the AT-rich domain of the beacon $M_{18}$, according to Altan-Bonnet et al., is quite ordinary for a DNA fragment of such a structure.

Nevertheless, $T_{st}$ of $(AT)_9$ and $A_{18}$ are equal to 26 и 22 °C, respectively. So low values cannot be explained by the sequential factor alone even bearing in mind the effects of the helix and the stacking stability. Besides, the range of their $T_{st}$ is 8 K, and the range of $T_m$ is even more – nearly 10 K (see fig. 5.2), although [GC] = 0 in AT-rich domains of all the three beacons.

It is evident from figure 5.2, that the beacon $A_{18}$ which has the smallest $T_{st}$ nevertheless, has the greatest $T_m$ since its AT-rich domain is A-tract. As already mentioned in Section 5.2, these structures are stabilized by three-center H-bonds between the chains. A great propeller twist emerging in this case additionally strengthens the stacking. Since H-bonds and stacking interactions open almost simultaneously at high temperatures, enhanced $T_m$ of A-tract seems quite reasonable.

Finally, low $T_{st}$ of the beacon $A_{18}$ can be explained by the possibility of a radial separation of the strands without pronounced disruption of the stack. This is suggested by the thermodynamic properties of $A_{18}$ which were analyzed in Section 5.3. However there is one more factor which is concerned with localization of nonlinear excitations in deformed DNA fragments. It can be called a "breather" factor. The contribution of this factor into a decrease in $T_m$ is maximum just for $A_{18}$, though for $(AT)_9$ and $M_{18}$ it seems to be essential too.

### 5.3.3. Breather factor

In the papers by Peyrard, Choi, Alexandrov and other investigations described in Section 4.2 the unstable DNA fragments were shown to be preferable places of localization of the nonlinear excitations energy. Investigation of the modified PBD model for moderate temperatures has revealed that some fragments of a heterogeneous DNA are characterized by much longer lifetimes of open states than the rest of the duplex [274–276]. As a rule, in these fragments the percent of AT-pairs is high and/or the average energy of stacking interactions between the bases is low.

Evidently, the most unstable place of any beacon is the fluorescent tagging point. First, the activation barrier of the bubble formation there should be reduced due to enhanced $\Delta S^{\dagger 0}$. Second, the fluorescent label *per se* can probably influence the DNA dynamics through steric hindrances for a bubble closing and $\Delta H^{\dagger 0}$ lowering (fluorophore and quencher may serve as "Brownian sails"). These mechanism is probably especially effective in the case of the beacon $(AT)_9$ with the least energy cost of nucleation and the maximum stacking preserving during bubble nucleation. Its $T_{st}$ is equal to 26 °C, which is less than $T_{st}$ of $M_{18}$ by 4 °C. These values are by 8.6 K less than the average $T_{st}$ of oligomers $L_{36}AS$ and $L_{60}B_{36}$ extrapolated to [GC] = 0, see above.

The second way of the nonlinear excitations' energy localization is caused by distortions of the DNA structure. A characteristic example is provided by curvature of the duplex axis in the AT-rich domain of the beacon $A_{18}$. This curvature is typical for A-tracts [285, 320, 348]. Enhancement of the DNA stretched oscillations (along complementary H-bonds) in bent region due to trapping of incoming breathers is described in detail in the paper by Ting и Peyrard [349]. The authors demonstrated that a growth of the amplitude of the stretched oscillations leads to a decrease of their frequency due to nonlinear nature of the Morse potential. It results in increasing the probability that each subsequent breather will be trapped by a "hot" bent DNA fragment. A similar effect was also demonstrated for the modified PBD model in which a distortion was introduced through dipole-dipole interactions. It is the energy localization at the center of A-tract that can be an additional cause of a reduced activation enthalpy required for a bubble nucleation in $A_{18}$.

Besides the main results obtained by Altan-Bonnet et al. there is one more feature of the beacon opening kinetics which deserves attention. The lower limit of the bubble relaxation time, according to FCS data, is in the range of $10^{-7}$–$10^{-6}$ s, see figure 4.12. This time is comparable with the characteristic time of the stacking breakdown in a one-chain polycytidylic acid which is 180–380 ns at 20 °C and 130–220 ns at 30 °C [335]. The similar timescale suggests a possibility of the formation of small open DNA states in which stacking interactions cannot break fast enough. If in this case a fluorophore is also separated from a quencher, such open states will make a certain (probably considerable) contribution into the fluorescent signal. On the other hand, due to high measurement inaccuracy, for small times, these openings may not be seen on the kinetic curves – such as those in figure 4.12.

As compared to the typical bubble relaxation times obtained in [213] the lifetime of such open states is very little. Actually, they are not "full-featured" denaturation bubbles, but merely large fluctuations.

This suggests the name of the fourth (probable) factor which may overestimate the registered probability of the beacon opening – fluctuation factor.

### 5.3.4. Fluctuation factor and assessment of its contribution

Actually, the analysis of this factor is not aimed at the explanation of the FCS data. Rather, it implies an attempt to bring together the experimental results and the data from the PBD model investigations. In the PBD model, $K_{d,bub}$ for a 2–3 basepair bubble nucleation is equal to 0.002–0.004 at 37 °C for DNA consisting of AT-pairs only [350]. These values are much closer to the FCS results than to the data of other experiments where $K_{d,bub}$ does not exceed $10^{-5}$ ([312], cited by [282]).

Let us suppose that retaining of stacking interactions in small openings can critically restrict access of the solvent molecules to the bases. Then, according to the data of most experiments, their concentration will be very small. At the same time modeling will show much higher values of this quantity since in this case a change in the distance between the bases in basepairs is registered.

In the remainder of Chapter 5 we will analyze some literature data so that to estimate:

1) characteristic number of bases within a specific open state with maintained stacking, which is formed as a result of small radial separaition of the DNA chains;

2) the fraction of open states of this type;

3) accessibility of imino groups of such bases to the molecules of the solution.

If this accessibility is, indeed, severely restricted, the discrepancy between the results obtained with the use of the PBD model and the experimental data will be explained unambiguously. On the other hand, in this case we have to modify the definition of the open state which was given in the Introduction. Actually, the only characteristic which enables one to determine distinctively an open state is opening of complementary H-bonds.

Let us estimate the probability that stacking interactions will be retained in the bubbles formed in the FCS experiments on the beacons. For the purpose look at figure 2.2 (Section 2.3) which illustrates the idea of the radial-torsional model by Barbi et al. [173]. The distance between the planes of neighboring bases $h'$ is equal to 3.4 Å. The distance between the points of their attachment to the sugar-phosphate backbone $L$ is, on the average, 4.7 Å. A small radial separation of complementary bases which corresponds to lengthening of the bond $r_n$, does not cause any noticeable conformation strains in the duplex. Simple calculation suggest that moderate reduction of $\varphi_n$ from 36° to 23° at constant $L$ leads to a motion of a base away from the DNA axis by 2 Å. This corresponds to $r_n$ increase by 4 Å in an individual pair. In this case a decrease of $\varphi_n$ partly compensates a reduction of the contact area of the bases and promotes retaining of stacking interactions.

Now let us estimate the distance over which the DNA chains should be spaced apart so that to exclude the contact of the fluorophore with the quencher. Figure 5.3. presents the structural formulae of the thymine bases in DNA to which these molecules are covalently bound. The lengthes of the hydrocarbonic bridges were fitted in such a way that the fluorophore and the quenching agent could form stack without preventing the nucleotide basepairs closing; quenching of the beacon fluorescence takes place due to formation of a stacking interaction between the fluorophore and the quencher [351].

**Fig. 5.3.** a) Structural formula of the fluorophore (carboxyrhodamine 6 G) covalently bound to the thymine base by a bond of 6 carbon atoms in a molecular beacon; b) Structural formula of the quencher (DABCYL) bound by the same bond.

Based on the size of these molecules we can conclude that for a fluorophore to be isolated from a quencher they should be spaced by approximately 7 Å. Therefore for moderate temperatures the total fluorescence should be sensitive to the kinetics of bubbles consisting of not less than 4–6 basepairs.

Theoretically, an open state with a retained stack can be stabilized due to entering of water molecules into the space between the open chains. Further, the viscosity of an aqueous “microcluster” between the DNA chains may probably be much higher than the viscosity of liquid water where the position of the molecules changes in the picosecond timescale [352, 353]. However even the most short-living open states registered by FCS relax for not less than 100 ns. Therefore even if the molecules which have located between the DNA chains stay there for hundreds of picoseconds they can alternate many times during the liferime of an open state. This testifies in favour of the accessibility of the DNA molecular groups, in particular, imino groups, to the solution compounds. This is the first indirect logical disproof of the significant contribution of “fluctuation factor”.

We can additionally estimate this accessibility relying on indirect data. In particular, one of the most important works in this field was carried out by Nonin et al. [317]. The authors investigated the dynamics of $^1$H exchange in terminal nucleotide pairs. It was shown that the terminal position of a nucleotide pair in itself is not a sufficient condition for $^1$H exchange to take place. On the other hand, the values of $K_d$ for the terminal bases which they obtained by NMR methods, were 1–2 orders of magnitude higher than the relevant $K_d$ calculated earlier with the use of calorimetry [354]. This suggests a key role of complementary H-bonds in terminal pairs: their opening make the protons of imino groups accessible for exchange irrespective of whether stacking interactions retain or not. This is the second indirect disproof.

Nevertheless, it is impossible to find out exactly how much the non-terminal nucleotide pairs differ from the terminal ones in this respect. However we can judge about their properties indirectly by comparing the kinetics of the catalytic $^1$H exchange with the kinetics of the exchange in the absence of an external catalyst. In the latter case the main way of exchange is the internal catalysis whose idea is described in Section 5.1. A more detailed analysis of its mechanism will enable us to demonstrate the characteristic features of the open states in the center of the duplex and to compare them with the terminal ones. Such an analysis is made in the next section, where the direct mathematical disproof of the prominent role of “fluctuation factor” is expanded.

## 5.4. Internal catalysis and accessibility of imino protons

In the case of internal catalysis an acceptor of an imino group proton is a nitrogen atom of a complementary base – $N_1$ of adenine or $N_3$ of cytosine. The transfer of $^1$H proceeds through an inner-sphere hydrogen-bonded complex. The complex consists of a pair of complementary bases and a water molecule between them. The key phases of the exchange are a consistent transfer of $^1$H and a rotation of $H_2O$ molecule, as is shown in scheme 6 [297].

$$\text{nu-H*}\cdots\underset{\text{H}_a}{\text{O}}\text{-H}_b\cdots\text{acc} \xleftrightarrow[\text{transfer}]{\text{concerted}} \text{nu}^-\cdots\text{H*-}\underset{\text{H}_a}{\text{O}}\cdots\text{H}_b^+\text{acc}$$

$$\xleftrightarrow{\text{rotation}} \text{nu}^-\cdots\text{H}_a\text{-}\underset{\text{H*}}{\text{O}}\cdots\text{H}_b^+\text{acc} \xleftrightarrow[\text{transfer}]{\text{back-}} \text{nu-H}_a\cdots\underset{\text{H*}}{\text{O}}\text{-H}_b\cdots\text{acc}$$

**Scheme 6.** Transfer of an imino proton to an acceptor via an inner-sphere hydrogen-bonded complex.

As we will show in what follows, an inner-sphere hydrogen-bonded complex is a full-featured open state.

The first evidence that a complementary base can act as a catalyst was obtained in the early works on $^1$H NMR [281, 284]. For example, the terminal thymine in oligomer

d(AATTGCAATT):d(AATTGCAATTT) was shown to be the slowest to exchange the imino proton in the absence of an external catalyst, because this base does not have the complementary one [284]. In the case of internal catalysis expression (5.9) takes the form:

$$\tau_{ex} = \tau'_{op} + \frac{1+10^{\mathrm{pK_{nu}} - \mathrm{pK_{acc}}}}{R_F K_d{}'}, \tag{5.14}$$

where $R_F$ is the frequency constant standing in place of the product $k_{coll}$ · [acc] [286]. A transition of a nucleotide pair into the state capable of exchanging $^1$H due to internal catalysis is not identical to a flip-out. In order to stress this we denote the equilibrium constant for this transition as $K_d'$, and the time of the closed state as $\tau_{op}'$. The rest denotations are the same as in expression (5.9). The complex structure implies $\alpha = 1$. As we will show in what follows, a complex nuH*··$H_2O$··acc is a sort of an open state which differs from a flip-out in its thermodynamic properties.

A big value of $R_F$ is compensated by a tiny probability of a successful $^1$H transfer since pK of $N_3$ in cytosine is equal to 4.2 and pK of $N_1$ in adenine is 3.7. As a result the values of the numerator in formula (5.14) for GC- and AT-pair are $1.6{\cdot}10^5$ and $10^6$, respectively. The only exclusion is the so-called acid catalysis which takes place in GC-pairs at low pH. Protonation of $N_7$ in guanine leads to a decrease in pK of its imino group from 9.4 to 7.2 and a decrease in the exchange time by more than 150-fold bringing it into proximity with $\tau_{ex}$ typical for high concentrations of an external catalyst [355].

One more peculiarity of an internal catalysis is its sensitivity to end effects. In the investigations by Nonin et al. it was shown that in the absence of an external acceptor the terminal bases exchange $^1$H slower than the bases in the center of the duplex [317]. In the latter ones, in view of stacking interactions on both the sides, small angular displacements leading to the formation of sustained nuH*··$H_2O$··acc complex are more frequent. For the terminal pairs, evidently, more pronounced fluctuations leading to flip-outs are more peculiar. Consistent $^1$H transfer there is hindered since the aqueous bridge appears to be too long or is absent at all.

It is impossible to calculate the exact value of $K_d'$ on the basis of the internal catalysis kinetics. This kinetics enables us only to find the product $R_F{\cdot}K_d'$ while the exact relation between $K_d'$ and $K_d$ is not known. However $K_d'$ can be estimated from the time of $^1$H transfer in the complex nuH*··$H_2O$··acc. Its limiting stage is rotation of the water molecule which requires an opening of one H-bond: it is known to proceed for several picoseconds [356]. Therefore, it may be assumed that at temperatures 10–40 °C the value of $R_F$ falls in the range of $10^{11}$–$10^{12}$ s$^{-1}$. This enables us to calculate characteristic $K_d'$ from the exchange rate in the absence of an external catalyst, in disregard for $\tau_{op}'$, since in this case $\tau_{op}' \ll \tau_{ex}$.

Typical values of $\tau_{ex}$ for [acc] = 0 were first obtained for some bases in oligonucleotides d(AATTGCAATT)$_2$ [284] and d(CGCGATCGCG)$_2$ [283]. The bases investigated are underlined in the formulae. If we reckon the bases from the end of the duplex, in the first oligomer they are denoted as $AT_3$, $AT_4$ and $GC_5$, while in the second one – as GC3, GC4 and AT5.

Table 5.4 lists $K_d'$ of these pairs at temperatures 15 and 20 °C for two values of $R_F$ ($10^{11}$ and $2{\cdot}10^{12}$ s$^{-1}$). They are compared with the relevant $K_d$. The second value of $R_F$ corresponds to the reciprocal of the shortest lifetime of a hydrogen bond in water (0.5 ps, see [352, 353]). The values of $\tau_{ex}$ for [acc] = 0 of oligomer d(CGCGATCGCG)$_2$ are obtained by digitization of the graph in figure 6 from paper [283]. The values of inaccuracy are not given since the main aim was an approximate estimate. Since the values of $K_d$ in that work are given only for 15 °C, we took the values of $K_d$ for 20 °C from paper [318] for the sake of comparison. There oligomer d(CGCGAATTCGCG)$_2$ possessing a similar strucure was studied. Nucleotide pairs $GC_3$, $GC_4$ and $AT_5$ used for comparison are underlined in its formula.

It is evident from table 5.4 that $K_d'$ differ from the relevant $K_d$ by an order of magnitude on the average. The vast difference for the pair $AT_3$ of oligomer d(AATTGCAATT)$_2$ can easily be explained by end-fraying which cause instability of the nuH*··$H_2O$··acc complex.

**Table 5.4.** Comparison of $K_d$ and $K_d'$ for different nucleotide pairs according to data of some investigations

| Basepair | T, °C | $K_d \cdot 10^6$ | $K_d' \cdot 10^6$, $R_F = 10^{11}$ | $K_d' \cdot 10^6$, $R_F = 2 \cdot 10^{12}$ | Ref. |
|---|---|---|---|---|---|
| $AT_3$ | 15 | 800 | 33 | 1.6 | [284] |
| $AT_4$ | 15 | 100 | 12 | 0.62 | [284] |
| $GC_5$ | 15 | 0.7 | $< 0.4$ | $< 0.02$ | [284] |
| $GC_3$ | 15 | 0.39 | 0.42 | 0.021 | [283] |
| $GC_3$ | 20 | – | 0.81 | 0.04 | [283] |
| $GC_3$ | 20 | 1.00 | – | – | [318] |
| $GC_4$ | 15 | 0.22 | 0.11 | 0.05 | [283] |
| $GC_4$ | 20 | – | 0.24 | 0.012 | [283] |
| $GC_4$ | 20 | 0.3 | – | – | [318] |
| $AT_5$ | 15 | 3.2 | 14 | 0.7 | [283] |
| $AT_5$ | 20 | – | 24 | 1.2 | [283] |
| $AT_5$ | 20 | 5.8 | – | – | [318] |

Formation of inner-sphere hydrogen-bonded complexes in the case of small angular displacements of the bases was confirmed by calculations performed by Giudice et al. [162]. Figure 5.4 illustrates nuH*··$H_2O$··acc complex formed by AT-pair. For such a complex to form the longitudinal axis of a base should deviate from the line parallel to complementary H-bonds in a closed duplex by a small angle – not larger than ±45° [162].

**Fig. 5.4.** Complex nuH*·· $H_2O$ ··acc formed by anguler displacement of AT-pair into a large groove [162]. Dashed lines indicate H-bonds which join H2O molecule with imino proton and with atom $N_1$ of adenine base.

Since a moderate angular displacement of the bases enables their interaction with water, a question expectedly arises of whether a catalyst molecule can get to the imino group in such a case? If the answer is "yes" – then a catalytic exchange of $^1$H from such "half open" states can lead to a certain misinterpretation of the experimental data. Indeed, since $K_d'$ is comparable with $K_d$, expression (5.9) takes on the form:

$$T_{ex} = \frac{1}{k_{ex} + k_{ex}'} = \left[ \left( \tau_{op} + \frac{1 + 10^{pK_{nu} - pK_{acc}}}{k_{coll} [acc] \alpha K_d} \right)^{-1} + \left( \tau_{op}' + \frac{1 + 10^{pK_{nu} - pK_{acc}}}{k_{coll} [acc] \alpha' K_d'} \right)^{-1} \right]^{-1}, \quad (5.15)$$

where $k_{ex}$ and $k_{ex}'$ are the rates of exchange from flip-outs and from hydrogen-bonded water-bridged complexes, $T_{ex}$ is the resultant exchange time. The value of $T_{ex}(0)$ that is $T_{ex}$ for $[acc]^{-1} = 0$ is equal to the apparent $\tau_{op}$ of a base. As distinct from atom N of a complementary base, a catalyst molecule has a restricted access to the "half open" imino group. For this reason the second term involves the specific accessibility parameter $\alpha'$ for the "half open" state.

For $\tau_{op}' \approx \tau_{op}$, a difference between $T_{ex}(0)$ and $\tau_{op}$ does not exceed 40 % even if $\alpha K_d = \alpha' K_d'$. A reduced ratio of $\tau_{op}'/\tau_{op}$ leads to a sharp decrease in the resultant exchange time. However, at $\tau_{op}'/\tau_{op} \leq 0.1$ this time ceases to change. Therefore a difference in 5.7 kJ/mol between the activation barrier for a flip-out and that for a "half opening" is quite sufficient for the latter to make a dominant contribution into the process of $^1$H exchange.

Since a small angular displacement of a base is an initial phase of a flip-out the inequality $\tau_{op}' \leq \tau_{op}$ is obviously fulfilled for any nucleotide pair irrespective of its nature and neighborhood. First, in the case of a small angular displacement, unlike in the case of a flip-out, a base partially retains stacking interactions with its neighbors. Therefore the values of $\Delta H^{\dagger 0}$ should differ for a flip-out and for a "half opening". Second, as it is evident from tables 5.1 and 5.2, in more than half of the flip-out cases $\Delta S^{\dagger 0} < 0$ which is a consequence of a complicated molecular dynamic trajectory of escape. The trajectory of a base motion in the case of a small displacement, on the contrary, should be very short. As a result, the activation barrier decreases since $\Delta S^{\dagger 0} > 0$ for typical simple trajectories.

For $\tau_{op}'/\tau_{op} \leq 0.1$, a deviation of $T_{ex}(0)$ from $\tau_{op}$ depends mainly on $\tau_{cl}$ – the lifetime of an open state in the case of a flip-out. For $\tau_{cl} \leq 2$ ns and $\alpha K_d$ exceeding $\alpha' K_d'$ by not less than an order of magnitude, $T_{ex}(0)/\tau_{op} \geq 0.8$. It is quite an admissible inaccuracy. However, according to the data of many experiments, $\tau_{cl}$ of some bases can be as high as 100 ns and more [28, 282, 299, 305, 318, 319, 325, 330, 331, 357]. For the inequality $T_{ex}(0)/\tau_{op} \geq 0.8$ to hold for such lifetimes of the open state $\alpha K_d$ should differ from $\alpha' K_d'$ by 1.5–2 orders of magnitude.

According to table 5.4, the mean ratio $K_d'/K_d$ for $R_F = 10^{12}$ s$^{-1}$ is equal to 0.1. This suggests that the condition $\alpha'/\alpha \leq 0.1$ is sufficient for the contribution made by the second term of expression (5.15) into $T_{ex}$ to be within reasonable errors. Unfortunately there are no experimental methods which would enable one to measure $\alpha'$. Besides, we do not know any theoretical works where this value would be estimated *in silico*.

The only possibility to assess $\alpha'$ is approximation of some experimental data by equation (5.15). In the work by Warmlander et al. it was shown that at high concentrations of ammonia the dependence of $T_{ex}$ on $[acc]^{-1}$ deviates from the linear form which it should have according to equation (5.9) [299]. The authors explained this phenomenon by the availability of two opening regimes which have different lifetimes of open and closed states. The mode in which the bases open at a low concentration of $NH_3$ was called *slow*, while the mode typical for high concentrations was called *fast*. Later on it was proved that high ammonia concentrations can affect the DNA structure additionally influencing the flip-out kinetics [305].

At the same time the results obtained by Warmlander et al. may be stiffly accurately described by equation (5.15). The object of investigation in [299] were oligomers containing A-tracts – elements of the structure possessing the highest $\tau_{cl}$ of the bases, up to several hundreds of nanoseconds [299, 305, 318, 319, 325, 330, 331]. Considerable $\tau_{cl}$ lead not only to a decrease in the value of $T_{ex}(0)/\tau_{op}$ but also to a growing increase in the slope of the function $T_{ex} = g([acc]^{-1})$ for $[acc]^{-1} > 0$.

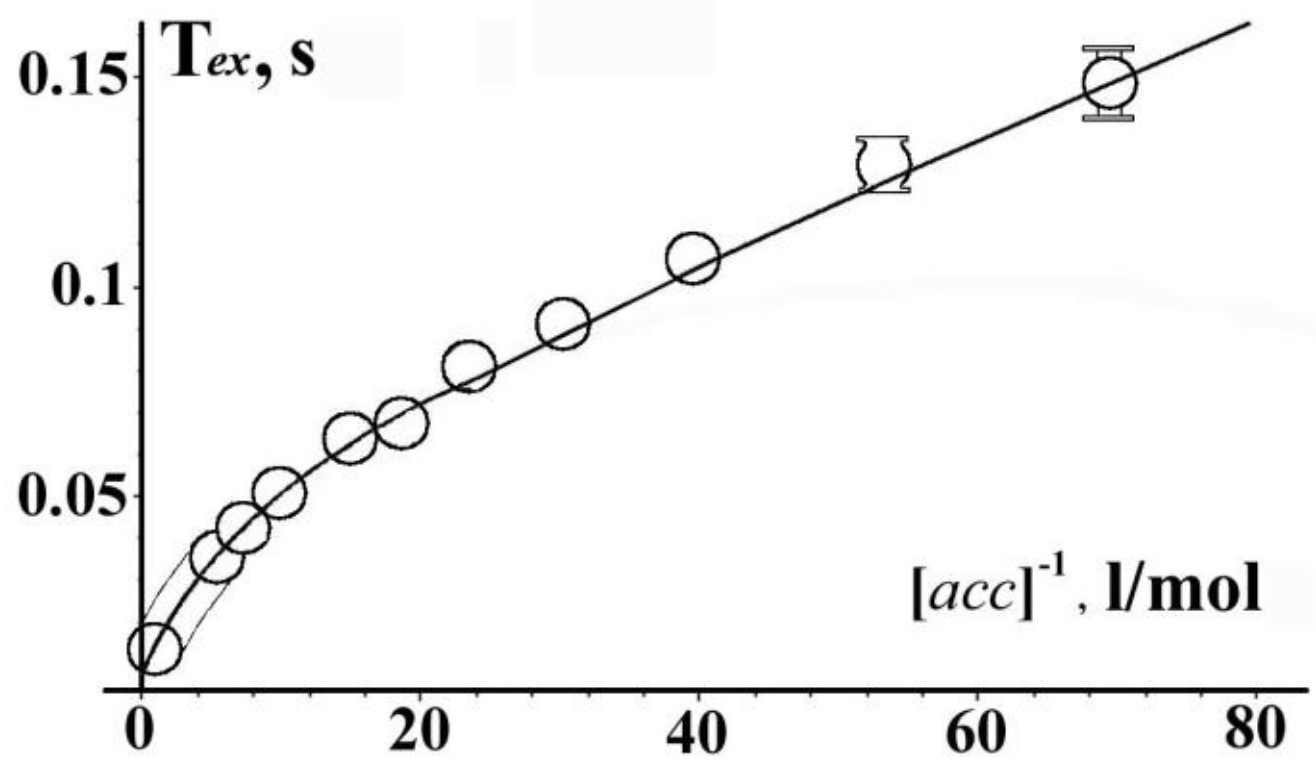

**Fig. 5.5.** Approximation of the experimental data by Warmlander et al. for $AT_6$ pair of oligomer $d(CGCGAATTCGCG)_2$ [299] by equation (5.15). The temperature is 15 °C, the catalyst is $NH_3$. The approximation parameters are: $\tau_{op}$ = 80 ms, $K_d = 1.87{\cdot}10^{-6}$, $K_d' = 3.16{\cdot}10^{-7}$. The points of the experimental graph are shown as empty circles which merge into a "passage" for small $[acc]^{-1}$. This is done for the sake of clarity.

We used equation (5.15) to describe the data for nucleotide pair $AT_6$ of oligomer $d(CGCGAATTCGCG)_2$ obtained by Warmlander et al. at temperatures 10, 15, 20 and 25 °C for two exchange catalysts – $NH_3$ and trimethylamine [299]. We also approximated the data on the pairs $GC_3$ and $GC_4$ of the same oligomer [299] (only for $NH_3$ as a catalyst). The example of such an approximation is shown in figure 5.5.

In none of the approximations performed we observed the ratio $\alpha' K_d'/(\alpha K_d)$ to be less than 0.1. This means that $\alpha'$ and $\alpha$ have a similar order of magnitude, as is evident from table 5.4. At the same time the ratio $K_d'/K_d \approx 0.1$ is typical for single-base opening, see above. Hence in the basepairs that are distant from the duplex ends, any opening of complementary H-bonds makes the imino groups accessible for $^1$H exchange. In this respect they are not different from the terminal pairs, see 5.3.4.

Based on the research findings presented in Chapter 5 we can make two main conclusions.

*First conclusion*. Many experiments on imino proton exchange are probably characterized by a considerable uncertainty in determining $\tau_{op}$ in the conditions of high catalyst concentration. The ratio $T_{ex}(0)/\tau_{op}$, depending on $\tau_{cl}$, falls in the range of 0.3–0.8. Moreover, the curved part of the graphs is located in the region of small values of $[acc]^{-1}$. As a result, the function $T_{ex} = g([acc]^{-1})$ looks like a straight line, though even at $\tau_{cl} > 50$ ns the apparent $T_{ex}(0)$ is far less than the true $\tau_{op}$.

From the thermodynamics viewpoint, the distorted results are probably concerned only with the enthropy contribution $T{\cdot}\Delta S^{\dagger 0}$ into the activation barrier $\Delta G^{\dagger 0}$. Actually the value of $\Delta S^{\dagger 0}$ of a flip-out should be far less for most of the bases. As an example we refer to the temperature dependencies of $\tau_{op}$ at "fast" and "slow" modes from the paper by Warmlander et al. [299]. For the equal values of $\Delta H^{\dagger 0}$, the activation enthropy for the "slow" regime was less than that for the "fast" one by 7 kJ/mol. Just this gap in entropy provided a considerable difference of $\tau_{op}$ in the experiment. This may suggest that the differences in the thermodynamic properties of the bubbles and flip-outs were underestimated in Section 5.3 (i.e., the vast majority of bases should have $\Delta S^{\dagger 0} \leq 0$).

*Second conclusion*. We have confirmed indirectly that an opening of complementary H-bonds between the DNA strands "automatically" makes the imino groups accessible to the molecules of the solution. Therefore any open DNA states can exchange $^1$H irrespective of whether stacking interactions are retained there or not. This fact is very important since it confirms the validity of the definition of the "open state" given in the Introduction. *An open state is any alteration of the DNA conformation which emerges as a result of a complete or* *partial opening of complementary H-bonds in one or several adjacent nucleotide pairs and makes the protons involved in these bonds accessible to the molecules of solution.*

Hence we have not managed to find out any peculiarities of the DNA duplex which would explain a considerable discrepancy between the probability of the open states obtained from PBD model investigations [274–276, 350] and that obtained in the experiment. Satisfactory explanations are found only for the particular case of the discrepancy between the results of FCS of molecular beacons and the data of other experimens, see Sections 5.3.1–5.3.3.

This suggests a necessity to improve the mechanical DNA models available so that they could better reproduce the experimental data. At the same time such an improvement should not result in the loss of their ability to take account of the transfer and nonlinear localization of the stretched oscillations' energy in DNA: it is importance of this process that a key feature that was demonstrated in this review. In the next Chapter and in the Conclusions we will suggest some ways of optimizing the mechanical approaches which should provide a better agreement between models and experiments.

## 6. GENERALIZATION OF DATA ON LOW-TEMPERATURE DUPLEX DYNAMICS

A wide variety of theoretical and experimental approaches used in DNA investigations provided a lot of data on the DNA dynamics and the factors on which it depends. The scope of this information is so large nowadays and the DNA behavior seems to be so complicated that any generalizations are difficult to make. Therefore the main aim of this work is to make a review of some theoretical and experimental works on DNA dynamics.

Comparison and analysis of some experimental results presented here had auxiliary purposes. For example, using the tables in Section 5.2 we demonstrated some regularities of flipping out of the bases in various contexts of the primary structure. The analysis made in Section 5.3 clearly demonstrated consistency of the experimental results obtained by basically different methods.

Nevertheless a large portion of the material refers to rather a narrow field – investigations of the DNA open states dynamics on the pre-melting interval. The scope of the data which are directly relevant to this field is relatively moderate. It makes their generalization quite a realistic objective. This Chapter just deals with this generalization. It is mainly aimed at systematizing the data which would make this work most useful for the researchers.

### 6.1. Number of basepairs directly involved in the bubble nucleation

When analyzing theoretical and experimantal data we believe that several basepairs open at a time during a bubble nucleation. There are some facts which confirm that the number of these pairs – let us denote it by $N$ – is much more than 1. Obviously, the condition $N > 1$ for a bubble partially results from the rigidity of the sugar-phosphate backbone. Besides, in this review we present some indirect experimental evidence in favour of this fact.

In Section 3.3 we describe the properties of the force barrier which prevents the onset of the mechanical unwinding of DNA, see [178, 258]. The existence of the barrier is explained by retaining of stacking interactions in the *fork* – boundary region between the single-stranded region and the native DNA duplex, see figure 3.4 [178]. Displacement of the fork region leads to opening of stacking interactions at its frontal end. As a result the system's enthropy increases. This partially compensates the free energy cost for the duplex unzipping but only in the case when the fork is already formed. At the very beginning of unzipping a fork is not formed yet and opening of the first few basepairs requires large energy cost.

A certain contribution into the height of the barrier is likely made by the necessity of a simultaneous opening of complementary H-bonds in the first several basepairs at the onset of force unzipping. Once again, the stiffness of sugar-phosphate backbone causes force transfer both to the terminal basepair and to its neighbor (neighbors). What's interesting is that the calculated value of the fork length (which is approximately 4 nucleotide pairs [178]) is in good agreement with our assessments of $N$.

The least length of a bubble sufficient for a fluorophore to be isolated from a quencher in the molecular beacons [213] is 4–6 pairs, see Section 5.3.4. A wide scatter of the bubble lifetimes suggests that the growth of an open region requires far less the free energy costs than the opening *per se*. Some contribution into this difference can also be made by the enthropy gain due to opening of the stacking in a bubble.

The condition $N > 1$ also provides a good explanation of the relative stability of GC-rich terminal domains of the molecular beacon $M_{18}$ [213]. This problem is considered in detail in Section 5.3.1. Obviously the range of enthalpies of a bubble nucleation depends greatly on $N$. This effect can also account for considerable differences between $K_{d,bub}$ observed in experiments performed by Choi et al. [29, 30].

One more indirect evidence that $N > 1$ is denaturation behavior of short DNA investigated by trapping of intermediate states [54, 55, 261] – see Sections 4.2 and 5.3.2. If a short oligonucleotide consists of an AT-rich domain restricted by two terminal GC-rich domains, then its critical length exceeds 20 basepairs. If it contains only two domains – AT-rich domain and GC-rich one – its critical length tends to zero.

Again, this phenomenon can easily be explained in view of rigidity of the sugar-phosphate backbone. Simultaneous radial opening of several adjacent basepairs causes positive supercoiling in the neighboring regions [13]. If a bubble is nucleated in the center of a short DNA, the supercoiling strain can be relieved at both its ends. This leads to destabilization and greatly facilitates full dissociation. When a bubble nucleates near an end of the duplex, the strain relaxes only at this end. As a result, the bubble's nucleation leads to an open terminal region that may readily close. The dependence of the critical length on the DNA primary stucture clearly demonstrates a close relation between the bubble dynamics and the thermally induced fluctuations of the torsional stress of the sugar-phosphate backbone. This fact additionally confirms simultaneous participance of several nucleotide pairs in the bubble nucleation.

It appears that the condition $N > 1$ is a characteristic whose consideration is crucial for the description of the experimentally observed duplex dynamics. Nevertheless, the only simple up-to-date model where it is taken into account is the Ising-like model by Kantorovitz et al. [272]. In our opinion it would also be appropriate to introduce phenomenologically the condition $N > 1$ into the mechanical models. In what follows we will point out some other features of the duplex dynamics on the basis of which the models of this group can be modified. Of course, even being essentially improved the models of this level will not allow reproducing adequately all the peculiarities of the DNA behavior. Nevertheless simple models enable one to see the properties of a heterogeneous duplex from different viewpoints. It is the whole of simple approaches that provide the picture which largely corresponds to the experiments. In this respect they are similar to photos of an unfamiliar item from different perspectives which enable one to judge about its shape.

### 6.2. Comparison of the approaches in the context of how they describe bubbles at moderate temperatures

First of all, let us choose the main features of the denaturation bubble dynamics in a heterogeneous DNA which were revealed *in vitro*. First, this is a considerable difference in the opening probabilities in different fragments of the duplex. Second, this is a pronounced effect of the local replacement of nucleotide pairs on the form of the instability profile of a heterogeneous DNA. This influence can extend to tens of basepairs from the place of the replacement. These properties were revealed by the experimental techniques described in Section 4.2; for details see [29, 30]. The third characteristic feature is the participance of several neighboring basepairs in the bubble nucleation. The proofs of this were given in the previous section.

Now bearing in mind these features and some other considerations let us point out the main criteria for estimating the theoretical approaches from the viewpoint of the denaturation bubbles description.

1. The investigation object of a model should be a relaxed duplex, i.e. its total supercoiling must be equal to zero. The experiments in which the main characteristics of denaturation bubbles were revealed were carried out on relaxed molecules. Therefore the failure of the model to meet this criterion renders assessment on all the rest ones meaningless.

2. A model should take into account the possibility of nucleation of a denaturation bubble restricted by closed fragments, i.e. which is not a fork at a duplex end. Besides, if an approach is mechanical one it should enable such stabilization of a bubble, provide for its lifetime to be much more than the period of a one stretched oscillation in the duplex.

3. A "distant effect" should be reproduced, which is the effect of local replacements of a basepair on the probability of a bubble formation in the regions distant from the point of the replacement. In this case the number of nucleotide pairs separating these regions should be greater by several fold than the averaged bubble length.

4. A ratio between the maximum and the minimum value of the ordinate on the instability profiles obtained in the model should be not less than 50, see Section 4.2. In the experiments it often exceeded 100 [29, 30].

5. Finally, an important role belongs to particular tasks for which the model was originally developed. Which simplifications were admissible and which were not in creating the model depends on it. The combination of the simplifications used, in turn, determines the adequacy of the bubbles description.

*Ising-like model by C. Benham.* The only approach which was specially developed to study local openings of a heterogeneous duplex is the model by C. Benham [33–35, 234–236, 243, 244]. The degree to which it meets the second, third and fourth criteria is much higher than in many models of this level. However this approach is not used to study a relaxed DNA, i.e. it does not meet the first criterion. The reasons why we do not consider the duplex behavior in the presence of an external torsional stress are given in the Introduction and in Section 3.2.

*Mechanical models*. For the mechanical models, the main selection criteria are the first and the second ones. In the angular models the duplex is relaxed, but the open states of adjacent nucleotide pairs are very different in their dynamics from the bubbles. Such states are rather unstable (intrinsic property of the angular models). For most radial models, the first condition is also met: even for radial-torsion approaches, a case with zero supercoiling may well be considered. Besides, the open states can be stabilized due to nonlinearity of the on-site potential that describes complementary H-bonds. Hence there are quite a lot of models which meet the first two criteria, see Section 2.4.

Nevertheless, the investigations of the bubble dynamics in the PBD model demonstrated that the necessary condition for the simulated and experimental data to agree is the finiteness of stacking interactions in the Hamiltonian ($\rho \neq 0$) [30]. Hence the PBD model and the radial-torsion approach by Barbi et al. with an anharmonic stacking potential [179] best meet the fourth criterion. In these cases the failure to meet the fifth criterion can be considered to be insignificant.

*Nearest-neighbor models.* An important problem of the nearest-neighbor models is the drawback in terms of meeting the third criterion. It is impossible to study the transfer and localization of the energy in these approaches, unlike the case of mechanical models. Nor is it possible to investigate the external torsional stress which provides distant effect in the Benham model.

The main application area of the nearest neighbor-approaches was phenomenological description of differential melting profiles of heterogeneous DNA or their calculation based on the nucleotide sequence, see Section 1.3. To that end, use was usually made of such techniques as Fixman-Friere algorithm [358], recursive algorithm by D. Poland [359] and some its modifications, see [53, 270]. Calculations were reduced to finding the probabilities of the open state for each individual base pair with regard to the state of the neighboring pairs. The only exclusion is the work by Gotoh and Tagashira where the stability of doublets was calculated [360].

Hence we can believe that the nearest-neighbor models deal not with denaturation bubbles but with the probability of individual openings. Therefore the instability profiles obtained within these models did not properly reflect the actual statistics of the bubbles, especially at low temperatures. The reasons of this inadequacy are most conveniently demonstrated via the analysis of the cooperativity parameter $\bar{\sigma}'$ whose esscence is described in Section 1.3. In the calculations of the melting profiles for heterogeneous DNA this parameter usually had a constant value equal to 0.000045 [345] which corresponds to $\bar{G}^s \approx 25$ kJ/mol at 300 K, see formula (1.12). It is evident from expression (1.13) that for such $\bar{\sigma}$ the stability of the terminal base pairs is on the average two orders of magnitude lower than that of the pairs in the center of the duplex. Therefore any heterogeneous DNA in the investigations of the nearest-neighbor models could unwind only from the ends.

Obviously, these results do not satisfy the second criterion. Since the base pairs in the center of the duplex can open only individually in this case, the fourth criterion is not satisfied either. For checking purposes we assessed the range of the probabilities of an open state in the nearest neighbor model which was investigated in the work by Wartell и Benight [345]. We have used Poland's algorithm [359]. Calculation was performed for some heterogeneous duplexes including DNA which were used in the bubble dynamics experiments [29, 30, 54, 55, 261]. The ratio between the maximum and the minimum ordinate of the instability profiles did not exceed 9, which approximately corresponds to the difference of thermodynamic parameters of individual pairs.

In most algorithms the formation probability for a bubble consisting of $N$ base pairs is calculated by averaging [270, 271, 361]. However in 2010 an approach was developed which enabled M. Kantorovitz et al. to calculate the probability of simultaneous opening of $N$ nucleotide pairs in the nearest neighbor model [272]. Actually, account was taken of the condition $N > 1$ which was lacking even in the mechanical approaches. The model results satisfied not only the second but also the fourth criterion. For example, in the 1.2 kb (1200 base pairs) CFTR fragment of a human genome, the calculation for $N = 5$ yielded the value of $K_{d,bub}$ in the range from $10^{-6}$–$10^{-5}$ to $10^{-3}$ at 37 °C [272]. A difference in the ordinate of the instability profiles is 2–2.5 orders of magnitude. It is quite consistent with the experimental data [29, 30]. The lower limit of $K_{d,bub}$ obtained in the calculations correlate very well with the average experimental value equal to $1.5{\cdot}10^{-5}$ at 37 °C [266]. Moreover, a special parameter effectively interrelating the open state probabilities of neighboring pairs enabled the distant effects to be reproduced.

Hence, a conformity with all the assessment criteria is possible not only for the mechanical approaches but also for the nearest-neighbor models. These groups of the models complement each other well: a comparison of the results of their investigation enables one to see the peculiarities of the DNA dynamics from different perspectives. Heterogeneity of $K_{d,bub}$ and distant effects can be reproduced both by modeling the transfer and nonlinear localization of the energy in the mechanical approaches and by introducing the condition $N > 1$ in the nearest-neighbor approach. However only the mechanical models allow to investigate the physical mechanisms underlying the DNA dynamics. Therefore, in our opinion modernization of the approaches of this type is more strategic.

One possibility of modernization is to introduce phenomenologically a local cooperativity coefficient into the PBD model. This coefficient must depend on $N$ and will produce a free energy barrier which must hinder opening of less than $N$ pairs at a time. The barrier should also hinder a fast closing of a bubble additionally improving the model's agreement with the experiments, see [213]. The work at such a modernization of the PBD model is planned for the near future. In the new model the average $K_{d,bub}$ should be far less than that in the classical PBD model [154] or in its variant with a heterogeneous stacking [274]. Therefore the simulations' results will be much closer to the experiment. Nevertheless it should be noted that by no means all sorts of experimental data are suitable for the comparison with the models when $T < T_m$. This problem is discussed in the next section.

### 6.3. Problem of standard thermodynamical parameters

Standard thermodynamical parameters of the nucleotide pairs opening were determined in calorimetric measurements long ago [362, 363, 354] (see also references in [53]). According to those data, the dissociation enthalpy of any oligonucleotide observed in vitro can easily be obtained by simple summation of the opening enthalpies of individual nucleotide pairs of this oligonucleotide [363]. Therefore, ln[$K_{d,bub}$] for $N$ basepairs should be approximately equal to the sum of logarithms $K_d$ for the individual basepairs opening minus the logarithm of the correction factor which takes account of the entropy growth and depends on $N$.

A similar method was used to calculate $K_{d,bub}$ in paper by Wartell and Benight [129]. Such calculations yield a vanishingly small probability of spontaneous nucleation of a bubble at moderate temperatures in virtually any nucleotide sequence. This becomes obvious if we compare the condition $N \geq 4$ with a typical $K_d$ of a flip-out (for which $N = 1$) which is approximately equal to $10^{-5}$ at $T = 30$ °C.

At first sight, the conclusions of calorimetric measurements are contradict with the data of FCS [213], S1 nuclease cleavage assays [29, 30] and, partly, with the hypothesis of the active participation of DNA in its specific interaction with proteins. However these contradictions can be easily explained if we take into consideration differences of the temperatures at which different experiments were carried out.

First, openings of H-bonds and stacking in DNA occur cooperatively at $T \approx T_m$. If a bubble is formed at $T < T_m$, stacking in individual chains may be partly maintained. Therefore a standard and activation enthalpy of a bubble nucleation can be considerably less than the values obtained for $T_m$. This problem is discussed in Section 5.3.

Second, Raman spectroscopy experiments have demonstrated that as $T$ increases to $T_m$ the stacking enthalpy enhances 20-fold [326]. This is additional evidence that the thermodynamic properties of the same open states are different at different temperatures.

Third, as we have repeatedly mentioned, $K_{d,bub}$ in some DNA fragments can be enhanced due to nonlinear localization of the stretched oscillations energy. Obviously, the energy is always localized in the "weakest" regions where the activation barrier of the bubble formation is the lowest. As the temperature grows the activation energy becomes sufficient for the activation barriers to be overcame in more regions. Therefore heterogeneity of $K_{d,bub}$ caused by localization of the energy should be most pronounced just at moderate temperatures.

This assumption has an interesting consequence. In small DNA as the temperature gradually increases, bubbles should open in approximately one and the same fragment consisting of $N$ nucleotide pairs. Therefore in oligomers such as those studied by the hairpin quenching techniques [54, 55, 261], a photometry signal at low $T$ is probably caused by the dynamics of the only bubble. Even so, the low-temperature part of a protometric profile and the similar fragment of the total fluorescence profile obtained by FCS of beacons [213] reflect kinetics of one and the same process.

Fourth, at moderate temperatures an important role can belong not only to localization of the energy of stretched oscillations but also to localization of the energy of angular fluctuations. As is shown in Section 5.4, as a result of angular displacements of the bases, metastable open states of DNA are formed which can probably promote both formation of the bubbles and flipping out of the bases. This problem should be considered in greater detail.

### 6.4. Localization of the energy of basepairs angular fluctuations

Apart from stretched oscillations along complementary H-bonds, the dyplex dynamics is significally affected by “angular” fluctuations of the bases. Their direction corresponds to the third degree of freedom in figure 5.1. Displacements of this type, like the radial ones, probably can transfer energy in DNA. In this case this transfer is studied with the use of angular models which were considered in detail in Section 2.1.

As already mentioned in Section 2.2, angular oscillations of the bases may demonstrate pronounced nonlinearity. Let us consider this problem in greater detail relying on the simulation results and experimental data given in Section 5.4. There a hydrogen-bonded complex $nuH^{*}\cdot\cdot H_2O\cdot\cdot acc$ is described, see figure 5.4. This complex is highly likely formed when the bases are displaced by 45° and more [162]. Its existence is confirmed by kinetic features of the internal catalysis of $^{1}H$ exchange, see Section 5.4.

It is evident from table 5.4 that kinetics of the internal catalysis in adjacent base pairs is mutually independent just like exchange in the presence of an external proton acceptor. Therefore, small angular displacements of the bases, as well as their flip-outs occur individually, non-cooperatively. At first sight this fact is inconsistent with simulation results since the usual length of angular solitons in the models is several base pairs – see for example [171, 172]. However in this case a discrepancy between the theory and the experiments merely means that the maximum amplitude of a soliton does not exceed 45°.

A basepair at the “top” of a soliton having the greatest angular displacement risks to bind to a water molecule to form a hydrogen-bonded $nuH^{*}\cdot\cdot H_2O\cdot\cdot acc$ complex. The probability of this event depends on the energy of H-bonds and stacking. As a result a solitary wave can arise in one place but promote formation of the $nuH^{*}\cdot\cdot H_2O\cdot\cdot acc$ complex in another place. Transition into a metastable state corresponding to this complex takes most of the soliton energy. The data of table 5.4 suggest that $\Delta H^0$ of the transition is rather high: for $GC_3$, $GC_4$ and $AT_5$ of $d(CGCGATCGCG)_2$ oligomer it is 90, 109 and 75 kJ/mol, respectively. These values are close to $\Delta H^0$ of a flip-out which for GC-pairs is within the range of 45–109 kJ/mol, and for AT-pairs (outside A-tracts) – in the range of 60–90 kJ/mol. The limit $\Delta H^0$ values of a flip-out are estimated relying upon the material from the works cited in tables 5.1 and 5.2.

On the one hand, large standard enthalpy prevents formation of hydrogen-bonded water-bridged complex. On the other hand, the bases included into this complex partly maintain stacking interactions with their neighbors. This facilitates the energy transfer from a next soliton to the same basepair: it may lead to its flipping out or other conformational changes. To some extent, this model reminds the mechanism of localization in the bent DNA region suggested by Ting and Peyrard [349].

A water-bridged complex where $H_2O$ binds imino group with an acceptor atom ($N_1$ of adenine or $N_3$ of cytosine) is far from the only type of metastable states which is formed when a basepair is bound to $H_2O$ molecule. In the theoretical ivestigation by Giudice et al. a great manifold of bonding variants was shown, see figure 12 in paper [162].

Obviously water-bridged complexes are important intermediates of a flip-out. At the same time both standard and activation parameters of a flip-out depend strongly on steric limitations. For example, $\Delta H^{\dagger 0}$ of a thymine flip-out in 5'-AAATAAA-3' fragment is approximately equal to 88 kJ/mol while in a similar 5'- AAATAGA-3' fragment $\Delta H^{\dagger 0}$ does not exceed 55 kJ/mol [331]. Steric effects are tightly bound to the ending part of the trajectory of a base’s passing into one of DNA grooves; their nature is described, for example, in the work by Coman and Russu [331]. As distinct from a flip-out, formation of a water-bridged complex is not concerned with steric limitations, see Section 5.4. The main factor which prevents the formation of such states is the high enthalpy of H-bonds and stacking.

Since any local transient denaturation of a duplex requires, first of all, an opening of complementary H-bonds, the dynamics of the open states of all the types should be interrelated. Any direct experimental evidence of such an interrelation is still lacking yet. Nevertheless, there are facts which indirectly suggest a close relation between the bubbles' behavior and the dynamics of the other open states. Relying on these facts we propose some mechanisms of interrelation of the DNA openings which will be considered in what follows.

### 6.5. Hypothesis of DNA open states interaction and the problem of bubble code

In analyzing the properties of molecular beacons an interesting peculiarity of their total fluorescence profiles, $I(T)$ was noticed having introduced apparent (fluorescent) value of equilibrium constant $K_{fl} = I(T)/(1 - I(T))$. For moderate temperatures $K_{fl}$ is indicative of $K_{d,bub}$, although $K_{fl} \neq K_{d,bub}$, see below. In all the beacons the temperature dependence of $\ln[K_{fl}]$ on the pre-melting interval essentially deviates from the quasilinear form. According to expression (5.13), the characteristics $\ln[K_{d,bub}]$, which determines the value of $\ln[K_{fl}]$ must depend linearly on $T^{-1}$. Moderate deviations are probable due to uncertainty of measurements at low temperatures. At the same time, the increase of $d(\ln[K_{fl}])/dT$ observed as $T$ decreased was rather sharp. As was mentioned in Section 5.3 this phenomenon became one of the reasons why the lower limit of the temperature interval on which we estimated the thermodynamic parameters from table 5.3 was raised. To make a qualitative comparison of the results of fluorescence spectroscopy with other data the melting profiles of oligomers $L_{48}AS$, $L_{42}B_{18}$ and $L_{60}B_{36}$ were digitized [54, 55], see Section 4.1. In all these profiles the temperature derivative of the signal turned out to be constant throughout the entire pre-melting interval.

The differences between the $I(T)$ behavior of beacons and the temperature dependence of the photometry signal for $L_{48}AS$, $L_{42}B_{18}$ and $L_{60}B_{36}$ may be caused by three factors.

The first factor is concerned with probable peculiarities of an interrelation between $K_{d,bub}$ and $K_{fl}$ . On the one hand there is no any evidence that these quantities are proportional. On the other hand $K_{fl}$ is obviously related with $K_{d,bub}$ by the formula

$$K_{fl} = j(T) \cdot K_{d,bub},$$

where $j(T)$ is a certain temperature function which takes account of the features of the molecular system fluorescence *per se*. It is unlikely that this function has a complicated form and influences the value of the registered signal significantly. General thermodynamical considerations suggest that if $j(T)$ is not constant it should be linear or exponential. Hence the dependencies $\ln[K_{fl}](T)$ and $\ln[K_{d,bub}](T)$ will be similar, at least qualitatively. A close relationship between $I(T)$ of the beacons and their photometric profiles is also suggested by Altan-Bonnet et al. themselves [213]. Therefore in what follows we will not speak about change in the signal but about change in $\dfrac{d\left(\ln\left[K_{d,bub}\right]\right)}{dT}$ or bubble opening enthaply. Moreover, $j(T)$ must not be complicated by means of bubbles' multiplicity, since characteristics $K_{fl}$ and $K_{d,bub}$ are related to one and the same process in oligonucleotides – opening of the only bubble.

The second factor responsible for change in $\dfrac{d\left(\ln\left[K_{d,bub}\right]\right)}{dT}$ may be an influence of a fluorophore and a quencher on the opening dynamics of a duplex. According to the results of FCS, the activation enthalpy of the reverse reaction – "closing" of a bubble – is constant for any beacon in the range of 25–60 °C [213]. Therefore an increase in $\Delta H^0$ on the pre-melting interval which takes place when $T$ decreases can be explained only by an increase in $\Delta H^{\dagger 0}$ of the forward reaction – opening. This effect can really take place if the influence of thermal fluctuations of a fluorophore and a quencher on the duplex stability enhances as $T$ grows.

The third factor is the difference of the temperature intervals on which the behavior of $\ln[K_{d,bub}]$ was investigated for:

1) beacons,

2) oligomers without fluorescent tagging.

For the former, the lowest temperature was 23–28 °C; for the latter, it exceeded 40 °C. At the same time a considerable change in $\frac{d\left(\ln\left[K_{d,bub}\right]\right)}{dT}$ for the beacon $M_{18}$ was observed when $T$ decreased to 35 °C. For $A_{18}$ and $(AT)_9$, this temperature was approximately 31 °C. Hence at the temperatures when $\frac{d\left(\ln\left[K_{d,bub}\right]\right)}{dT}$ changes, the value of $K_{d,bub}$ may be under the photometry detection limit yet.

Experimental data available do not allow determining precisely how much this effect depends on the presence of a fluorophore and a quencher. However we can make an approximate estimate of this influence. We need simply determine apparent $\Delta H^0$ of the beacons opening at different temperatures and compare them with the similar values for oligomers without fluorescent tagging.

Figure 6.1. illustrates the dependence of $\ln[K_{fl}]$ on the reciprocal temperature for the beacon $M_{18}$. The experimental data were kindly presented by O. Krichevsky. We thought the value of $K_{d,bub}$ to be approximately equal to $K_{fl}$, therefore $\Delta H^0$ were called "apparent". The estimation of $\Delta H^0$ was made on the intervals of 28–36 and 54–62 °C: in the figure they are denoted by Roman numerals I and II. The values of $\frac{d\left(\ln\left[K_{d,bub}\right]\right)}{dT}$ on these intervals were calculated by the least square method. The relevant legs in figure 6.1 clearly demonstrate a very small value of the error of this method on both the intervals.

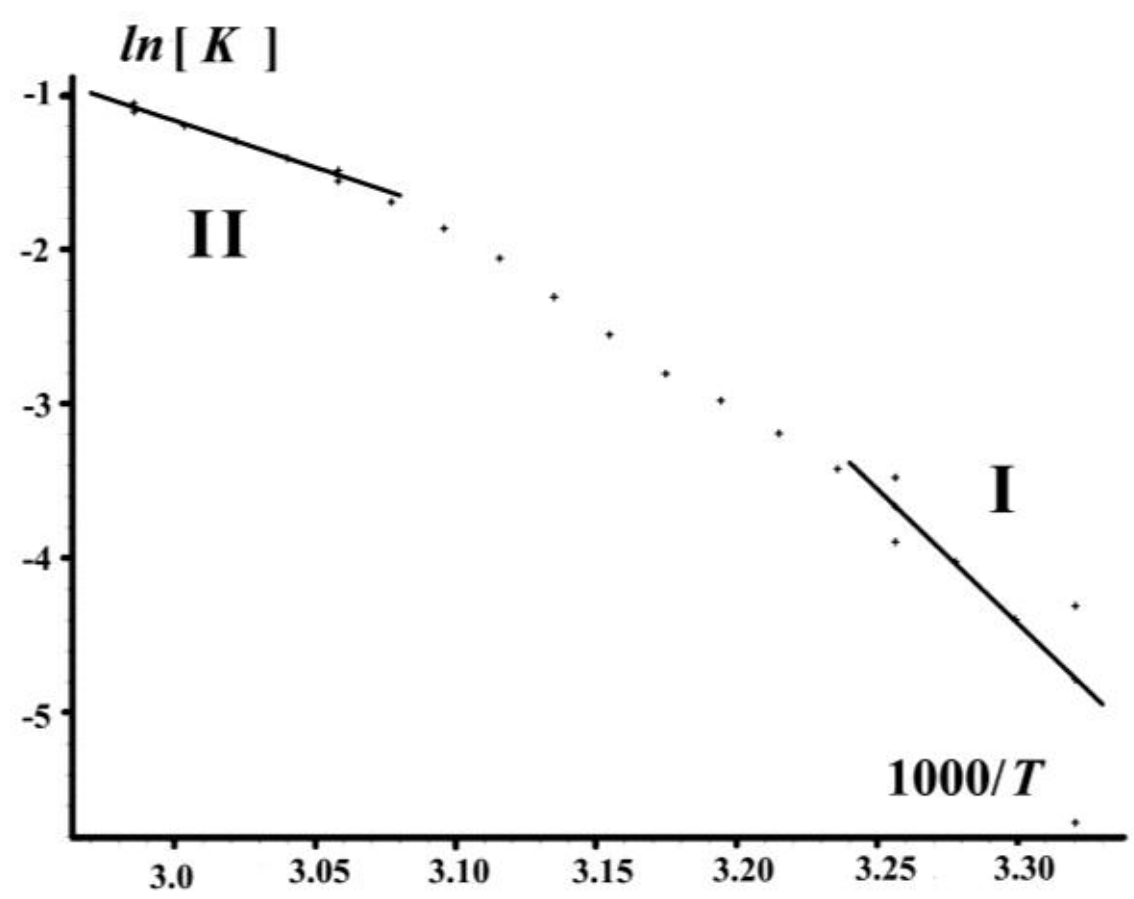

**Fig. 6.1.** Dependence of $\ln[K_{d,bub}]$ of the beacon $M_{18}$ on $T^{-1}$ which illustrates a change in $\Delta H^{\circ}$ in a narrow temperature range and a scale of the measurement incertainty with regard to taking the logarithm.

The normalized value of the measurement inaccuracy $I(T)$ in the experiment did not exceed 0.005. In the figure it is given on a logarithmic scale for the temperatures which are boundaries of the intervals I and II. This is intended to demonstrate that the considerable difference of $\frac{d\left(\ln\left[K_{d,bub}\right]\right)}{dT}$ is not a logarithmation artifact.

The values of $\Delta H^0$ on the intervals I and II are calculated with the use of the well-known expession of chemical thermodynamics:

$$\Delta H^0 = RT_1T_2\frac{d\left(\ln\left[K_{d,bub}\right]\right)}{dT},$$

where $R$ is the universal gas constant. Approximate values of $\Delta H^0$ for the intervals were equal to 135 and 50 kJ/mol, respectively. On the intermediate interval, 36–54 °C (for which the activation parameters were calculated, see table 5.3), $\Delta H^0$ was approximately 90 kJ/mol.

The standard opening enthalpies of the beacon $(AT)_9$ were calculated for the intervals 24–29 and 31–44 °C. For the first one, $\Delta H^0$ was not less than 130 kJ/mol, while for the second one, it did not exceed 90 kJ/mol. A similar difference in the enthalpies was also observed for the beacon $A_{18}$: their approximate values on the intervals 23–30 and 30–54 °C were 95 and 55 kJ/mol, respectively. Since $I(T)$ of the beacons $(AT)_9$ and $A_{18}$ are obtained by digitization, the reliability of their estimation is lower than that for the beacon $M_{18}$. Nevertheless, the average error of the estimation by the least square method on the "low-temperature" intervals for $(AT)_9$ and $A_{18}$ turned out to be as small as that on the interval I for the beacon $M_{18}$.

The apparent $\Delta H^0$ values for the opening of oligonucleotides $L_{42}B_{18}$ and $L_{60}B_{36}$ on the pre-melting interval did not exceed 130 kJ/mol. This value for $L_{48}AS$ turned out to be much higher – approximately 200 kJ/mol. Apparently, in most fragments of a native DNA the bubble enthalpies are in the limits of these values. This is indicated indirectly by the similarity between $\Delta H^0$ of the opening in $L_{42}B_{18}$ and $L_{60}B_{36}$ and the enthalpy of the beacon opening at reduced temperatures.

The most probable mechanism of why $\Delta H^0$ of the beacon opening reduces when $T$ grows is that the interaction of the open states is considerably intensified in the point of fluorescent tagging. Fluorophore and quencher are bound not to the sugar-phosphate backbone but to the thymine bases of the neighboring pairs in the beacon center, see fig. 4.11 and 5.3. Thermal fluctuations of the fluorophore and the quencher should reduce the activation barrier of a flip-out of these pairs and make longer the lifetime of an open state due to enhancement of $\Delta S^{\dagger 0}$.

At low temperatures the flip-out events in these AT-pairs are very seldom and occur independently of one another. However as $T$ grows, the efficiency of their interaction probably increases. As is shown above, at least $N$ adjacent basepairs open simultaneously in the course of bubble nucleation. A preliminary opening of H-bonds and stacking due to angular displacements in some of these pairs leads to a considerable decrease of $\Delta H^{\dagger 0}$ of a bubble opening. Since in this case a bubble nucleates in a less ordered fragment, this effect is partly compensated by a reduction of $\Delta S^{\dagger 0}$ which we do observe in the beacons.

This mechanism implies that a reduction of $\Delta H^{\dagger 0}$ in oligomers which do not contain a fluorophore and a quencher should take place at higher temperatures as compared to beacons. However as stated above, the bubble enthalpy in $L_{48}AS$, $L_{42}B_{18}$ and $L_{60}B_{36}$ was constant throughout the entire pre-melting interval. Therefore in these nucleotides the flip-out dynamics influences the bubble nucleation only slightly over the whole pre-melting interval.

Unfortunately, the experimental data available do not allow judging about interrelation between the opening processes of different types in a non-modified, native DNA. Besides, such an interaction is probably far from the only mechanism controlling the temperature dependence of $K_{d,bub}$. In all the oligomers considered which do not have a fluorescent tagging there are rather long fragments which consist of only AT-pairs. Nevertheless, as is shown above, the values of their apparent $\Delta H^0$ can differ by dozens of kJ/mol!

Extrapolation to low temperatures demonstrates that $K_{d,bub}$ of $L_{48}AS$ oligonucleotide does not exceed $0.5 \cdot 10^{-5}$ at 20 °C. This is quite consistent with the early estimates of the bubble opening probability in DNA ([312], cited by [282]), see Section 5.2. However for $L_{42}B_{18}$ and $L_{60}B_{36}$, extrapolation to 20 °C yields the values of $K_{d,bub}$ equal to $0.5 \cdot 10^{-3}$ and $10^{-3}$, respectively.

Considerable differences between the termodynamic properties of these oligomers and those of $L_{48}AS$ can be caused by two reasons.

The first factor is concerned with the middle position of AT-rich domains in $L_{42}B_{18}$ and $L_{60}B_{36}$. In Section 6.1 we describe the influence of the bubble opening on the torsional stress in neighboring duplex regions. A release of the stress of positive supercoiling at both the ends of oligomers $L_{42}B_{18}$ and $L_{60}B_{36}$ promotes the emergence of the total negative supercoiling of the duplex which retains for some time after the bubble closing. At this time DNA is destabilized and $\Delta H^0$ of the bubble is reduced. Considering that in $L_{48}AS$ molecule the torsional stress may release only at one end, it is far less susceptible to such a destabilization.

We suppose that the lifetime of the transient total negative supercoiling caused by the release of the torsional stress at the DNA ends can exceed the bubble lifetime *per se*. The conceivable mechanism of maintaining the supercoiling is concerned with the structure of an H-bond. According to some reports, an H-bond has not one but two energy minima [364]. Its scheme is shown in figure 6.2.

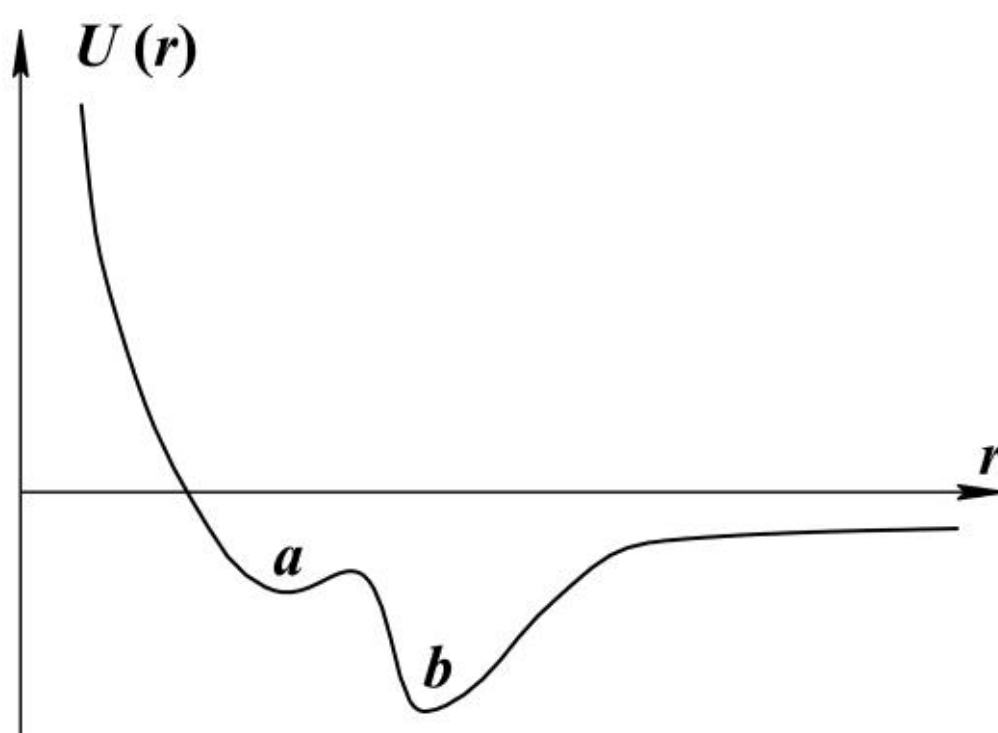

**Fig. 6.2.** Scheme of a bistable potential for H-bond. $U(r)$ which is the potential energy of the interatomic interaction is determined by $r$ – the distance between the atoms.

The first minimum denoted by the letter *a* is determined only by the distance between the atoms. The second minimum denoted by the letter *b* depends, in addition, on the angle between an H-bond and a covalent bond O–H. It is rather narrowly focused [364]. It is felt that an increase in the water density upon heating is just associated with the transition of H-bonds between its molecules into *a* minimum [ibid].

The negative supercoiling stress leads to a reduction of the distance between the points where the bases are attached to the sugar-phosphate backbone. In view of bistability of the H-bond potential, a relaxation can take place both in the case of its opening and in the case of a transition from minimum *b* to minimum *a*. Such a transition occuring in some H-bonds of neighboring nucleotide pairs leads to a descrease in $\Delta H^{\dagger 0}$ of the bubble opening. Since the torsional coiling fluctuations are intensified as DNA is heated this mechanism can probably even partially cause a nonlinear change of $\ln[K_{d,bub}]$ with a rise in $T$.

The second reason of why apparent $\Delta H^0$ of a bubble in $L_{48}AS$ differs from a relevant value in oligomers $L_{42}B_{18}$ and $L_{60}B_{36}$ is that in AT-rich domains of the latter there may be a fragment of the primary structure which contains a “bubble code”. According to the results of the experiments described in Section 4.2, the fragments of a heterogeneous DNA with the highest $K_{d,bub}$ do not always have the smallest total $\Delta G^0$ of stacking and H-bonds. It seems that $K_{d,bub}$ can depend greatly on the structural features of a local duplex fragment. In turn, the features are determined by the nucleotide sequence of this fragment. A striking example of such dependence is the behavior of the beacon $A_{18}$, see Section 5.3. Since a separation of $A_{18}$ strands at low $T$ requires lesser extent of the stacking decay than in other beacons the activation barrier of a bubble formation there is rather low. The possibility of successive opening of H-bonds and stacking in the course of the bubble opening breaks the activation barrier of this reaction into several smaller ones. This breaking of the activation barrier is typical for many biochemical reactions. In particular, a similar mechanism plays a great role in enzymatic catalysis [365].

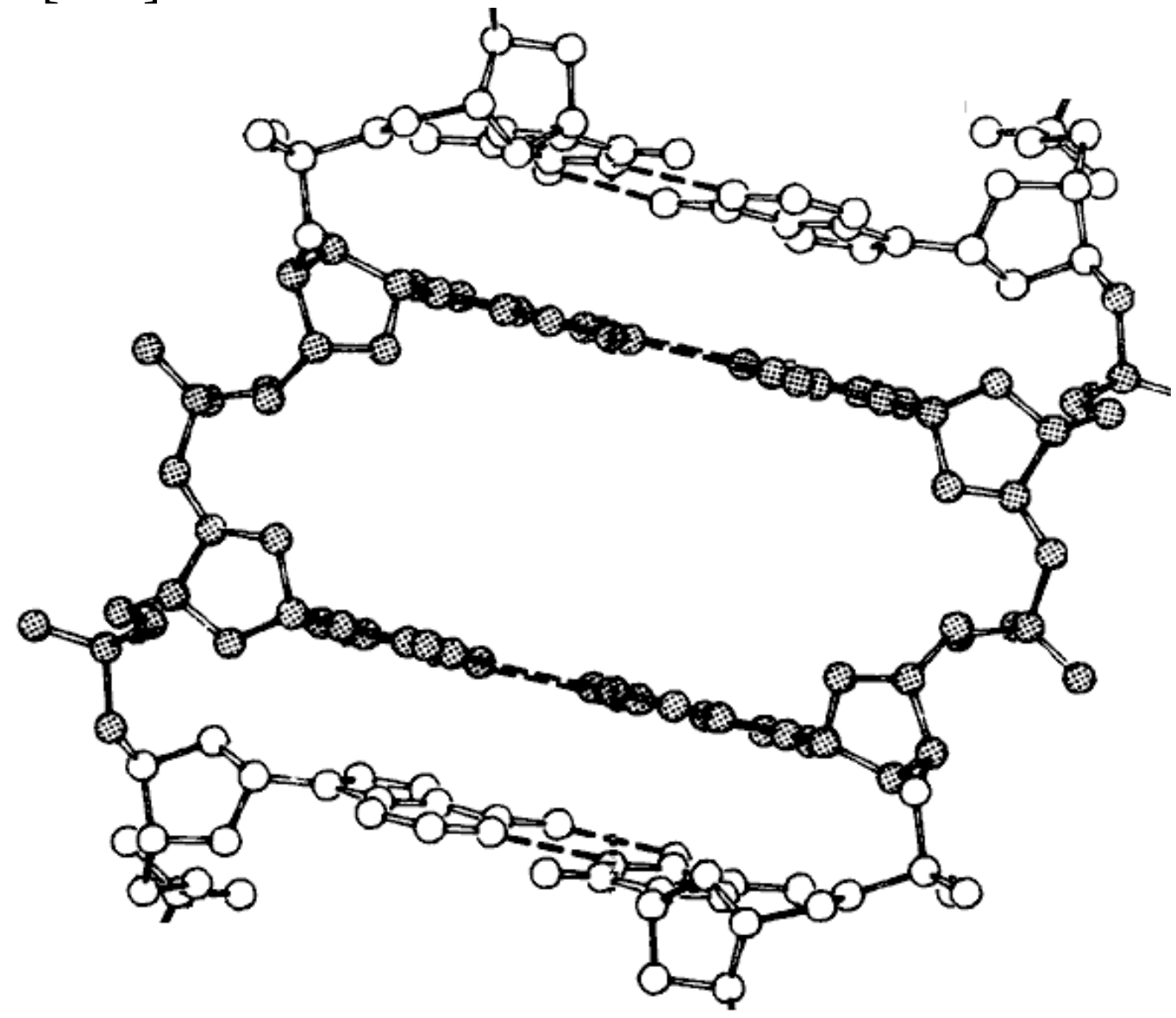

**Fig. 6.3.** Local conformational transition in DNA known as β-premelton. Nucleotide pairs with stacking opened on one side are painted gray.

Apparently, a local context of a primary structure not only determines the form of the opening activation barrier. It seems to influence the the interaction between the angular basepair fluctuations and the stretched oscillations. The above-considered contribution of the fluorescence label dynamics into the kinetics of the beacon opening is a clear (though an "overacted") example of this possibility. In a non-modified DNA even at 50 °C neither $K_d$, nor $K_d'$ exceed $10^{-4}$, therefore the probability of their interaction at the pre-melting temperatures is very low. On the other hand it is possible that in some contexts of the DNA primary structure the angular dynamics of an individual base can greatly affect both $K_d$ and $K_d'$ of the adjacent basepairs and increase $K_{d,bub}$ of the fragment where they resides.

A significant role in the opening dynamics can also belong to local conformation changes which are not open states. In 1983 Sobell and Banerjee described the so-called "β-premeltons" – elongations of DNA which arise as a result of stacking decay between neighboring base pairs [366]. In this case H-bonds are retained completely. The structure of a β-premelton is shown in figure 6.3.

States of this type are partially stabilized by a change in the conformation of the sugar-phosphate backbone [214]. It is possible that $H_2O$ intercalate deeper into the duplex as compared to the native region: the distance between the base pairs in a premelton is quite sufficient even for intercalation of such substances as actinomycin [214, 366].

The existence of conformational transitions which do not lead to a guplex opening is also confirmed by some calculation data. Bouvier and Grubmuller compared the results of their simulations with some features of the opening dynamics which were described in private messages of their colleagues involved in NMR investigations of DNA. According to Bouvier and Grubmuller the states with the following properties can form in a duplex [163]:

1) high activation barrier – approximately 67 kJ/mol, that is several kJ/mol higher than the average $\Delta G^{\dagger 0}$ of a flip-out, see tables 5.1 and 5.2,

2) rather small $\Delta G^0$ varying from 8.3 to 12.5 kJ/mol,

3) inability to exchange $^1$H with the molecules of a solvent. The last property proves rigorously the intactness of H-bonds, see Section 5.4.

Hence, investigation of the "bubble code" problem should be closely related to the study of an interaction of the open states with one another and with other conformation changes as well. The complexity of this problem is not so much a close interrelation of conformation fluctuations as too few statistics of relevant experimental data. It can be said that the investigations of the DNA dynamics are in their early stages as yet though the DNA structure was decoded more than 60 years ago.

## CONCLUSIONS

We discussed investigations of thermodynamic and kinetic properties of the DNA open states which were carried out in the world over the past few decades. The main attention was given to investigations of the duplex opening processes at temperatures much less than melting temperatures $T_m$. Having analyzed some literature data we have shown that under these conditions open states of three types can be formed: denaturation bubbles, flip-outs and hydrogen-bonded water-bridged complexes. Classification into the types was performed according to two conditional criteria. The first criterion is the number of opening basepairs denoted by letter $N$. For a hydrogen-bonded complex or a flip-out, as a rule, $N = 1$. Nucleation of a bubble, on the contrary, is associated with simultaneous opening of H-bonds in 3–4 or more basepairs. The proofs of this fact are described in Section 6.1.

The second criterion is closely related to the first one. This is a character of molecular dynamics trajectory of the opening. Indirect qualitative characteristics of this criterion is activation thermodynamic parameters of the opening, see Section 5.2. As it is evident from table 5.3, $\Delta S^{\dagger 0}$ of the bubble opening is much higher than zero. The activation enthropy of a flip-out averaged over the bases out of A-tracts is, on the contrary, in the vicinity of zero and lower, see tables 5.1 and 5.2. The flip-outs in A-tracts and in the vicinity of the duplex ends are much closer to bubbles as to their activation parametrs. This is due to a possibility of a simultaneuos flip-out in two neighboring base pairs [307]. Besides, the analysis of NMR experiments which we made in Section 5.4, demonstrates that the values of $\Delta S^{\dagger 0}$ in A-tracts are probably overestimated, i.e. the thermodynamic difference is even more distinct.

Since the ranges of the thermodynamic parameters partially overlap the second criterion is less strict than the first one. However just this criterion along with the analysis of end-effects on imino proton exchange in the case of internal catalysis has enabled a flip-out to be separated from a hydrogen-bonded nuH$^*$··$H_2O$··acc complex. The contribution of the duplex half open states into $^1$H exchange processes is considered in detail in Section 5.4. The characteristic features and thermodynamic peculuarities of a hydrogen-bonded water-bridged complex are described in Sections 5.4 and 6.4. The activation enthropy of its formation is much higher than zero – in this respect this complex is similar to a bubble nucleation. Though the experimental material on the open states is very poor, further investigations in this field are very important. In particular, experimental data on the dynamics of hydrogen-bonded water-bridged complexes can make a great contribution into the development of angular mechanical models of DNA. This problem will be discussed below.

A common feature of the theoretical approaches considered here is their level. In these models each basepair is described by a few variables – from one to four. The approaches of this level are divided into Ising-like models and mechanical ones. Each of these groups, in turn, is divided into several subgroups. In what follows we will generalize the role of each group in theoretical investigations of the thermodynamic properties of the DNA open states. Let us start with the Ising-like models.

*Nearest-neighbor models.* The main problem for which the approaches of this group were developed is to investigate the dependence between the fine structure of the melting profiles of a heterogeneous duplex and its nucleotide sequence. At present a phenomenological description of the DNA melting curves is the main field of application of most nearest-neighbor models, see Section 1.3. Another applied problem for which these approaches were designed is to compute the instability profiles of long heterogeneous DNA in order to find promoter sites of the genome. However in this case the computed profiles are far from the picture observed in a real DNA since formation of bubbles is impossible in the models of this subgroup, see Section 6.2. The only exclusion is the model by Kantorovich et al. which deals not with the stability of individual base pairs but just with the probability of simultaneous opening of several adjacent base pairs [272]. Taking account of the condition $N > 1$ which is a key characteristics of a bubble enabled the authors to reproduce considerable heterogeneity of $K_{d,bub}$ in a native DNA which contains promoter sites.

Notwithstanding their phenomenological nature, the nearest-neighbor models made a great contribution into the investigations of the properties of a cooperative transition in DNA in the course of melting. The parameters of these models were determined from experiments which were carried out on short heterogeneous duplexes at temperatures close to $T_m$, see Section 6.3. The length of the oligomers investigated did not exceed 20 base pairs, therefore they could unwind only from the ends without forming a bubble in the middle. Nevertheless the parameters fitted have allowed to describe precisely the melting profiles of DNA consisting of several hundreds of base pairs as well [53]. This fact is important since a high cooperativity in the nearest-neighbor models leads to a difference in the stability of terminal and central regions of the duplex, see Section 6.2. This effect suggests a significant role of end-effects at high temperature and confirms a sharp increase in the opening cooperativity of DNA secondary structure in the vicinity of $T_m$.

At the same time the melting profiles of DNA consisting of more than 1000 base pairs are described by these models much worse than the profiles of shorter duplexes, see fig. 10–18 and 20–22 in the review by Wartell and Benight [53]. One of the reasons may be a strong dependence of the opening probability of each nucleotide pair on the state of the adjacent pairs. Overevaluated cooperativity partially compensates the failure to comply with the condition $N > 1$ preventing too easy opening of the pairs but at the same time prohibit formation of a bubble in the middle of the duplex. One more reason why for long DNA the results are inconsistent with the experiment may be an inaccurate estimate of the contribution of large denaturated fragments. This shortcoming is absent in Ising-like approaches of another subgroup – Poland-Scheraga type (PS-type) models.

*PS-type models.* As distinct from the nearest-neighbor models, the approaches of this subgroup are not phenomenological. They were initially developed to investigate the physical mechanisms of the phase transition which takes place in the course of DNA melting, see Section 1.2. Investigation of the nature of a phase transition in quasi-one-dimensional systems is one of the most important problems of modern theoretical physics. Uniqueness of the PS-type approaches is in their suitability for theoretical investigations of entropy of very large bubbles – “loops”. PS-type models allow studying the dependence of the loop’s specific configurational enthropy on its size, energy, complementary bonds and other factors. Just with the use of PS-type models it was shown that one of the reasons why the heat capacity sharply increases at $T_m$ is that the specific enthropy of the loops sharply decreases upon growth of their size. Besides, it was demonstrated that for a heterogeneous DNA, this effect is less pronounced than for a homogeneous one [107], see Section 1.2. The enthropy contribution of single-stranded fragments into the dynamics of a micromechanical DNA denaturation is also investigated, see Section 3.2.

Hence, Ising-like models made a considerable contribution into investigations of the thermodynamics of the DNA duplex opening at temperatures close to $T_m$. However on the pre-melting interval the properties of the open states can be quite different, see Sections 5.3 and 6.5. The character of activation barriers of a bubble nucleation and cooperativity of its growts/reduction may depend strongly not only on the DNA primary structure but also on *T*. Such effects are very difficult to study in simple models, hardly can they be described by Ising-like models. Much more promising in this respect are radial and radial-torsional mechanical models. Their main advantage is that they allow studying the processes of transfer and nonlinear localization of the basepair stretched oscillations' energy. They are the processes that have a great influence on the open states dynamics. The models of these groups are also suitable from the viewpoint of phenomenological optimization which is aimed at a more adequate description of the temperature effects on the kinetics of a bubble nucleation and cooperativity of its growth/reduction. Not less important is the development of angular approaches which are suitable for modeling the transfer of the energy of angular fluctuations of the bases and its influence on the kinetics of nuH*··$H_2O$··acc complexes, see Section 6.4.

Let us generalize the role of each subgroup of the mechanical approaches in DNA investigations and shape possible prospects of their development.

*Radial and radial-torsional approaches.* One of the most important experimental facts investigated in the models of this type is the slip-and-stick character of transversal force-induced unzipping of DNA, see Section 3.1. This phenomenon was first described in Ising-like models which were studied by methods of equilibrium statistical physics, see Sections 3.2 and 3.3. Nevertheless, a slip-and-stick transition of the system between the minima of the free energy in these approaches takes place only in the case of a heterogeneous DNA. For a homogeneous duplex, the possibility of such behavior was first demonstrated just in a mechanical model, see Section 3.3. The origin of a non-uniform resisting force of a homogeneous DNA during unzipping is a nonlinear localization of the energy in its random fragments which leads to facilitation of their opening. In a heterogeneous DNA the position of unstable fragments depends on the primary structure and localization of the energy should enhance the depth of the energy minima.

Obviously at a qualitative level the slip-and-stick micromechanical unzipping can be reproduced in many mechanical models described in Sections 2.3 and 2.4. In this case the main requirement to the approach is that complementary H-bonds should be described via a nonlinear potential. Even in the early radial models it was shown that this requirement is the main condition for the energy localization [151, 152]. However, as it was demonstrated in the PBD model, not less important is the nonlinearity of the potential which describes stacking interactions. First of all, this condition ($\rho \neq 0$) is the key requirement for adequate description the first order transition both in the radial and the radial-torsional models, see Sections 2.2 and 2.3. Moreover, the anharmonicity of the stacking potential is crucial for an adequate description of the instability profiles of a heterogeneous DNA at moderate temperatures, see Sections 4.2. and 6.2.

As has been shown in Section 4.2, the methods for investigation of mechanical models should also comply with some requirements. Thus, passing on from Langevin dynamics to the study of the equilibrium properties of the PBD model leads to a considerable loss of the agreement with the experimental data. This is clearly seen from the results obtained by Van Erp et al. and other researchers [264, 267–269]. They failed to reproduce either a considerable heterogeneity of the instability profile or the distant effect of mutation observed *in vitro* [29, 30]. The data obtained by de los Santos et al., who used a simplified Langevin approach are also illustrative [367]. In order to save computing time the authors greatly magnified the friction coefficient: the data obtained turned out to be far from the experiments too.

Strictly speaking, high heterogeneity of $K_{d,bub}$ observed in experiments is not adequately described even in the PBD model [29, 30, 212, 273]. The same shortcoming is shared by the modernized PBD model where the stacking potential depends on the nucleotide sequence [274]. Therefore, in our opinion, an effective search for active sites of DNA in simple mechanical models requires their further optimization. Let us outline its possible paths.

The main point of the modernization should be introduction of an additional energy barrier which would hinder opening of minor than $N$ adjacent base pairs. The question of whether the condition of $N > 1$ is necessary is considered in Sections 6.1 and 6.2. As is shown in Section 5.4, the discrepancy between the mean $K_{d,bub}$ obtained in the PBD model and the experimental results cannot be explained by any peculiarities of the techniques. Therefore we should optimize the model phenomenologically by correct enhancing the energy barrier of a bubble nucleation.

Moreover, it makes sense to investigate the dependence of $N$ *per se* on the local ratio between GC- and AT-pairs. As we have mentioned in Section 6.2, a "local cooperativity" can be introduced into a mechanical model through the condition of $N > 1$. At the same time, as it is shown in Section 4.1, a pronounced dependence of the melting cooperativity of DNA areas on their basepair ratio is an experimental fact. However at moderate temperatures the kinetics of the duplex opening depends on its primary structure greater than in the vicinity of $T_m$. Therefore in this modernization one should rely on a great amount of experimental data.

Another way of the models improvement is taking account of flip-outs and hydrogen-bonded water-bridged complexes. The material of Section 5.2 demonstrates noncooperativity of single basepair openings and mutual independence of their kinetics. The data of Section 6.5 indirectly suggest that these processes have an effect on the bubble dynamics at increased temperature. A flip-out can be introduced into radial and radial-torsion models as a short-term decrease of the well depth of the on-site potential which describes complementary H-bonds in a basepair. The easiest way to "control" the flip-outs is through the stochastic function. Its parameters can be derived from the thermodynamic properties of single basepair openings or fitted by comparing with the experiments. This will allow taking into account not only different $K_d$ of AT- and GC-paits, but also the kinetic features of A- and G-tracts as well as the end-effects. Similarly, we can introduce the kinetics of hydrogen-bonded water-bridged complexes into the models.

These ways of modernization are suitable for both radial and radial-torsional models. Among radial models, the PBD variant with heterogeneous sequence-dependent stacking potential is suitable [274]. This model is well developed and is widely used to calculate the instability profiles of native DNA, see Section 4.2. Another example of a promising approach is Toda-Lennard-Jones model described in Section 2.4. The most promising radial-torsion approach is the model by Barbi et al. with an anharmonic stacking potential [179], see Section 2.3.

Since in the radial-torsional models the helical geometry of DNA is taken into account, they have wider possibilities than the radial ones. Evidently, this advantage will retain after the modernization of the approaches of both the groups. For example, the radial models do not allow investigations of thermal fluctuations of supercoiling and the strain relaxation at the ends of a short duplex described in Section 6.1. Another important niche of radial-torsional approaches can be investigation of the influence of a bubble on the dynamics of the regions separated from it by a helix pitch distance [347], see Section 5.3.2. Finally, the models of this group enable one to describe more adequately the duplex distortions in A-tracts or regions of DNA-protein interaction. This would provide better information on the energy localization in distorted DNA regions. The possibility of such effects was first shown in radial approaches, see Section 5.3.3.

On the other hand, radial models allow a much more simple parametrization, and their investigation requires less computational burden. Therefore if DNA is relaxed, it is much more convenient to study its low-temperature dynamics in the models of this group. An apparent advantage of radial models is a possibility of a direct comparison with many experimental data.

Simplicity and convenience of radial models as well as a significant development potential of radial-torsional models can play an important role in the investigations of the "bubble code" problem. However, the angular models described in Section 2.1 are an equally useful tool for investigation of the duplex nonequilibrium dynamics. Distribution of the energy of angular oscillations among the bases is one of the key factors determining the order in which stacking interactions and H-bonds are opened during bubble nucleation, see Section 6.5. Therefore angular approaches are not less actual than radial and radial-torsional ones.

*Angular approaches.* As distinct from the models which involve radial degrees of freedom, angular models do not allow separation of DNA strands. This restricts the area of their application by the low-temperature DNA dynamics. For the models of this group, the problem of comparison of the computational data with the experiment is particularly acute. In our opinion, the best way to solve it is to develop some specific techniques and carry out the corresponding experiments. A promising variant is to investigate the kinetics of imino proton exchange in the absence of an external catalyst in oligonucleotides with a specific primary structure. These are DNA consisting mainly of GC-pairs but having a heterogeneous center which contains AT-pair (pairs).

Duplexes of such a structure are suitable for investigation of how the transfer and localization of the angular displacements energy influences the kinetics of the hydrogen-bonded water-bridged complexes. A possible mechanism of such an influence is described in Section 6.4. The $K_d'$ values of different nucleotide pairs can be determined by quantum chemical calculations. Comparing these data with experiments should allow estimation of the contribution of soliton excitations into the kinetics of nuH*··$H_2O$··acc complexes. Regularities revealed in such investigations can be extremely useful for further development and parametrization of angular models.

*Applicability limits of the models of the level considered.* Though simple theoretical approaches play a great role in the investigations of the DNA dynamics, some specific features of the molecule behavior cannot be described with the models of this level. An example is provided by DNA denaturation under the action of a small force directed along the molecule's axis, see Sections 3.1 and 3.3. Stabilizing local changes in the conformation which take place due to thermal noise, the external force has little effect on H-bonds. However accumulation of a considerable number of conformational changes of the sugar-phosphate backbone leads to cooperative decay of the system of these bonds. To describe such phenomena, models of rather a high level are required.

Another example is provided by basepairs flipping out. Its properties are considered in Chapter 5. This process is described phenomenologically in the nearest-neighbor models [280], however simple approaches do not enable investigations of its physical nature. An escape of a base from the Watson–Crick helix is often hindered by steric limitations. Therefore the flip-out trajectory depends strongly on the context of the primary structure where the base located. So, the trajectory may be very complicated. The main method to investigate the openings of individual nucleotide pairs *in silico* is molecular dynamical calculations [161–163, 368, 369].

Theoretical investigations of the bubble code problem also require approaches of a higher level along with simple models. A total $\Delta G^{\dagger 0}$ of H-bonds and stacking interactions in DNA fragment completely determines its $K_{d,bub}$ only in the vicinity of $T_m$. At moderate temperatures $K_{d,bub}$ depends additionally on a number of factors described in Section 6.5. First, this is a complex interrelation of angular and radial oscillations of the bases. Second, this is the influence of torsional fluctuations on the duplex dynamics. Third, this is the possibility of breaking the activation barrier of a bubble formation into some lower barriers in some contexts of the nucleotide sequence. Fourth, this is a probable contribution of some local changes of the duplex conformation which are not open states.

Hence, the scope of simple approaches and the development prospects of each model are limited particularly by the complexity of the DNA structure. Nevertheless, parallel development of several model groups in complex with new experimental investigations of the duplex dynamics has enormous potential. Comparing the behavior of fundamentally different models in similar conditions we can get much more information on the mechanisms responsible for the DNA duplex opening. A striking example is the "bubble code" problem. Investigation of the optimized models of all the above-described groups, provided their proper parametrization, will considerably reduce the necessity to deal with the models of a higher level. In this case a great role can also belong to the approaches specialized for the study of metastable states in which DNA remains closed.

Generally speaking, any model which enables one to describe at least one characteristic of the duplex dynamics is worth further developing and the results of its investigations deserve experimental testing. The ways of optimizing should be chosen for each model individually. Modification of an approach should not lead to a loss of its main advantage – ability to describe some or other characteristic of the DNA behavior. Therefore the above-discussed ways of modernization are not necessarily appropriate for all the models described in Chapter 2. Comparison of the calculated data with the experiments also requires some caution. The experimental data should be not only up-to-date, but also be as close as possible to the aspect of the DNA behavior which is investigated with the model. In other words, the comparison of a model and an experiment should be as direct as possible. The case of angular models demonstrates that sometimes even designing and carrying out of special experiments is required.

In sum, it can be stated that the variety of simple theoretical and experimental approaches in investigations of the DNA open states in many respects compensates the complexity of the structure and behavior of this molecule. Apparently just simple techniques should play the main role in further investigations of the functional behavior of most biological macromolecules. Collaborative work of research teams at complex development of the simple approaches is an excellent alternative to resource-intensive tasks on the investigation of complex DNA models.

## ACKNOWLEDGEMENTS

One of the authors (A. Sh.) is thankful to V.M. Komarov and N.S. Fialko for valuable consultations and expresses particular gratitude to O.M. Krichevsky for experimental material and to A.V. Teplukhin for constructive criticism of some details in Chapter 5. The authors are thankful to the Russian Foundation for Basic Research for support of this work (projects №13-07-00331, №13-07-00256, №12-07- 00279, №12-07-33006).